\documentclass{jfm}

\usepackage{newtxtext}
\usepackage{newtxmath}
\usepackage{natbib}
\usepackage{hyperref}
\usepackage{comment}
\hypersetup{
    colorlinks = true,
    urlcolor   = blue,
    citecolor  = black,
}

\newcommand{\RomanNumeralCaps}[1]
\linenumbers
\usepackage[usenames, dvipsnames]{xcolor}
\usepackage{subcaption}	% Allows using subcaption inside figure environment

\usepackage{tikz, pgfplots}
\usetikzlibrary{decorations.markings,arrows.meta,calc,bending}
\usepackage{xifthen}

\usepackage{bm}

\definecolor{C0}{HTML}{1F77B4} 
\definecolor{C1}{HTML}{FF7F0E}
\definecolor{C2}{HTML}{2CA02C}
\definecolor{C3}{HTML}{D62728}
\definecolor{C4}{HTML}{9467BD}
\definecolor{C5}{HTML}{8C564B}
\definecolor{C6}{HTML}{E377C2}
\definecolor{C7}{HTML}{7F7F7F}
\definecolor{C8}{HTML}{BCBD22}
\definecolor{C9}{HTML}{17BECF}

\tikzset{
     >={Latex[length=.2cm]}
}
\tikzset{
    mylab/.style={label={[xshift=.03\tw,yshift=+.5em]below:\footnotesize{\textbf{(#1)}}}}
}
\tikzset{
    mylab1/.style 2 args ={label={[xshift=#2,yshift=.5em]below:\footnotesize{\textbf{(#1)}}}}
}
\providecommand\bk[1]{\left(#1\right)}
\providecommand\vv\mathbf

\usepackage{cleveref}
\newcommand{\fig}[1]{\cref{#1}}   % figure at middle of sentence
\newcommand{\Fig}[1]{\Cref{#1}}   % Figure at starting of sentence
     \crefname{figure}{figure}{figures}
\let\ig\includegraphics
\let\tw\textwidth

\providecommand{\Rey}{\mathit{Re}}

\renewcommand{\vec}[1]{\boldsymbol{#1}}
\providecommand{\gvec}[1]{\boldsymbol{#1}}
\providecommand{\mat}[1]{\bm{\mathsf{#1}}}

\providecommand{\eq}[1]{Eq.~\eqref{#1}}
\providecommand{\Eq}[1]{Equation~\eqref{#1}}

\providecommand{\q}{\Psi}             % quantity to of interest
\providecommand{\var}{\Phi}           % variable to reconstruct
\providecommand{\cau}{\var_{I}}   % variable to reconstruct
\providecommand{\res}{\var_{R}}   % variable to reconstruct
\providecommand{\dcau}{\var_{I,\Delta}}   % variable to reconstruct
\providecommand{\dres}{\var_{R,\Delta}}   % variable to reconstruct
\providecommand{\Q}{\gvec{\Psi}}    % quantity to of interest
\providecommand{\Var}{\gvec{\Phi}} % variable to reconstruct
\providecommand{\Cau}{\Var_{I}}   % variable to reconstruct

\providecommand{\Res}{\Var_{R}}   % variable to reconstruct

\providecommand\method{IND}
\providecommand\amethod{aIND}

\providecommand{\chh}{h}            % channel height
\providecommand{\taux}{\tau_x}      % streamwise wall-shear stress fluctuation
\providecommand{\tauxf}{\tau_{x,+}} % future streamwise wall-shear stress fluctuation

\title{Informative and non-informative decomposition of turbulent flow fields}
\author{Gonzalo Arranz\aff{1} \and Adri\'an Lozano-Dur\'an\aff{1,2}}

\affiliation{
  \aff{1}Department of Aeronautics and Astronautics, Massachusetts Institute
  of Technology, Cambridge, MA 02139, USA
  \aff{2}Graduate Aerospace Laboratories, California Institute of Technology, Pasadena, CA 91125, USA
}

\begin{document}

\maketitle

\abstract Not all the information in a turbulent field is relevant for
understanding particular regions or variables in the flow.  Here, we
present a method for decomposing a source field into its informative
$\Cau(\vec{x},t)$ and residual $\Res(\vec{x},t)$ components relative
to another target field. The method is referred to as informative and
non-informative decomposition ({\method}). All the necessary
information for physical understanding, reduced-order modelling, and
control of the target variable is contained in $\Cau(\vec{x},t)$,
whereas $\Res(\vec{x},t)$ offers no substantial utility in these
contexts. The decomposition is formulated as an optimisation problem
that seeks to maximise the time-lagged mutual information of the
informative component with the target variable while minimising the mutual
information with the residual component.
The method is applied to extract the informative and residual
components of the velocity field in a turbulent channel flow, using
the wall-shear stress as the target variable. We demonstrate the
utility of {\method} in three scenarios: (i) physical insight of the
effect of the velocity fluctuations on the wall-shear stress, (ii)
prediction of the wall-shear stress using velocities far from the
wall, and (iii) development of control strategies for drag reduction
in a turbulent channel flow using opposition control.
In case (i), {\method} reveals that the informative velocity related
to wall-shear stress consists of wall-attached high- and low-velocity
streaks, collocated with regions of vertical motions and weak spanwise
velocity. This informative structure is embedded within a larger-scale
streak-roll structure of residual velocity, which bears no information
about the wall-shear stress. In case (ii), the best-performing model
for predicting wall shear stress is a convolutional neural network
that uses the informative component of the velocity as input, while
the residual velocity component provides no predictive
capabilities. Finally, in case (iii), we demonstrate that the
  informative component of the wall-normal velocity is closely linked
  to the observability of the target variable and holds the essential
  information needed to develop successful control strategies.

%%%%%%%%%%%%%%%%%%%%%%%%%%%%%%%%%%%%%%%%%%%%%%%%%%%%%%%%%%%%%%%%%%%%%%%%%%%%%%%%
%% INTRODUCTION
%%%%%%%%%%%%%%%%%%%%%%%%%%%%%%%%%%%%%%%%%%%%%%%%%%%%%%%%%%%%%%%%%%%%%%%%%%%%%%%%
\section{Introduction}\label{sec:intro}

% Intro: Reynolds decomposition and goal
Since the early days of turbulence research, there have been multiple
attempts to decompose the flow into different components to facilitate
its physical understanding, control its behaviour and devise
reduced-order models. One of the earliest examples is the Reynolds
decomposition~\citep{reynolds1895}, which divides the velocity field
into its mean and fluctuating components. More sophisticated
approaches rapidly emerged aiming at extracting the coherent structure
of the flow through correlations and structure
identification~\citep{robinson1991, panton2001, adrian2007, smits2011,
  mckeon2017, jimenez2018}.  This interest is justified by the hope
that insights into the dynamics can be gained by analysing a subset of
the entire flow, while the remaining incoherent flow plays only
  a secondary role in understanding the overall dynamics. In this
work, we introduce a method to decompose turbulent flow fields into
informative and non-informative components, referred to as {\method},
such that the informative component contains all the useful
information for physical understanding, modelling, and control with
respect to a given quantity of interest.

% Coherent structures: local in space
The quest to divide turbulent flows in terms of coherent and
incoherent motions has a long history, tracing back to the work of
\citet{theodorsen1952}, and has been a subject of active research
since the pioneering experimental visualisations of \citet{kline1967}
and the identification of large-scale coherent regions in mixing
layers by \citet{brown1974}. Despite this rich history, the field
still lacks consensus about the definition of a coherent structure due
to the variety of interpretations proposed by different researchers.
One of the initial approaches to distinguish turbulent regions was the
turbulent/nonturbulent discriminator circuits introduced by
\citet{corrsin1954}. Since then, single- and two-point correlations
have become conventional tools for identifying coherent regions within
the flow~\citep[e.g.,][]{sillero2014}. 
The development of more
sophisticated correlation techniques, such as the linear stochastic
estimation~\citep{adrian1988} 
(together with its extensions~\citep{tinney2006,baars2014,encinar2019}), 
and the characteristic-eddy
approach~\citep{moin1989}, has further improved our understanding of
the coherent structure of turbulence.  An alternative set of methods
focuses on decomposing the flow into localised regions where certain
quantities of interest are particularly intense.  The first attempts,
dating back to the 1970s, include the variable-interval time average
method~\citep{blackwelder1976} for obtaining temporal structures of
bursting events and its modified version, the variable-interval space
average method~\citep{kim1985}, for characterising spatial rather than
temporal structures.  With the advent of larger databases and
computational resources, more refined techniques have emerged to
extract three-dimensional, spatially localised flow structures. These
include investigations into regions of rotating fluid~\citep[e.g.,
  vortices][]{moisy2004, alamo2006}, motions carrying most of the
kinetic energy~\citep[e.g., regions of high and low velocity streaks
  by][]{hwang2018, bae2021}, and those responsible for most of the
momentum transfer in wall turbulence~\citep[e.g., quadrant events and
  uniform momentum zones by][]{meinhart1995, adrian2000, lozano2012,
  lozano2014, wallace2016, silva2016}.

% Coherent structures: global in space
The methods described above offer a local-in-space characterisation of
coherent structures, in contrast to the global-in-space modal
decompositions of turbulent flows~\citep{taira2017, taira2020}. One of
the first established global-in-space methods is the proper orthogonal 
decomposition (POD)~\citep{lumley1967}, wherein the flow is decomposed into a 
series of eigenmodes that optimally reconstruct the energy of the field. This
method has evolved in different directions, such as space-only
POD~\citep{sirovich1987}, spectral POD~\citep{towne2018}, and
conditional POD~\citep{schmidt2019}, to name a few. Another popular
approach is dynamic mode decomposition (DMD)~\citep{schmid2010,
  schmid2011}, along with decompositions based on the spectral
analysis of the Koopman operator~\citep{rowley2009,
  mezic2013}. Similar to POD, various modifications of DMD have been
developed, e.g., the extended DMD~\citep{williams2015}, the
multi-resolution DMD~\citep{kutz2016}, and the high-order
DMD~\citep{leclainche2017} [see \citep{schmid2022} for a review]. 
POD and DMD methods do not explicitly account for nonlinear interactions.
To overcome this, extensions to detect quadratic nonlinear interactions based
on the bispectrum have also been developed \citep{baars2014,schmidt2020bi}.
Another noteworthy modal decomposition approach is empirical mode decomposition, 
first proposed by \citet{huang1998} and
recently used in the field of fluid
mechanics~\citep[e.g.,][]{cheng2019}. While the methods above are
purely data-driven, other modal decompositions, such as resolvent
analysis and input-output analysis, are grounded in the linearised
Navier-Stokes equations~\citep{trefethen1993, jovanovic2005,
  mckeon2010}. It has been shown that POD, DMD, and resolvent analysis
are equivalent under certain conditions~\citep{towne2018}. Recently,
machine learning has opened new opportunities for nonlinear modal
decompositions of turbulent flows~\citep{brunton2019}.

% limitations
% -- are there more limitations?
The flow decomposition approaches presented above, either local or
global in space, have greatly contributed to advancing our knowledge
about the coherent structure of turbulence. Nonetheless, there are
still open questions, especially regarding the dynamics of turbulence,
that cannot be easily answered by current methodologies. Part of these
limitations stem from the linearity of most methods, yet turbulence is
a nonlinear system. A more salient issue perhaps lies in the fact that
current methods (with exceptions, such as the extended 
POD~\citep{boree2003}) tend to focus on decomposing source variables without
accounting for other target variables of interest.  In general, it is
expected that different target variables would require different
decomposition approaches of the source variable. For example, we might
be interested in a decomposition of the velocity that is useful for
understanding the wall-shear stress. Hence, the viewpoint adopted here
aims at answering the question: What part of the flow is relevant to
understanding the dynamics of another variable?  In this context,
coherent structures are defined as those containing the useful
\emph{information} needed to understand the evolution of a target
variable.

% information theory
The concept of information alluded above refers to the Shannon
information~\citep{shannon1948, cover2006}, i.e., the average
unpredictability in a random variable. The systematic use of
information-theoretic tools for causality, modelling, and control in
fluid mechanics has been recently discussed by
\citet{lozanoduran2022}. \citet{betchov1964} was one of the first
authors to propose an information-theoretic metric to quantify the
complexity of turbulence. Some works have leveraged Shannon
information to analyse different aspects of two-dimensional turbulence
and energy cascade models~\citep{cerbus2013, materassi2014,
  granero-belinchon_thesis, shavit2020, lee2021,
  tanogami2024}. Information theory has also been used for causal
inference in turbulent flows~\citep{liang2016, lozano2019b, wang2021,
  lozanoduran2022, martinez2023}, and reduced-order
modelling~\citep{lozano2019b}. The reader is referred to
\citet{lozanoduran2022} for a more detailed account of the
applications of information-theoretic tools in fluid mechanics.

% outline
This work is organised as follows: The formulation of the flow
decomposition into informative and non-informative components is
introduced in Section~\ref{sec:formulation}: we first discuss
the exact formulation of {\method} in Sections~\ref{sec:met:genformulation} 
and \ref{sec:met:target}, followed by its numerically tractable approximation,
{\amethod}, in Section~\ref{sec:met:approx}.
Section~\ref{sec:results} demonstrates the application of the method
to the decomposition of the velocity field, using wall-shear stress in
a turbulent channel flow as the target variable. This decomposition is
leveraged for physical understanding, prediction of the wall-shear
stress using velocities away from the wall via convolutional neural
networks, and drag reduction through opposition control. Finally,
conclusions are presented in Section~\ref{sec:conclusions}.

%%%%%%%%%%%%%%%%%%%%%%%%%%%%%%%%%%%%%%%%%%%%%%%%%%%%%%%%%%%%%%%%%%%%%%%%%%%%%%%%
%% METHODOLOGY
%%%%%%%%%%%%%%%%%%%%%%%%%%%%%%%%%%%%%%%%%%%%%%%%%%%%%%%%%%%%%%%%%%%%%%%%%%%%%%%%
\section{Methodology}\label{sec:formulation}

%%%%%%%%%%%%%%%%%%%%%%%%%%%%%%%%%%%%%%%%%%%%%%%%%%%%%%%%%%%%%%%%%%%%%%%%%%%%%%%%
%% METHODOLOGY - Decomposition of the source variable
%%%%%%%%%%%%%%%%%%%%%%%%%%%%%%%%%%%%%%%%%%%%%%%%%%%%%%%%%%%%%%%%%%%%%%%%%%%%%%%%
\subsection{{\method} of the source variable}\label{sec:met:genformulation}

\begin{figure}
    \centering
    \ig[width=\tw]{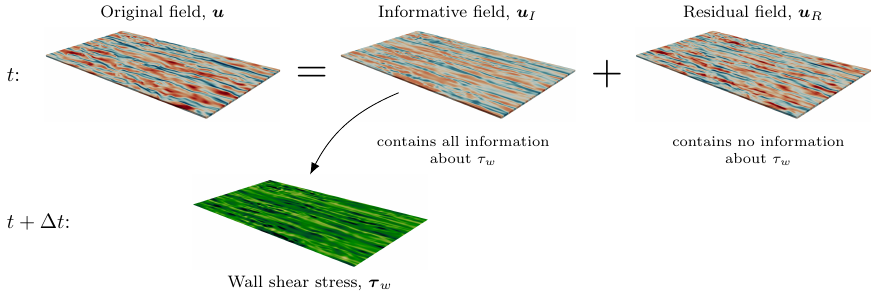}
    \caption{Schematic of {\method} applied to wall-bounded turbulent
      flow.  The source variable is the velocity fluctuations in the
      fluid volume $\vec{x}\in\Omega_{\vec{u}}$ at $t$, and the target
      variable is the wall shear stress vector at the wall at
      $t+\Delta t$.  For the sake of visualisation, only the streamwise
      component of the velocity fluctuations and the wall shear stress
      is displayed.  The velocity fluctuations at time $t$ are
      decomposed into their informative and residual components to the
      wall shear stress in the future, $t + \Delta t$.\label{fig:method}}
\end{figure}

% notation
Let us denote the \emph{source variable} by $\Var(\vec{x},t)$ with
$\vec{x} \in \Omega_{\Var}$ and the \emph{target variable} by
$\Q(\vec{x},t)$ with $\vec{x} \in \Omega_{\Q}$ where $\vec{x}$ and $t$
represent the spatial and time coordinates, respectively.  For
example, in the case of a turbulent channel flow, the source variable
could be the velocity fluctuations defined over the entire domain,
$\Var(\vec{x},t) = \vec{u}(\vec{x},t)$, and the target variable could
be the shear stress vector at every point over one of the walls,
$\Q(\vec{x},t) = \vec{\tau}_w(\vec{x},t)$, as shown in \fig{fig:method}.
We seek to decompose $\Var(\vec{x},t)$ into two independent
contributions: an \emph{informative contribution} to the target variable in the 
future, $\Q_+ = \Q(\vec{x},t+\Delta T)$ with $\Delta T \geq 0$, and a
\emph{residual term} that conveys no information about $\Q_+$ (i.e.,
the non-informative component):
\begin{equation}\label{eq:VarCauRes}
  \Var(\vec{x},t) = \Cau(\vec{x},t) + \Res(\vec{x},t),
\end{equation}
where $\Cau$ and $\Res$ are the informative and residual
contributions, respectively. The decomposition is referred to as
Informative/Non-informative Decomposition or {\method}.
 
% definition of informative
To find a decomposition of the form shown in \eq{eq:VarCauRes}, we
need to introduce a definition of \emph{information}. We rely on the
concept of Shannon information~\citep{shannon1948}, which quantifies
the average information in the variable $\Q_+$ as
\begin{equation}\label{eq:H}
        H(\Q_+) = - \sum_{\vec{S} \in \mathcal{S}} p_{\Q_+} (\Q_+ = \vec{S})
        \log p_{\Q_+}( \Q_+ = \vec{S} )  \geq 0,
\end{equation}
where $H(\Q_+)$ is referred to as the Shannon entropy or information
of $\Q_+$, $p_{\Q_+}(\Q_+ = \vec{S})$ denotes the probability of
$\Q_+$ being in the state $\vec{S}$, and $\mathcal{S}$ represents the
set of all possible states of $\Q_+$.  The remaining information in
$\Q_+$, after discounting for the information in $\Var$, is measured
by the conditional Shannon information:
\begin{equation}\label{eq:Hcond}
	H(\Q_+|\Var) = - \sum_{\vec{S} \in \mathcal{S}} \sum_{\vec{R} \in \mathcal{R}}
	p_{\Q_+,\Var}(\vec{S},\vec{R}) \log
    \frac{p_{\Q_+,\Var}(\vec{S},\vec{R})}{p_{\Var}(\vec{R})} \geq 0,
\end{equation}
where $p_{\Q_+,\Var}$ is the joint probability distribution of $\Q_+$
and $\Var$, $\vec{R}$ is a particular state of $\Var$, and
$\mathcal{R}$ is the set of all possible states of $\Var$.
The difference between \eq{eq:H} and \eq{eq:Hcond} quantifies the
amount of shared information between the variables
\begin{equation}\label{eq:Tflux}
	I(\Q_+; \Var) = H( \Q_+ ) - H( \Q_+ | \Var ),
\end{equation}
and is referred to as the mutual information between $\Q_+$ and
$\Var$.  The condition $H(\Q_+) \geq H(\Q_+|\Var)$ --known as
\emph{information can't hurt}~\citep{cover2006}-- guarantees that 
$I(\Q_+; \Var)$ is always non-negative. The mutual information is
equal to 0 only when the variables are independent, i.e.,
$p_{\Q_+,\Var}(\vec{S},\vec{R}) = p_{\Q_+}(\vec{S}) p_{\Var}(\vec{R})$
for all possible states $\vec{S} \in \mathcal{S}$ and $\vec{R} \in
\mathcal{R}$.

% conditions for optimization problem
We are now in a position to define the conditions that $\Cau$ and
$\Res$ must satisfy.  
First, the informative contribution should maximise $I(\Q_+;\Cau)$
from \eq{eq:Tflux}, which is achieved when
\begin{equation}\label{eq:cond2}
  I(\Q_+ ; \Cau) = H(\Q_+),
\end{equation}
namely, $\Cau$ contains all the information in $\Q_+$. \Eq{eq:cond2} can be
rewritten using \eq{eq:Tflux} as 
$$H(\Q_+|\Cau) = 0,$$
which is mathematically equivalent to expressing $\Q_+$ as a function of $\Cau$, 
namely, $\Q_+ = \boldsymbol{\mathcal{F}}(\Cau)$.
Secondly, the residual term, $\Res$, and the informative term, $\Cau$, 
must be independent, which requires
\begin{equation}\label{eq:cond1}
    I(\Res;\Cau)=0.
\end{equation}
This also ensures that the residual component has no information 
about $\Q_+$, namely $I(\Res;\Q_+) = 0$, since $I(\Res;\Q_+) <
I(\Res;\Cau)$.  The previous inequality is known as the
\emph{data-processing inequality}, and states that no transformation
of a variable can increase its information content, which can only
remain the same or decrease~\citep[Theorem 2.8.1]{cover2006}.
In addition, since $\Res$ and $\Cau$ are statistically independent
from \eq{eq:cond1}, the equality
\begin{equation}\label{eq:energy}
	%\|\Var\|^2 = \|\Cau^\text{opt}\|^2 + \|\Res^\text{opt}\|^2.
	\|\Var\|^2 = \|\Cau\|^2 + \|\Res\|^2,
\end{equation}
is satisfied.
If $\Var$ contains no information about $\Q_+$, then
$\|\Cau\|^2/\|\Var\|^2 \simeq0$ and $\|\Res\|^2/\|\Var\|^2 \simeq 1$.
Conversely, if $\Var$ exclusively contains all the information
necessary to understand $\Q_+$, then $\|\Cau\|^2/\|\Var\|^2 = 1$.
Note that, in general, $\Cau$, $\Res$ and $\boldsymbol{\mathcal{F}}$
are functions of $\Delta T$, which has been omitted here for the sake
of simplicity in the notation.

Since the Shannon information is based on the joint probability
distribution of the variables, rather than their specific values,
there may exist many functions that satisfy
Eqs.~\eqref{eq:cond2} and \eqref{eq:cond1}.  To identify a unique
solution, we impose that the informative field $\Cau(\vec{x},t)$
  is smooth.  Note that, assuming $\Var(\vec{x},t)$ is smooth, the previous 
  condition also implies that the residual field must be smooth. 
 
In summary, the necessary conditions that {\method} satisfies are:
\begin{itemize}
  \item The source variable is decomposed as the sum of the
    informative and the residual contributions: $\Var = \Cau +
    \Res$~\eqref{eq:VarCauRes}.
  \item The informative field contains all the information about the target
    variable in the future: $I(\Q_+ ; \Cau) = H(\Q_+)$~\eqref{eq:cond2}.
  \item The informative and residual components share no information:
    $I(\Res ; \Cau) = 0$~\eqref{eq:cond1}.
  \item The informative field is smooth.
\end{itemize}

%%%%%%%%%%%%%%%%%%%%%%%%%%%%%%%%%%%%%%%%%%%%%%%%%%%%%%%%%%%%%%%%%%%%%%%%%%%%%%%%
%% METHODOLOGY - Decomposition of the target variable
\subsection{{\method} of the target variable}
\label{sec:met:target}

Alternatively, we can seek to decompose the target variable as $\Q =
\Q_{I} + \Q_{R}$, where $\Q_{I}$ and $\Q_{R}$ are, respectively, the
informative and residual components of $\Q$ with respect to $\Var_{-}
= \Var(\vec{x}, t - \Delta T)$, with $\Delta T > 0$. 
% The problem is formulated as
%
The constraints to be satisfied are:
\begin{align*}\label{eq:problem_target}
  I( \Var_{-};\Q_{I} ) &= H(\Var_{-}), & I(\Q_R;\Q_I)&=0,
\end{align*}
together with the smoothness of $\Q_{I}$.
In this case, $\Q_I$ corresponds to the part of $\Q$ that can explain
the source variable $\Var$ in the past, while $\Q_R$ is the remaining
term, which is agnostic to the information in the source variable.

%%%%%%%%%%%%%%%%%%%%%%%%%%%%%%%%%%%%%%%%%%%%%%%%%%%%%%%%%%%%%%%%%%%%%%%%%%%%%%%%
%% METHODOLOGY - Decomposition of the target variable
\subsection{Approximate {\method}}
\label{sec:met:approx}

We frame the conditions of {\method} described in \S\ref{sec:met:genformulation} 
as a minimisation problem.
To that end, several assumptions are adopted.
First, Eqs.~\eqref{eq:cond2} and \eqref{eq:cond1} require calculating high-dimensional 
joint probability distributions, which might be impractical due to limited
data and computational resources. 
The curse of high-dimensionality
comes from both the high dimensionality of $\Var$ and $\Q$ and the
large number of points in $\vec{x}$.  To make the problem tractable,
we introduce the approximate {\method} or {\amethod} for short.
First, the source and target variables are restricted to be scalars,
$\var$ and $\q$, respectively. Second, we consider only two points in
space: $\var(\vec{x}, t)$ and $\q_+(\vec{x}-\Delta \vec{x}, t + \Delta
T)$, where $\vec{x}$ and $\Delta \vec{x}$ are fixed. This reduces the
problem to the computation of two-dimensional joint probability
distributions, which is trivially affordable in most cases,
even enabling the use of experimental data.

% I( ; ) as penalization factor
Another difficulty arises from the constraint in \eq{eq:cond1}, which
depends on the unknown probability distribution of the variable
$\res = \var - \cau$, which adds to the complexity of the optimization problem.
To alleviate this issue, we seek to minimise $I(\res; \cau)$ 
rather than include it as a hard constraint. 

Finally, provided that $\var$ and $\q_+$ are smooth, minimising 
$\| \var - \cau \|^2$ ensures that $\cau$ is smooth too.
Therefore, we include the mean square error as a penalisation term in the
minimisation problem.
Thus, the formulation of the {\amethod} is posed as
\begin{equation}\label{eq:problem_scalar}
	%\hat{\var}_I := 
    %\arg \min_{\cau} \quad \| \var - \cau \|^2 + \gamma
  \arg \min_{\cau,\mathcal{F}} \quad I(\res;\cau) + \gamma\| \var - \cau
  \|^2
        \quad \mathrm{s.t.}\quad \q_{+} =
        \mathcal{F}(\cau), %\ \Delta \vec{x} = \Delta \vec{x}^{\max},
\end{equation}
where $\gamma \geq 0$ is a regularisation constant and $\res = \var - \cau$.
\Eq{eq:problem_scalar} is solved by assuming that the mapping
$\mathcal{F}$ is invertible over a given interval. This allows replacing $\cau(t)
= \mathcal{F}^{-1}(\q_+(t))$ over that interval in \eq{eq:problem_scalar} and
solving for $\mathcal{F}^{-1}$ using standard optimization techniques.
More details about the solution of \eq{eq:problem_scalar} are provided in
Appendix~\ref{app:imp:bijective}.
\Eq{eq:problem_scalar} yields the informative and residual components
for a given $\vec{x}$, $\Delta \vec{x}$, and $t$, denoted as
$\dcau(\vec{x},t; \Delta \vec{x})$ and $\dres(\vec{x},t; \Delta
\vec{x})$, together with the mapping $\mathcal{F}$.  
We can find the best approximation
to {\method} by selecting the value of $\Delta \vec{x}$ that maximises
the informative component. To that end, we introduce the relative
energy of $\dcau$ as
\begin{equation}
    E_I(\Delta \vec{x}; \vec{x}, \Delta T ) =  \frac{\|\dcau\|^2}{\|\var\|^2}.
\end{equation}
High values of $E_I$ define the \emph{informative region} of $\dcau$
over $\q_+$ and constitute the information-theoretic generalisation of
the two-point linear correlation (see Appendix~\ref{app:gaus}).  We
define $\Delta \vec{x}^{\max}$ as the shift $\Delta \vec{x}$ that
maximises $E_I$ for a given $\vec{x}$ and $\Delta T$.  Hence, we use
$\Delta \vec{x} = \Delta \vec{x}^{\max}$ for {\amethod} and simply
refer to the variables in this case as $\cau$ and $\res$.
During the optimisation, we ensure that $2I(\cau;\res) < 0.03H(\cau,\res)$ 
to guarantee that $\cau$ and $\res$ are independent, and that \eq{eq:energy} 
holds.
We also assess \emph{a posteriori} that $I(\res; \q_+)$ remains small for all 
$\vec{x}$ (see Appendix~\ref{app:validation_tauw}).

Finally, we list below the main simplifications of aIND with
  respect to the general IND framework:
\begin{itemize}
    \item The source and the target variable are restricted to be scalars.
    \item The constraint in \eq{eq:cond1} is cast as the minimisation term in \eq{eq:problem_scalar}.
    \item The minimisation problem in \eq{eq:problem_scalar} is
      computed for two points in space. The closest approximation to
      IND is achieved by selecting the value of $\Delta \vec{x}$ that
      maximises the magnitude of the informative component.
    \item \eq{eq:problem_scalar} is solved by assuming that the
      mapping $\mathcal{F}$ is invertible over a given interval.
\end{itemize}
Despite the simplifications above, aIND still
successfully recovers the exact analytical solution in the
validation cases presented in Appendix~\ref{app:validation}, even
outperforming correlation-based methods such as LSE and EPOD.

%%%%%%%%%%%%%%%%%%%%%%%%%%%%%%%%%%%%%%%%%%%%%%%%%%%%%%%%%%%%%%%%%%%%%%%%%%%%%%%%
%% VALIDATION
%%%%%%%%%%%%%%%%%%%%%%%%%%%%%%%%%%%%%%%%%%%%%%%%%%%%%%%%%%%%%%%%%%%%%%%%%%%%%%%%
\subsection{Validation}\label{subsec:validation}

% intro
The methodology presented in \S\ref{sec:met:genformulation} and its
numerical implementation (Appendix~\ref{app:imp:bijective}) have been
validated with several analytical examples.  In this section, we
discuss one of these examples that also illustrates the use and
interpretation of the {\method}.

\begin{figure}
    \centering
      \begin{subfigure}{.65\tw}
        \begin{tikzpicture}
            \begin{footnotesize}
            \node[anchor=south west] (fig) at (0,0)
            {\ig[width=\tw]{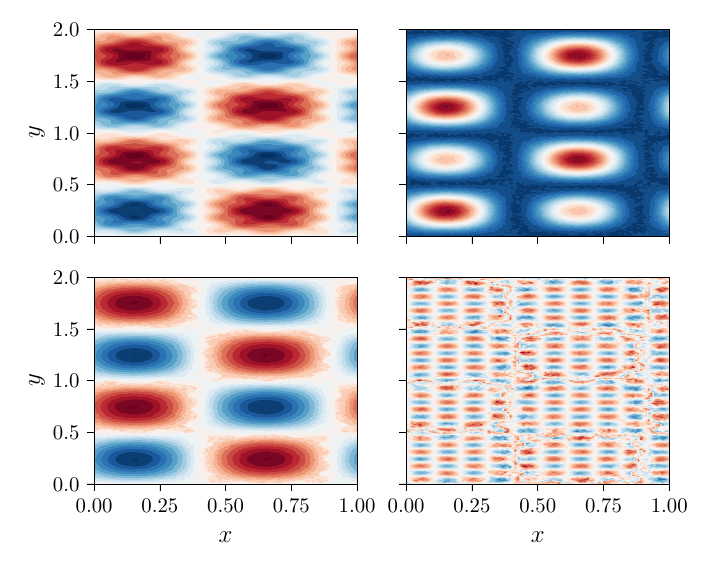}};
            \begin{scope}[x=(fig.south east), y=(fig.north west)]
                \node[fill=white,opacity=.1,text opacity=1.] at (.46,.9) {(a)}; 
                \node[white] at (.90,.9) {(b)}; 
                \node[fill=white] at (.46,.47) {(c)}; 
                \node[fill=white] at (.90,.47) {(d)}; 
                \node[fill=white] at (.33,.06) {$x$}; 
                \node[fill=white] at (.76,.06) {$x$}; 
                \node[fill=white,rotate=90] at (.06,.75) {$y$}; 
                \node[fill=white,rotate=90] at (.06,.33) {$y$}; 
                \phantomcaption\label{fig:valsou}
                \phantomcaption\label{fig:valtar}
                \phantomcaption\label{fig:valcau}
                \phantomcaption\label{fig:valres}
            \end{scope}
            \end{footnotesize}
        \end{tikzpicture}
    \end{subfigure}~
    %    \begin{subfigure}{.35\tw}
        \begin{subfigure}{.37\tw}
        \begin{tikzpicture}
            \node[anchor=south west] (fig) at (0,0) 
            {\ig[width=\tw]{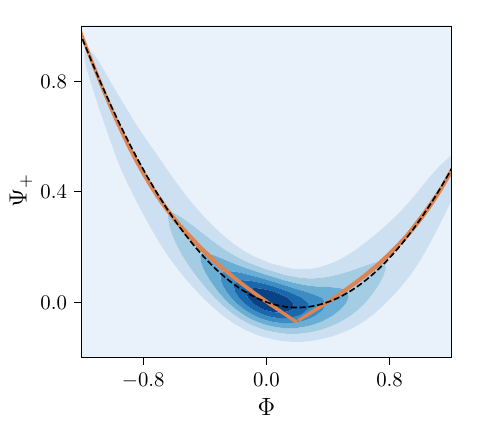}};
            \begin{scope}[x=(fig.south east), y=(fig.north west)]
                \node at (.85,.85) {(e)}; 
                \node[fill=white] at (.57,.06) {$\var$}; 
                \node[fill=white,rotate=90] at (.05,.55) {$\q_+$}; 
            \end{scope}
            \draw[white] (0,-1) circle (.0001);
        \end{tikzpicture}
        \phantomcaption\label{fig:valfun}
    \end{subfigure}
	\caption{Validation of {\amethod} for the system in
          \eq{eq:val1}. (a,b,c,d) Snapshots of $\var$, $\q_+$,
          $\var_I$ and $\var_R$, respectively. The contours range from 
          the minimum value (dark blue) to the maximum value (dark red) for each
          quantity; these correspond to $\var,\var_I \approx [-2, 2]$,
          $\q_+ \approx [-1.1,1.4]$, and $\var_R \approx [-0.35, 0.35]$.
          (e) Contours of the joint probability $(\var,\q_+)$ from
          (white) lower to (blue) higher probability.  Analytical
          solution $\mathcal{F}^\text{exact}(\cau) = 0.5\cau^2 -
          0.2\cau$ (dashed black) and numerical solution
          $\mathcal{F}(\var_I)$ (orange).\label{fig:val}}
\end{figure}
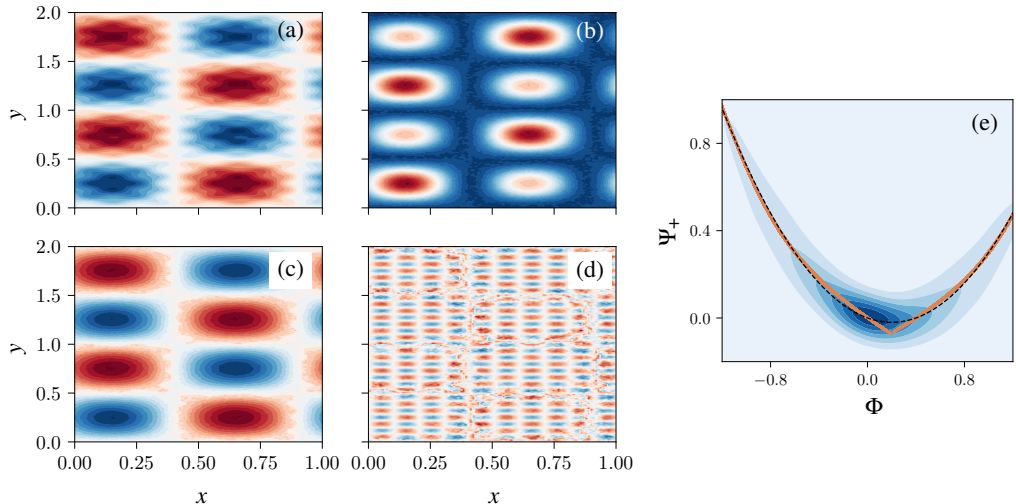

Consider the source and target fields:
\begin{align}\label{eq:val1}
    \text{source:}& \quad \var(\vec{x},t) = f(\vec{x},t) + g(\vec{x},t),\\
    \text{target:}& \quad \q_+(\vec{x},t) = \q(\vec{x},t+1) = 
    0.5 f(\vec{x},t)^2 - 0.2f(\vec{x},t)  + \epsilon(\vec{x},t),
\end{align}
where
\providecommand{\bk}[1]{\left(#1\right)}
\begin{align*}
    f(\vec{x},t) &= 2 \sin\bk{ 2 \pi x - 2 t }\sin\bk{ 2\pi y }, \\
    g(\vec{x},t) &= \frac{1}{5} \sin({ 7 \sqrt{2} \pi x - 0.1 t })
    \sin({ 8\sqrt{3} \pi y - 0.5 t}).
\end{align*}
The source field is a combination of the streamwise travelling wave,
$f$, and the lower amplitude, higher wavenumber travelling wave,
$g$. The target is a function of $f$ and $\epsilon$, where the
latter is a random variable that follows the pointwise normal
distribution with zero mean and standard deviation ($\sigma$)
equal to 0.1: $\epsilon(\vec{x},t) \sim
\mathcal{N}(0,\sigma)$. Snapshots of $\var$ and $\q$ are shown in
\fig{fig:valsou,fig:valtar}, respectively.

% exact solution
For $\Delta T = 1$ and values of $\sigma \to 0$, the analytical
solution of the {\method} is
\begin{equation}
    \cau^\text{exact} = f, \quad \res^\text{exact} = g,
\end{equation}  
where the mapping to comply with $H(\q_+ | \cau^\text{exact}) = 0$ is
$\mathcal{F}^\text{exact}(\cau) = 0.5\cau^2 - 0.2\cau$, and the
residual term satisfies the condition
$I(\cau^\text{exact};\res^\text{exact}) = 0$, since the variables are
independent.

% results
The results of solving the optimisation problem using {\amethod},
denoted by $\var_I$, $\var_R$, and $\mathcal{F}$ are displayed in
\fig{fig:valcau,fig:valres,fig:valfun}.  It can be observed that
$\var_I$ approximates well the travelling wave represented by
$\var_I^\text{exact}=f$.  The small differences between $\var_I$ and
$\var_I^\text{exact}$, also appreciable in $\var_R$, are localised at
values of $f \approx 0.2$ and can be explained by the small
discrepancies between $\mathcal{F}$ and $\mathcal{F}^\text{exact}$ at
the inflection point as seen in \fig{fig:valfun}.  These are
mostly a consequence of $\epsilon$ and the numerical implementation
(see Appendix~\ref{app:imp:bijective}), and they diminish as
$\sigma\rightarrow 0$.
Additional validation cases, together with a comparison of aIND with EPOD 
and LSE, can be found in Appendix~\ref{app:validation}.

%%%%%%%%%%%%%%%%%%%%%%%%%%%%%%%%%%%%%%%%%%%%%%%%%%%%%%%%%%%%%%%%%%%%%%%%%%%%%%%%%%
% RESULTS
%%%%%%%%%%%%%%%%%%%%%%%%%%%%%%%%%%%%%%%%%%%%%%%%%%%%%%%%%%%%%%%%%%%%%%%%%%%%%%%%%%
\section{Results}\label{sec:results}

% General setup
We study the {\amethod} of the streamwise ($u$), wall-normal ($v$) and
spanwise ($w$) velocity fluctuations in a turbulent channel flow using
as target the streamwise component of the shear stress at the wall,
$\taux(x,z,t) = \rho\nu \partial U(x,0,z,t) /\partial y$, 
where $\rho$ is the fluid density, $\nu$ is the kinematic viscosity, 
$U$ is the instantaneous streamwise velocity
and $x$, $y$ and $z$ are the streamwise, wall-normal, and spanwise
directions, respectively. The wall is located at $y=0$. The data are
obtained from direct numerical simulation in a computational domain of
size $8 \pi \chh \times 2 \chh \times 4 \pi \chh$ in the streamwise,
wall-normal, and spanwise directions, respectively, where $\chh$
represents the channel half-height. The flow is driven by a constant
mass flux imposed in the streamwise direction. The Reynolds number,
based on the friction velocity $u_\tau$, is $\Rey_\tau = u_\tau \chh /
\nu \approx 180$. Viscous units, defined in terms of $\nu$ and
$u_\tau$, are denoted by superscript $*$. The time step is fixed at
$\Delta t^* = 5 \cdot 10^{-3}$, and snapshots are stored every $\Delta
t_s^* = 0.5$. A description of the numerical solver and computational
details can be found in \citet{lozano2020}.

% aIND
The source and target variables for {\amethod} are
\begin{align}\label{eq:ST_tau}
  \text{source}:& \ u(\vec{x},t), \ v(\vec{x},t) \ \text{or} \ w(\vec{x},t), \\
  \text{target}:& \ \tauxf = \taux(x-\Delta x^{\max}_\square,z - \Delta z^{\max}_\square,t+\Delta T),
\end{align}
where $\square=u$, $v$ or $w$.  The {\amethod} gives
\begin{align}
  u(\vec{x},t) &= u_I(\vec{x},t) + u_R(\vec{x},t), \\
  v(\vec{x},t) &= v_I(\vec{x},t) + v_R(\vec{x},t), \\
  w(\vec{x},t) &= w_I(\vec{x},t) + w_R(\vec{x},t), 
\end{align}
where the informative and residual components are also a function of
$\Delta T$. 
We focus our analysis on $\Delta T^* \approx 25$ unless
otherwise specified. This value corresponds to the time shift at which
$H(\tauxf | \taux )/H(\tauxf) \lesssim 0.03$, meaning
that $\tauxf$ shares no significant information with its past.
For $\Delta T^* > 25$, the value of $H(\tauxf(\Delta T) | \taux )$ gradually
diminishes towards 0 asymptotically.
This value is similar to
the one reported by \citet{zaki2021}, who found using adjoint methods
that wall observations at $\Delta T^* \approx 20$ are the most
sensitive to upstream and near-wall velocity perturbations.  The shift
$\Delta \vec{x}_\square^{\max} = [\Delta x^{\max}_\square,\Delta
  z^{\max}_\square]$ for $\square=u$, $v$ or $w$ is computed by a
parametric sweep performed in Appendix~\ref{sec:optDx}.  Their values
are a function of $y$, but can be roughly approximated by $\Delta
\vec{x}_u^{\max}/\chh \approx [-1, 0]$, $\Delta \vec{x}_v^{\max}/\chh
\approx [-1.2, 0]$ and $\Delta \vec{x}_w^{\max}/\chh \approx [-0.8,
  \pm0.15]$.  Due to the homogeneity and statistical stationarity of
the flow, the mapping $\mathcal{F}$ is only a function of $y$ and
$\Delta T$. The validity of the approximations made in the {\amethod}
is discussed in Appendix~\ref{app:validation_tauw}, where it is shown
that the residual component of $u$ contains almost no information
about the future wall-shear stress.
For the interested reader, we also include the relative energy field,
$E_I(\Delta \vec{x}; \vec{x}, \Delta T^* = 25)$, of the three velocity
components in Appendix~\ref{sec:optDx}.

%%%%%%%%%%%%%%%%%%%%%%%%%%%%%%%%%%%%%%%%%%%%%%%%%%%%%%%%%%%%%%%%%%%%%%%%%%%%%%%%
%% RESULTS - CAUSALITY WALL SHEAR - AVERAGE FIELD
%%%%%%%%%%%%%%%%%%%%%%%%%%%%%%%%%%%%%%%%%%%%%%%%%%%%%%%%%%%%%%%%%%%%%%%%%%%%%%%%
\subsection{Coherent structure of the informative and residual components of ${\bf u}$ to $\taux$}
\label{sec:res:coherent}

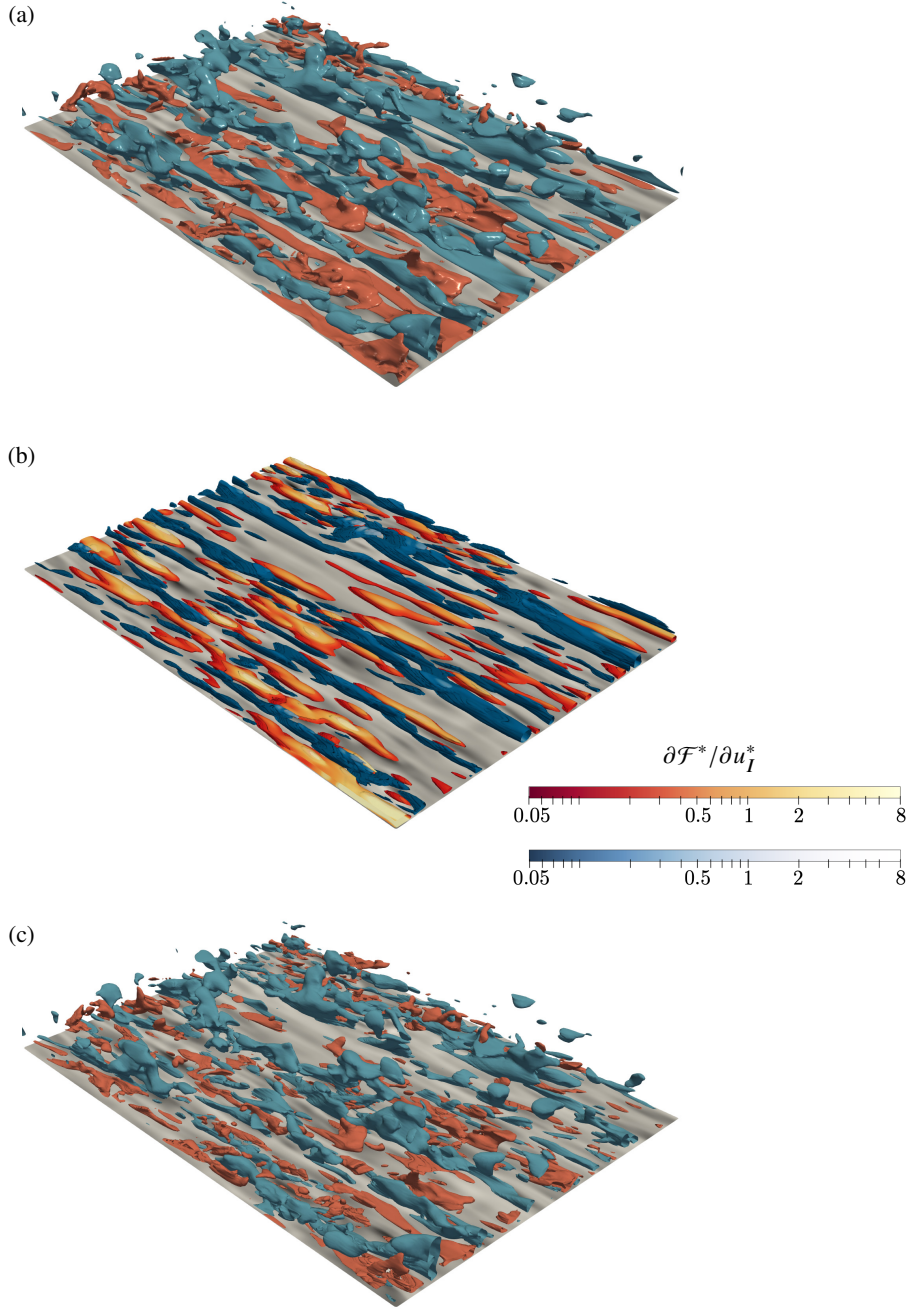
\begin{figure}
    \centering
    \begin{subfigure}{\tw}
        \begin{tikzpicture}
        \begin{footnotesize}
        \node[inner sep=0pt] (fig) at (0,0) 
            {\ig[width=.65\tw,trim=4.5cm 13cm 5cm 8cm,clip] 
            {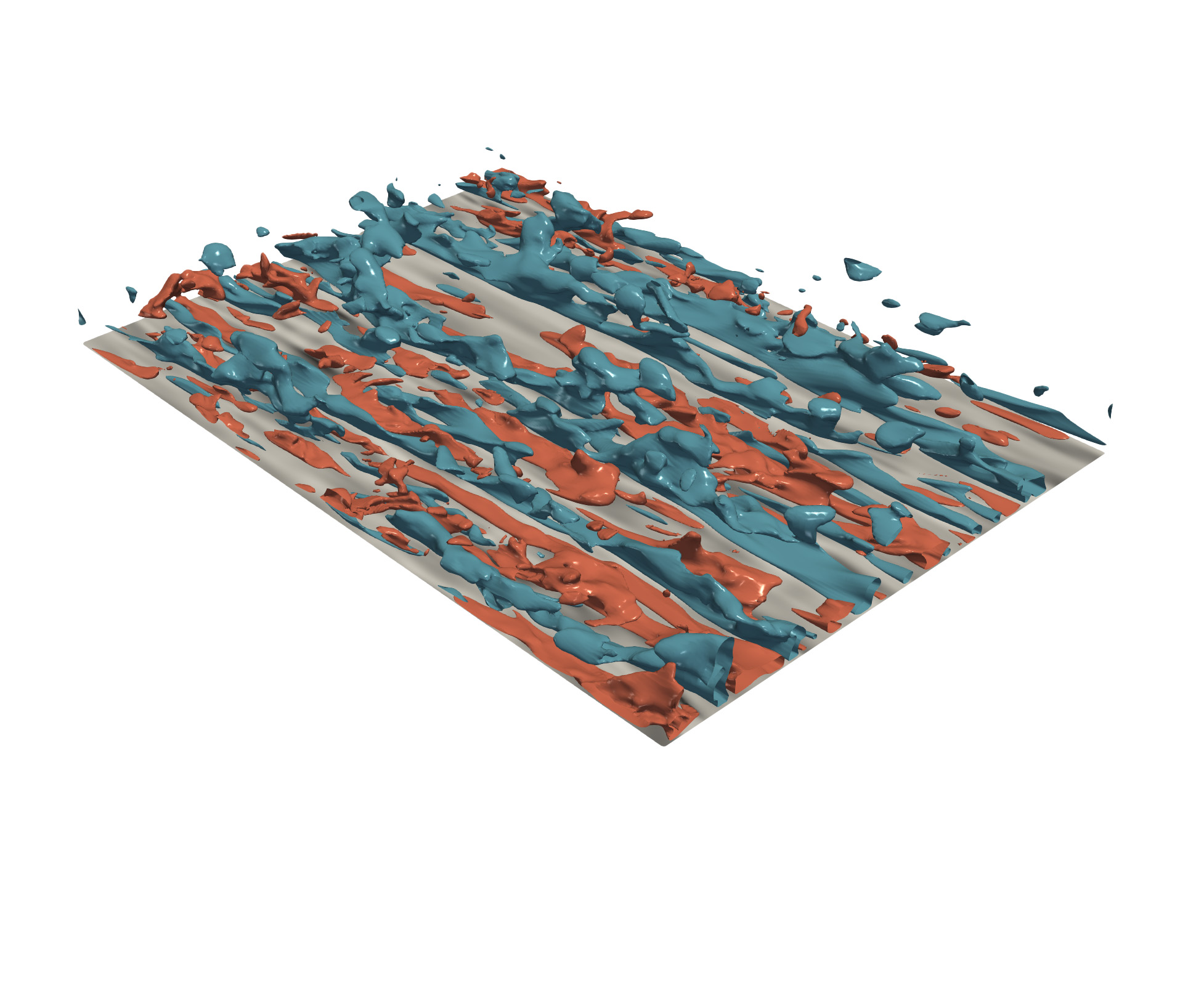}};
        \node[anchor=north] at (fig.north west) {(a)}; 
        \fill[white] (-20em,-10em) circle (.1pt);
        \end{footnotesize}
        \end{tikzpicture}
    \phantomsubcaption\label{fig:instOri}
    \end{subfigure}\hfill
    \begin{subfigure}{\tw}
        \begin{tikzpicture}
        \begin{footnotesize}
         \node[inner sep=0pt] (fig) at (0,0) 
            {\ig[width=.65\tw,trim=4.5cm 13cm 5cm 8cm,clip] 
            {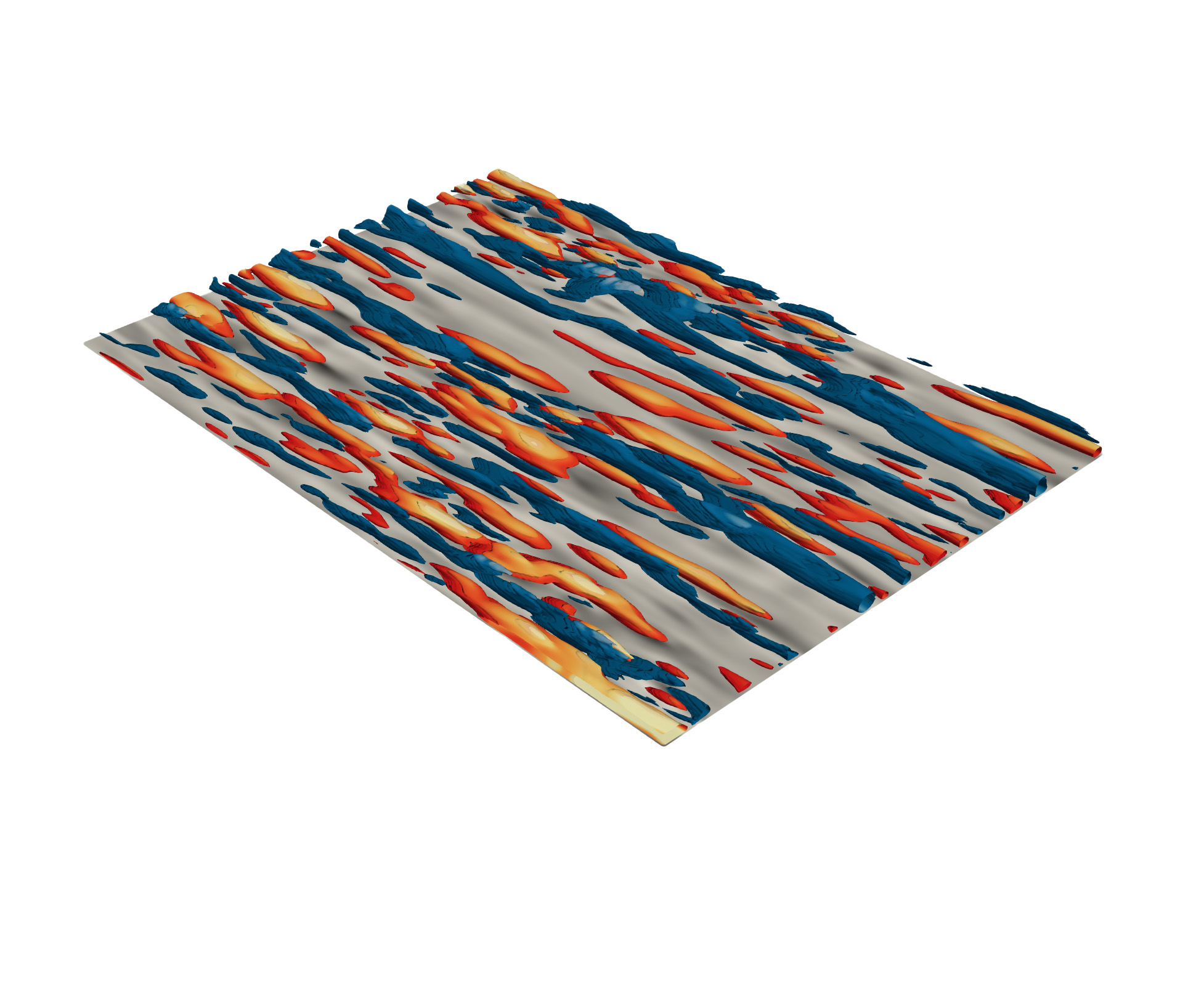}};
        \node[anchor=north] at (fig.north west) {(b)};
        \node[inner sep=0,xshift=1em] (fig1) at (fig.south east) 
            {\ig[width=.4\tw,trim=0 0cm 0 0cm,clip]{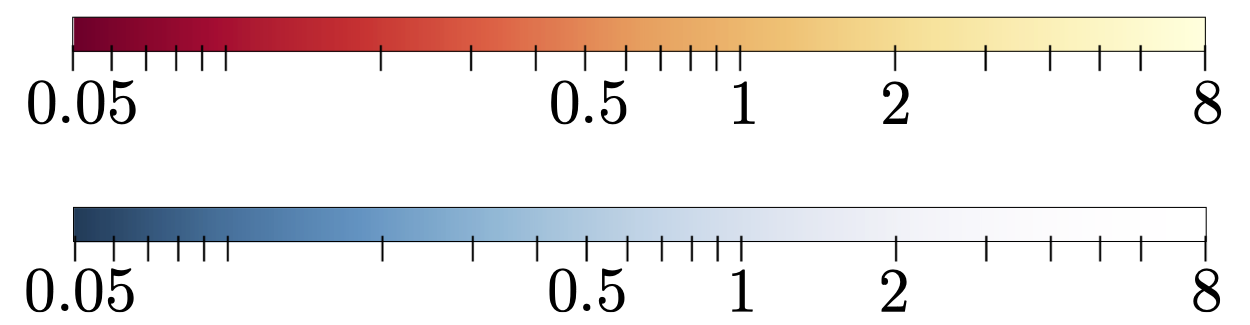}};
        \node[fill=white,anchor=south] at (fig1.north) 
            {$\partial \mathcal{F}^*/\partial u_I^*$};
        \fill[white] (-20em,0) circle (.1pt);
        \end{footnotesize}
        \end{tikzpicture}
    \phantomsubcaption\label{fig:instInf}
    \end{subfigure}
    \begin{subfigure}{\tw}
        \begin{tikzpicture}
        \begin{footnotesize}
        \node[inner sep=0pt] (fig) at (0,0) 
            {\ig[width=.65\tw,trim=4.5cm 13cm 5cm 8cm,clip]
            {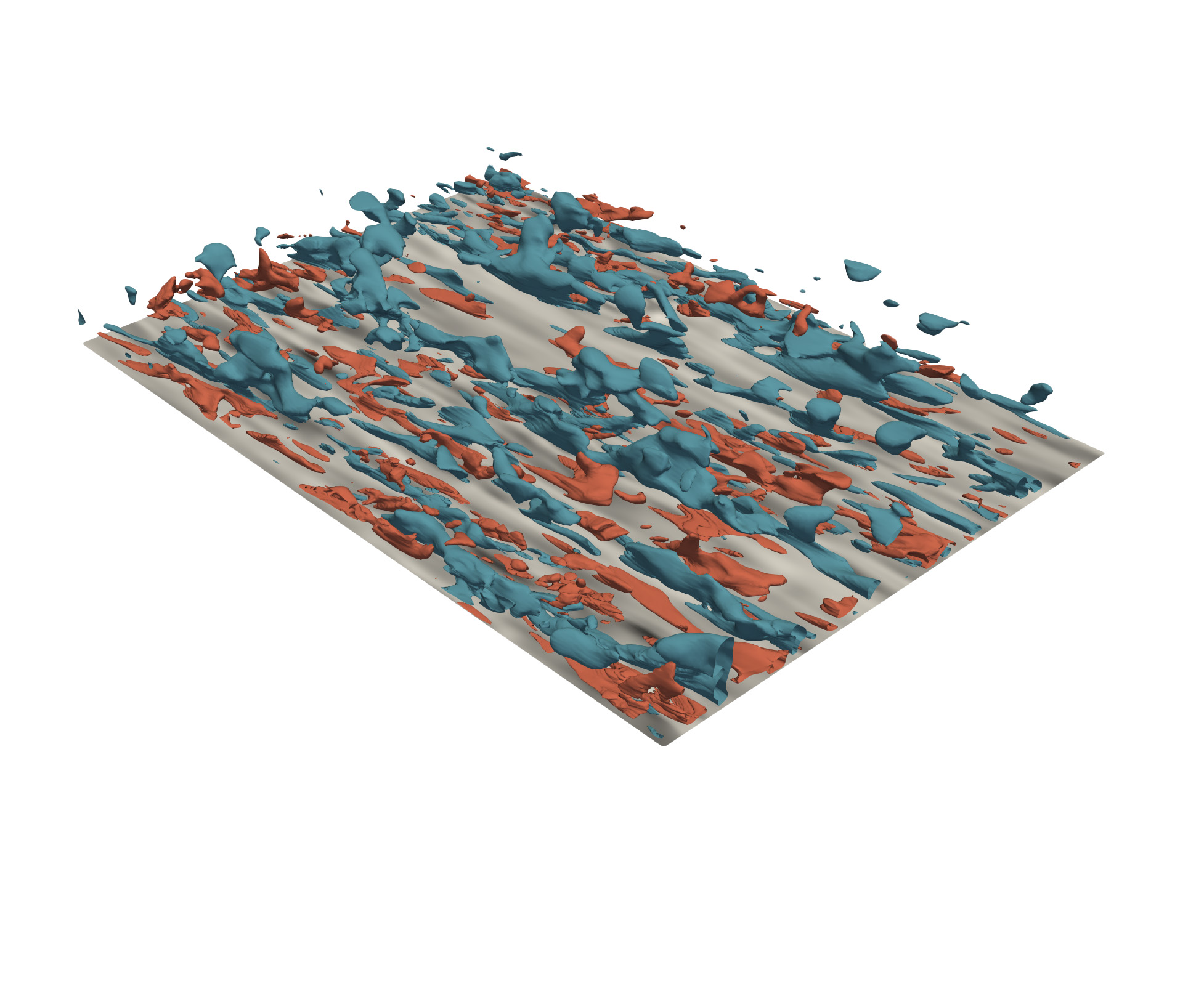}};
        \node[anchor=north] at (fig.north west) {(c)};
        \fill[white] (-20em,0) circle (.1pt);
        \end{footnotesize}
        \end{tikzpicture}
    \phantomsubcaption\label{fig:instRes}
    \end{subfigure}\hfill
    \caption{Instantaneous flow structures.  (a) Iso-contours of
      the streamwise velocity fluctuations, $u$; (b) iso-contours
      of the informative streamwise velocity fluctuations, $u_I$;
      and (c) iso-contours of the residual streamwise velocity
      fluctuations, $u_R$.  In (a) and (c) the iso-contours
      correspond to: (blue) $u^* \approx -2.7$, (red) $u^* \approx
      2.7$; and in (b): (blue) $u_I^* \approx -1.8$ and (orange)
      $u_I^* \approx 1.8$.  The wall is coloured by the instantaneous
      wall shear stress at $\Delta T$, from (white) $\taux^*
      \approx 0.5$ to (black) $\taux^* \approx 2$.\label{fig:cau_flow}}
\end{figure}

We start by visualising the instantaneous informative and residual
components of the flow.  We focus on the streamwise component, as it
turns out to be the most informative to $\taux$, as detailed below.
\Fig{fig:instOri} displays iso-surfaces of $u(\vec{x},t)$, revealing
the alternating high- and low-velocity streaks attached to the wall
along with smaller detached regions.  The informative and residual
components, $u_I(\vec{x},t)$ and $u_R(\vec{x},t)$, are shown in
\fig{fig:instInf,fig:instRes}, respectively.  The structures in $u_I$
exhibit a similar alternating pattern as in the original field, with
the high- and low-velocity streaks located roughly in the same
positions as $u(\vec{x},t)$. These structures are also attached to the
wall but do not extend as far as the streaks in the original field,
especially for $u_I(\vec{x},t) > 0$.  In contrast, the residual field
$u_R(\vec{x},t)$ lacks most of the elongated streaks close to the wall
but resembles $u(\vec{x},t)$ far away, once the flow bears barely no
information about $\tauxf$.

\begin{figure}
    \centering
    \begin{subfigure}{.33\tw}
        \begin{tikzpicture}
            \node[anchor=south west] (fig) at (0,0) 
            {\ig[width=\tw]{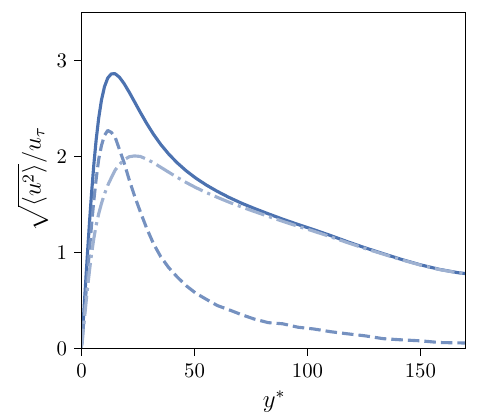}};
            \begin{scope}[x=(fig.south east),y=(fig.north west)]
                \node[scale=.9,fill=white] at (.55,.05) {$y^*$};
                \node[scale=.9,fill=white,rotate=90] at (.07,.55)
                {$\langle{u_\square^*}^2\rangle^{1/2}$};
                \node at (.24,.88) {(a)};
            \end{scope}
        \phantomcaption\label{fig:cau_flow_stats_rms_u}
        \end{tikzpicture}
    \end{subfigure}~
    \begin{subfigure}{.33\tw}
        \begin{tikzpicture}
            \node[anchor=south west] (fig) at (0,0) 
            {\ig[width=\tw]{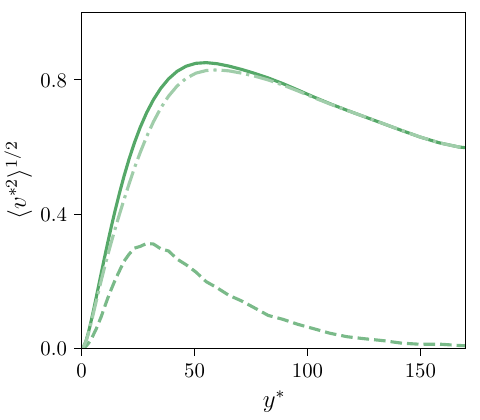}};
            \begin{scope}[x=(fig.south east),y=(fig.north west)]
                \node[fill=white] at (.55,.05) {$y^*$};
                \node[scale=.9,fill=white,rotate=90] at (.04,.55)
                {$\langle{v_\square^*}^2\rangle^{1/2}$};
                \node at (.24,.88) {(b)};
            \end{scope}
        \phantomcaption\label{fig:cau_flow_stats_rms_v}
        \end{tikzpicture}
    \end{subfigure}~
    \begin{subfigure}{.33\tw}
        \begin{tikzpicture}
            \node[anchor=south west] (fig) at (0,0) 
            {\ig[width=\tw]{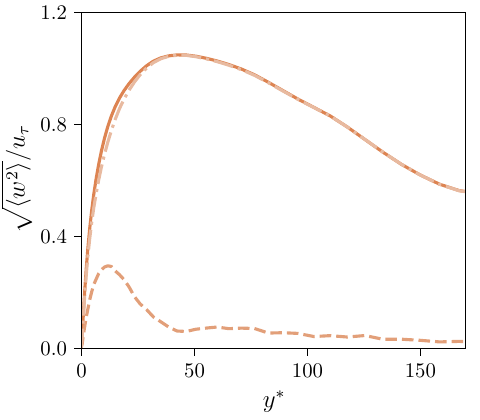}};
            \begin{scope}[x=(fig.south east),y=(fig.north west)]
                \node[scale=.9,fill=white] at (.55,.05) {$y^*$};
                \fill[white] (.02,.45) rectangle (.1,.7);
                \node[scale=.9,rotate=90] at (.06,.55)
                {$\langle{w_\square^*}^2\rangle^{1/2}$};
                \node at (.24,.88) {(c)};
            \end{scope}
        \phantomcaption\label{fig:cau_flow_stats_rms_w}
        \end{tikzpicture}
    \end{subfigure}~
    \caption{Root-mean-squared turbulence intensities of the (a)
      streamwise, (b) wall-normal, and (c) spanwise velocity
      components.  (solid) original flow field; (dashed) informative
      flow field; and (dash-dot) residual
      field.  \label{fig:cau_flow_stats_rms}}
\end{figure}

% urms
\Fig{fig:cau_flow_stats_rms} displays the root-mean-squared turbulence
intensities as a function of the wall distance. Note that, from
  the minimised term in \eq{eq:problem_scalar},
%by construction, 
$\langle u^2 \rangle(y) = \langle u_I^2 \rangle(y) +
\langle u_R^2 \rangle(y)$ (similarly for the other components).  From
\fig{fig:cau_flow_stats_rms_u}, we observe that $\langle u_I^2
\rangle^{1/2}$ is predominantly located within the region $y^* \leq
50$.  This finding aligns with our earlier visual assessments from
\fig{fig:cau_flow}. The residual component $\langle u_R^2
\rangle^{1/2}$ also has a strong presence close to the wall, although
it is shifted towards larger values of $y$. Interestingly, about half
of the streamwise kinetic energy in the near-wall region originates
from $\langle u_R^2 \rangle$, despite its lack of information about
$\tauxf$.  This phenomenon is akin the inactive motions in wall
turbulence~\citep[e.g.][]{townsend1961, jimenez2008, deshpande2021}
with the difference that here inactive structures are interpreted as
those that do not reflect time variations of the wall-shear stress.
Another interesting observation is that $\langle u_I^2 \rangle^{1/2}$
peaks at $y^* \approx 10$, which is slightly below the well-known peak
for $\langle u^2 \rangle^{1/2}$, whereas $\langle u_R^2 \rangle^{1/2}$
peaks at $y^* \approx 30$.  This suggests that the near-wall peak of
$\langle u^2 \rangle^{1/2}$ is controlled by a combination of active
and inactive motions as defined above.

% vrms and wrms
The root-mean-squared velocities for the cross flow are shown in
\fig{fig:cau_flow_stats_rms_v,fig:cau_flow_stats_rms_w}. The
informative component of the wall-normal velocity $\langle v_I^2
\rangle^{1/2}$ is predominantly confined within the region $y^* \leq
70$, although its magnitude is small. The residual component, $\langle
v_R^2 \rangle^{1/2}$, is the major contributor to the wall-normal
fluctuations across the channel height. The dominance of $\langle
v_R^2 \rangle^{1/2}$ has important implications for control strategies
in drag reduction, which are investigated in
\S\ref{sec:res:control}. A similar observation is made for $\langle
w^2 \rangle^{1/2}$, with $\langle w_I^2 \rangle^{1/2}$ being
negligible except close to the wall for $y^* < 40$.

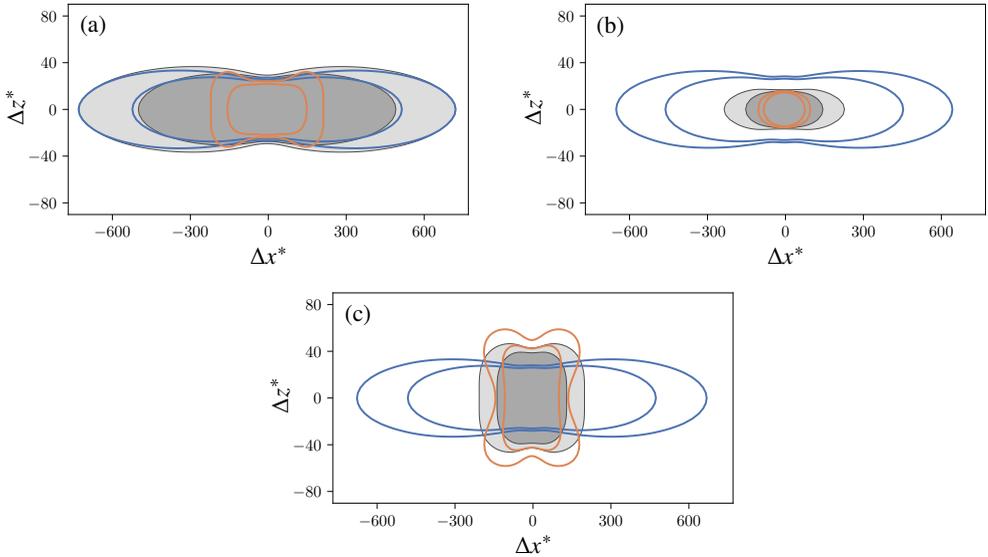
\begin{figure}
    \begin{tikzpicture}
        \node[anchor=south west] (f1) at (0,0)
        {\ig[width=.49\tw]{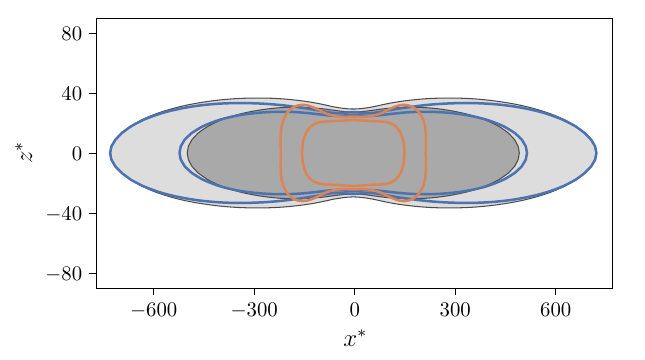}};
        \node[anchor=south west] (f2) at (f1.south east)
        {\ig[width=.49\tw]{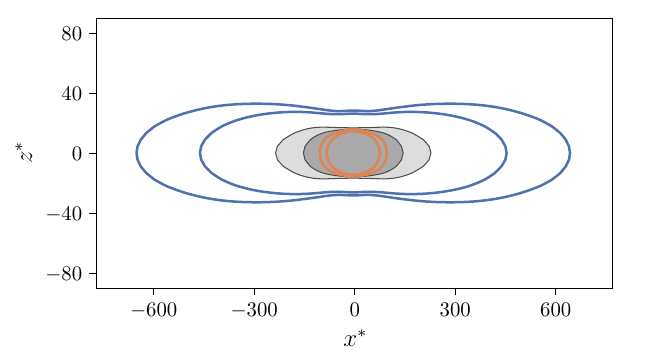}};
        \node[anchor=south west] (f3) at (3.5,-3.8)
        {\ig[width=.49\tw]{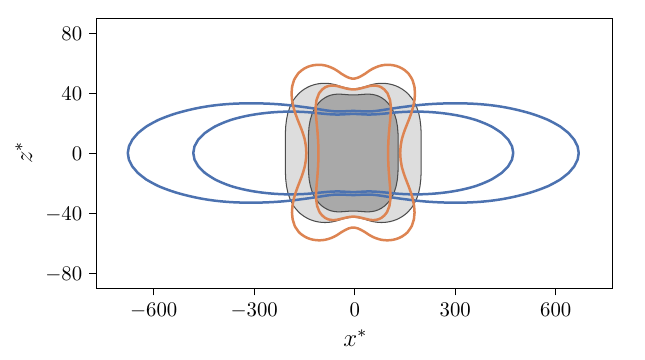}};
        \begin{scope}[x=(f1.south east),y=(f1.north west)]
            \node[fill=white] at (.55,.08) {$\Delta x^*$};
            \node[fill=white,rotate=90] at (.06,.58) {$\Delta z^*$};
            \node at (.21,.85) {(a)};
        \end{scope}
        \begin{scope}[shift={(f2.south west)},x=(f2.south east),y=(f2.north west)]
            \node[fill=white] at (.55,.08) {$\Delta x^*$};
            \node[fill=white,rotate=90] at (.06,.58) {$\Delta z^*$};
            \node at (.21,.85) {(b)};
        \end{scope}
        \begin{scope}[shift={(f3.south west)},x=(f3.south east),y=(f3.north west)]
            \node[fill=white] at (.55,.08) {$\Delta x^*$};
            \node[fill=white,rotate=90,inner sep=6pt] at (.06,.58) {$\Delta z^*$};
            \node at (.21,.85) {(c)};
        \end{scope}
        \phantomsubcaption\label{fig:corr_y15_u}
        \phantomsubcaption\label{fig:corr_y15_v}
        \phantomsubcaption\label{fig:corr_y15_w}
    \end{tikzpicture}
    \caption{Auto-correlation coefficient of the velocity fluctuations
      in the $y^*=15$ plane. (a) Streamwise component, (b) wall normal
      component, and (c) spanwise component.  (Grey) Original field,
      (blue) informative field, and (orange) residual field.  The
      contours correspond to $C_{\square\square} = [0.05, 0.1]$.
    \label{fig:corr_y15}}
\end{figure}

% Correlations
The statistical coherence of the informative and residual velocity in
the wall-parallel plane is quantified with the two-point
auto-correlation
\begin{equation}
    C_{\phi\phi}(\Delta x, \Delta z; y_\text{ref}) =
    \frac{\langle \phi(x,y_\text{ref},z,t) \phi(x+\Delta x, y_\text{ref}, z+\Delta z,t)
    \rangle}{\langle \phi(x,y_{\text{ref}},z,t)^2 \rangle},
\end{equation}
where $\phi$ is any component of the velocity field, and
$y_\text{ref}^* = 15$.  The auto-correlations are shown in
\fig{fig:corr_y15} for the total, informative, and residual components
of the three velocities.  The shape of informative structure is
elongated along the streamwise direction for the three correlations
$C_{u_Iu_I}$, $C_{v_Iv_I}$, and $C_{w_Iw_I}$. The results for $u$,
shown in \fig{fig:corr_y15_u}, reveal that $u_I$ closely resembles the
streaky structures of $u$ in terms of streamwise and spanwise
lengths. On the other hand, $u_R$ consists of more compact and
isotropic eddies in the $(x,z)$-plane. \Fig{fig:corr_y15_v} shows that
$v_I$ captures the elongated motions in $v$, which represents a small
fraction of its total energy, whereas the shorter motions in $v$ are
contained in $v_R$. A similar conclusion is drawn for $w$, as shown in
\fig{fig:corr_y15_w}, where both $w$ and $w_R$ share a similar
structure, differing from the elongated motions of $w_I$. The emerging
picture from the correlations is that informative velocities tend to
comprise streamwise elongated motions, whereas the remaining residual
components are shorter and more isotropic. The differences between the
structure of $v$ and $w$ with their informative counterparts are
consistent with the lower intensities of $v_I$ and $w_I$ discussed in
\fig{fig:cau_flow_stats_rms}.
It should be noted that the shape of the structures depends on the target
variable and they may differ for a different target quantity. 
For example, wall pressure fluctuations have been linked to more
isotropic structures in the streamwise direction by several authors
\citep{schewe1983,johansson1987,kim1987,ghaemi2013}.
The aIND may provide insights in this regard, as it has been noted in the
literature that at least quadratic terms are needed to capture the interaction
between the velocity and the wall pressure \citep{naguib2001,murray2003}.

\begin{figure}
    \begin{tikzpicture}
        \node[anchor=south west] (f1) at (0,0) 
        {\ig[height=4.5cm,trim=0 0 .5cm 0,clip]
        {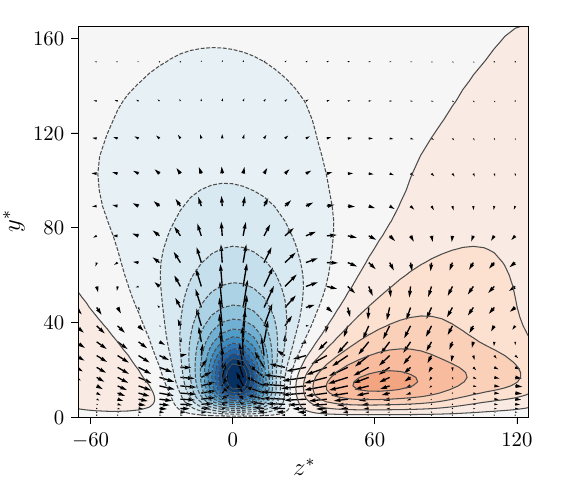}};
        \phantomsubcaption\label{fig:averagestreakO}
        \node[anchor=south west] (f2) at (f1.south east)
        {\ig[height=4.5cm,trim=1.2cm 0 .5cm 0,clip]
        {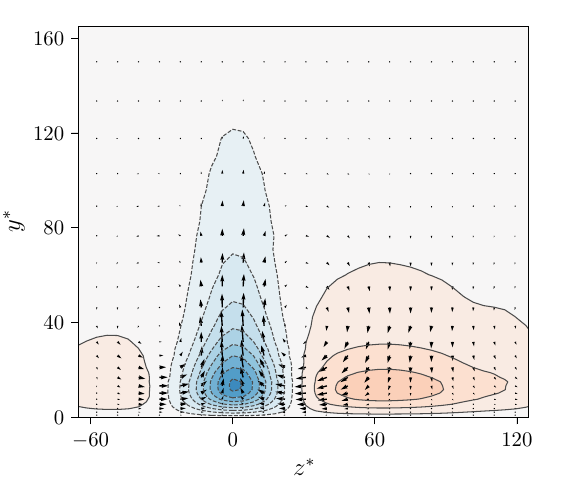}};
        \phantomsubcaption\label{fig:averagestreakI}
        \node[anchor=south west] (f3) at (f2.south east)
        {\ig[height=4.5cm,trim=1.2cm 0 .5cm 0,clip]
        {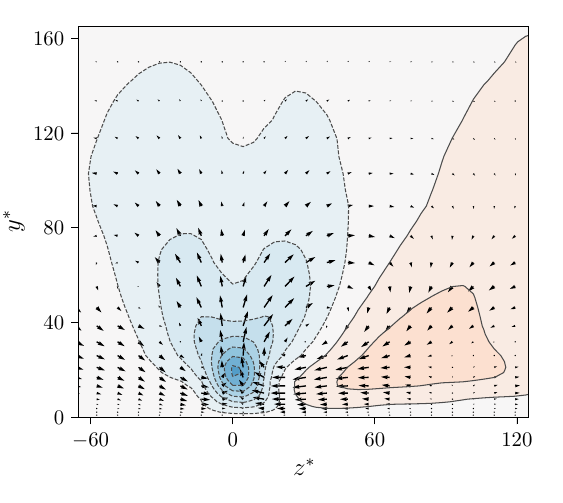}};
        \phantomsubcaption\label{fig:averagestreakR}
        \begin{scope}[shift={(f1.south west)},x=(f1.south east),y=(f1.north west)]
            \node[fill=white] at (.57,.08) {$\Delta z^*$};
            \node[fill=white,rotate=90] at (.04,.58) {$y^*$};
            \node at (.22,.87) {(a)};
        \end{scope}
        \begin{scope}[shift={(f2.south west)},x=(f2.south east),y=(f2.north west)]
            \node[fill=white] at (.52,.08) {$\Delta z^*$};
            \node at (.11,.87) {(b)};
        \end{scope}
        \begin{scope}[shift={(f3.south west)},x=(f3.south east),y=(f3.north west)]
            \node[fill=white] at (.52,.08) {$\Delta z^*$};
            \node at (.11,.87) {(c)};
        \end{scope}
    \end{tikzpicture}
    \caption{Conditionally-averaged flow in the $(y,z)$-plane centred
      about low-velocity streaks.  (a) Original field, (b) informative
      field, and (c) residual field.  The contours range from (blue)
      $-5 u_\tau$ to (red) $5 u_\tau$.  The arrows display the
      in-plane velocity components ($v_\square$ and $w_\square$).
    \label{fig:averagestreak}}
\end{figure}
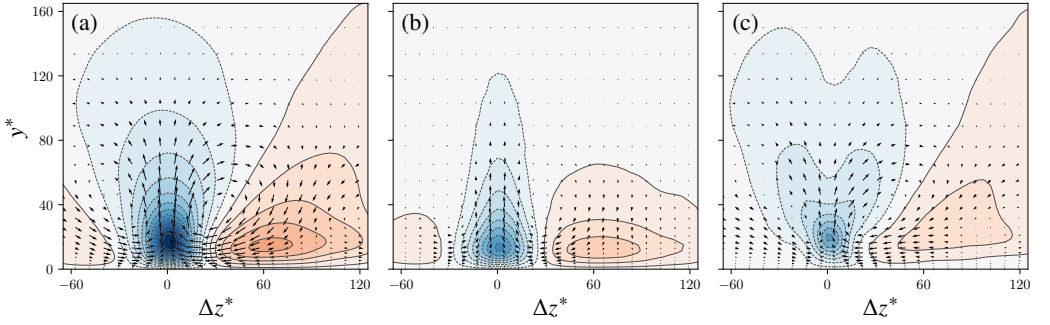

% conditional-averaged flow (method)
We now analyse the average coherent structure of the flow in the
$(y,z)$-plane. It is widely recognised in the literature that the most
dynamically relevant energy-containing structure in wall turbulence
comprises a low-velocity streak accompanied by a collocated
roll~\citep[e.g.][]{kline1967,kim1987,lozano2012, farrell2012}.  A
statistical description of this structure can be obtained by
conditionally averaging the flow around low-velocity streaks.
To this end, low-velocity streaks were identified by finding local
minima of $u$ at $y^* = 15$. For each streak, a local frame of
reference was introduced with axes parallel to the original $x$,
  $y$, and $z$ coordinates.  The origin of this local frame of
  reference is at the wall, such that its $y$-axis is aligned with the
  local minimum of $u$. The $z$-axis, denoted by $\Delta z$, points
  towards the nearest local maximum of $u$. This orientation ensures
  that any nearby high-speed streak is located in the region $\Delta z
  > 0$. Then, the conditional average flow was computed by averaging
$[u, v, w]$ over a window of size $\pm h$.
The resulting
conditionally-averaged flow in the $(y,z)$-plane is shown in 
\fig{fig:averagestreakO}.  This process was repeated for the
informative and residual velocity fields using the same streaks
previously identified for $u$. The conditionally-averaged informative
and residual velocities are shown in \fig{fig:averagestreakI,fig:averagestreakR}, 
respectively.

% conditional-averaged flow (results)
The conditional average velocity is shown in
\fig{fig:averagestreakO}, which captures the structure of the
low-/high-velocity streak pair and the accompanying roll
characteristic of wall-bounded turbulence. The informative velocity
(\fig{fig:averagestreakI}) is dominated by streak motions,
although these are smaller than the streaks of the entire
field. The informative wall-normal velocity is present mostly within
the streaks, while the informative spanwise component is active close
to the wall in the interface of the streak.  Conversely,
\fig{fig:averagestreakR} shows that the residual velocity
contains the large-scale streaks and the remaining spanwise motions.
The emerging picture is that the informative component of the velocity
contributing to the wall shear stress consists of smaller near-wall
streaks collocated with vertical motions (i.e., sweeps and ejections),
and spanwise velocity at the near-wall root of the roll.  This
informative structure is embedded within a larger-scale streak-roll
structure of residual velocity, which bears no information about the
wall-shear stress.

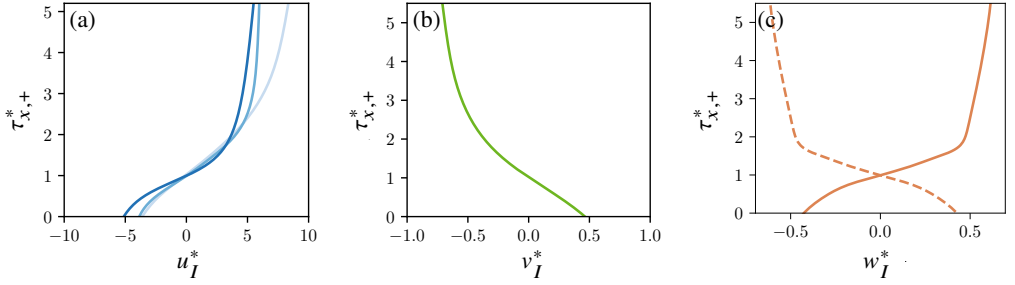
\begin{figure}
\centering
\begin{subfigure}{.33\tw}
    \begin{tikzpicture}
    \node[anchor=south west] (fig) at (0,0)
    {\ig[width=\tw]{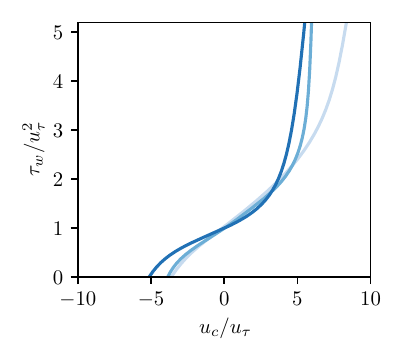}};
    \begin{scope}[x=(fig.south east), y=(fig.north west)]
        \node[fill=white,rotate=90] at (0.08,.58) {$\tauxf^*$};
        \fill[white] (.48,.08) rectangle (.64,.15);
        \node[fill=white] at (.56,.10) {$u_I^*$};
        \node at (.26,.86) {(a)};
    \end{scope}
    \end{tikzpicture}
\phantomsubcaption
\label{fig:ch_functions_u}
\end{subfigure}~
\begin{subfigure}{.33\tw}
    \begin{tikzpicture}
    \node[anchor=south west] (fig) at (0,0)
    {\ig[width=\tw]{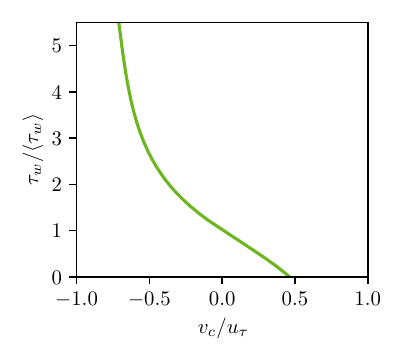}};
    \begin{scope}[x=(fig.south east), y=(fig.north west)]
        \node[fill=white,rotate=90] at (0.08,.58) {$\tauxf^*$};
        \fill[white] (.48,.08) rectangle (.64,.15);
        \node[fill=white] at (.56,.10) {$v_I^*$};
        \node at (.26,.86) {(b)};
    \end{scope}
    \end{tikzpicture}
\phantomsubcaption
\label{fig:ch_functions_v}
\end{subfigure}~
\begin{subfigure}{.33\tw}
    \begin{tikzpicture}
    \node[anchor=south west] (fig) at (0,0)
    {\ig[width=\tw]{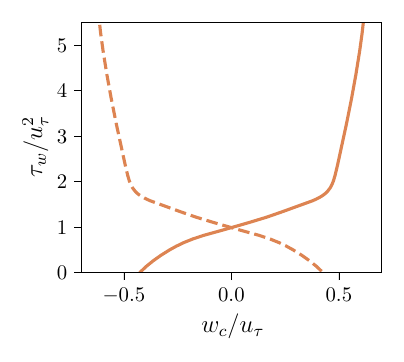}};
    \begin{scope}[x=(fig.south east), y=(fig.north west)]
        \node[fill=white,rotate=90] at (0.08,.58) {$\tauxf^*$};
        \fill[white] (.48,.08) rectangle (.64,.15);
        \node[fill=white] at (.56,.10) {$w_I^*$};
        \node at (.26,.86) {(c)};
    \end{scope}
    \end{tikzpicture}
\phantomsubcaption
\label{fig:ch_functions_w}
\end{subfigure}
\caption{Mapping functions of the informative contributions of
(a) streamwise, (b) wall-normal, and (c) spanwise velocity 
fluctuations to the streamwise wall shear stress 
for $\Delta T^* = 25$. 
(a) Also shows the effect of the time lag $\Delta T$ in the
mapping $\taux = \mathcal{F}(u_I)$.  Line colors correspond
to (dark blue) $\Delta T^* \approx 25$; (cyan) $\Delta T^*
\approx 10$; and (light blue) $\Delta T^* \approx 5$.
In (c), solid line corresponds to $+\Delta z^{\max}_w$ and
dashed line to $-\Delta z^{\max}_w$. \label{fig:ch_functions}}
\end{figure}

%  mapping functions: Fu and Delta T
We close this section by analysing the mappings $\tauxf =
\mathcal{F}_u(u_I)$, $\tauxf = \mathcal{F}_v(v_I)$, $\tauxf =
\mathcal{F}_w(w_I)$ obtained from the constraints $H(\tauxf | u_I) =
0$, $H(\tauxf | v_I) = 0$, and $H(\tauxf | w_I) = 0$, respectively.
The mapping are depicted in \fig{fig:ch_functions} at the wall-normal
position where the energy for $u_I$, $v_I$, and $w_I$ is maximum,
namely, $y^* \approx 8$, $19$, and $6$, respectively (see
Appendix~\ref{sec:optDx}).  \Fig{fig:ch_functions_u} reveals an almost
linear relationship between $u_I$ and $\tauxf$ within the range of $0
\leq \tauxf^* \leq 2$. Negative values of $u_I$ align with $\tauxf^* < 1$, 
while positive values of $u_I$ correspond to $\tauxf^* > 1$. This
is clear a manifestation of the proportionality between streak
intensity and $\taux$, such that higher streamwise velocities
translate into higher wall shear stress by increasing $\partial
U/\partial y$. However, the process saturates and a noticeable change
in the slope occurs for larger values of $\tauxf$, leading to $u_I$
values which are relatively independent of $\tauxf$. This finding
indicates that $u_I$ provides limited information about high values of
$\tauxf$ at the timescale $\Delta T^*=25$. In other words, minor
uncertainties in $u_I$ result in significant uncertainties in $\tauxf$
after $\Delta T$.

% effect of Delta T
The effect of $\Delta T$ on $\mathcal{F}_u(u_I)$ is also analysed in
\fig{fig:ch_functions_u}.  The main effect of decreasing $\Delta T^*$
is to decrease the slope of $\mathcal{F}_u(u_I)$ for $u_I^* > 5$.
This result reveals that there exists a time horizon beyond which it
is not possible to predict extreme events of wall shear stress from
local fluctuations.  Hence, extreme values of the wall shear stress
can be attributed to almost instantaneous high fluctuations of the
streamwise velocity.  The latter is in agreement with
\citet{guerrero2020}, who linked extreme positive wall shear stresses
with the presence of high-momentum regions created by quasi-streamwise
vortices.

%  mapping functions: Fv
The mapping of $v_I$ is shown in \fig{fig:ch_functions_v},
which demonstrates again a nearly linear, albeit negative,
relationship between $v_I$ and $\tauxf$ in the range $0 \leq \tauxf^*
\leq 2$.  Positive values of $v_I$ are indicative of $\tauxf^*<1$,
whereas negative values imply $\tauxf^*> 1$.  Note that changes in the
value of $\tauxf$ encompass either $u_I>0$ and $v_I<0$ or $u_I<0$ and
$v_I>0$, revealing a connection between the dynamics of $\tauxf$ and
the well-known sweep and ejection motions in wall-bounded
turbulence~\citep{wallace1972, wallace2016}.  The mappings also show
that excursions into large wall shear stresses are caused by
sweeps. Analogous to $u_I$, the value of $v_I$ remains approximately
constant for $\tauxf^*>2$. Beyond that threshold, $v_I$ provides no
information about $\tauxf$.

%  mapping functions: Fw
The mapping of $w_I$ presents two maxima ($\pm\Delta z_w^{\max}$) due to
the spanwise symmetry of the flow. The results for each maximum, shown
in \fig{fig:ch_functions_w}, are antisymmetric with respect to
$w_I$.  Similarly to $u_I$ and $v_I$, there is an almost linear
relationship between $w_I$ and $\tauxf$ in the range $0 \leq \tauxf^*
\leq 2$.  For $+\Delta z_w^{\max}$, negative values of $w_I$ indicate
$\tauxf^*<1$, whereas positive values are linked to $\tauxf^*>1$.  The
opposite is true for $-\Delta z_w^{\max}$. Low values of $\tauxf$ are
connected to low $u_I$ and positive (negative) values of $w_I$ for
$+\Delta z_w^{\max}$ ($-\Delta z_w^{\max}$).  This outcome is
consistent with the conditional average flow from
\fig{fig:averagestreak}, where it was shown that the information
transfer between $w_I$ and $\tauxf$ is mediated through the bottom
part of the roll structure that accompanies high/low velocity
streaks. The saturation of the influence of $w_I$ to intense values of
the wall shear stress is again observed for $\tauxf^* \gtrsim 2$.

% meaning of derivative
The information provided by the mappings can be embedded into the
instantaneous coherent structures. In \fig{fig:instInf}, the
$u_I(\vec{x},t)$ structures are coloured by the local value of
$\partial \mathcal{F}/\partial u_I$. As discussed above, this metric
serves as a measure of the uncertainty in the wall shear stress as a
function of $u_I$. Low values of $\partial \mathcal{F}/\partial u_I$
are associated with low uncertainty in $\tauxf$. This implies that
small changes in $u_I$ result in small changes in $\tauxf$.  On the
other hand, high values of $\partial \mathcal{F}/\partial u_I$ are
associated with high uncertainty in $\tauxf$, such that small
variations in $u_I$ result in large changes in $\tauxf$.
Interestingly, \fig{fig:instInf} shows that low-speed streaks
--associated with ejections-- are connected to low uncertainty values
for $\taux$ along their entire wall-normal extent. On the contrary,
the high-speed streaks of $u_I$, linked to extreme events, carry
increasing uncertainty in $\taux$ (indicated by the light yellow
colour) as they move further away from the wall.

%%%%%%%%%%%%%%%%%%%%%%%%%%%%%%%%%%%%%%%%%%%%%%%%%%%%%%%%%%%%%%%%%%%%%%%%%%%%%%%%
%% RESULTS - RECONSTRUCTION

\subsection{Reduced-order modelling: reconstruction of the wall-shear stress from $u$}
\label{sec:res:tau:recons}

% intro
We evaluate the predictive capabilities of the informative and
residual components of the streamwise velocity fluctuations to
reconstruct the wall shear stress in the future. The main aim
of this section is to illustrate that when $u$ is used as the input
for developing a model, the resulting model exclusively utilizes
information from $u_I$, while $u_R$ is disregarded.

Two scenarios are considered. In the first case, we devise a model for
the pointwise, temporal forecasting of $\tauxf$ using pointwise data
of $u$. In the second scenario, the spatially two-dimensional
wall-shear stress is reconstructed using $u$ data from a wall-parallel
plane located at given distance from the wall. 

% pointwise reconstruction
First, we discuss the pointwise forecasting of $\tauxf$ using
pointwise data of $u$.  We aim to predict the future of the wall
shear stress at one point at the wall,
$\tauxf=\taux(x_0,z_0,t+\Delta T)$, where $x_0$ and $z_0$ are fixed, and the 
time lag is $\Delta T^* = 25$.  Three models are considered using as input 
$u(\boldsymbol{x}_0, t)$, $u_I(\boldsymbol{x}_0, t)$ and 
$u_R(\boldsymbol{x}_0, t)$, respectively, where $\boldsymbol{x}_0
= [x_0+\Delta x_u^{\max}, y_\text{ref}, z_0]$ and $y_\text{ref}^*
\approx 10$. 
The data is extracted from a simulation with the same set-up and friction 
Reynolds number as in \S\ref{sec:results} but in a smaller computational domain 
($\pi\chh \times 2\chh \times \pi/2\chh$). Note that all the points $[x_0,z_0]$ 
are statistically equivalent and can be used to train the model.

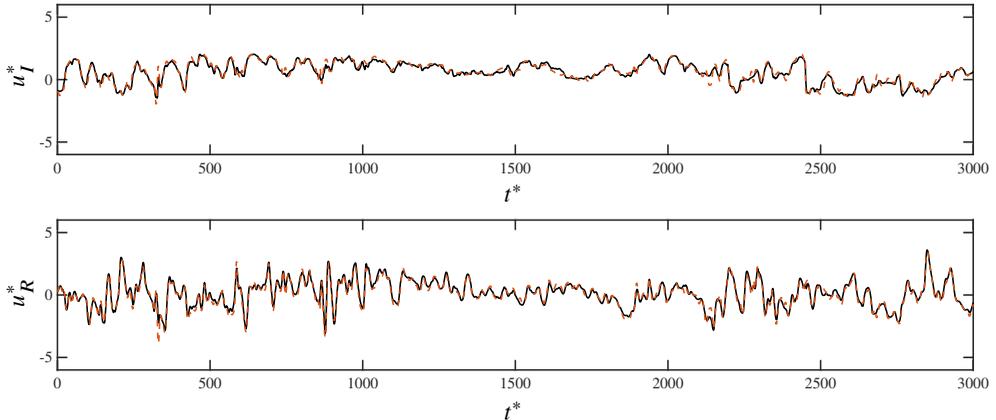
\begin{figure}
  \centering
  \begin{tikzpicture}
    \node[anchor=south west] 
    (fig) at (0,0) {\ig[width=0.95\tw]{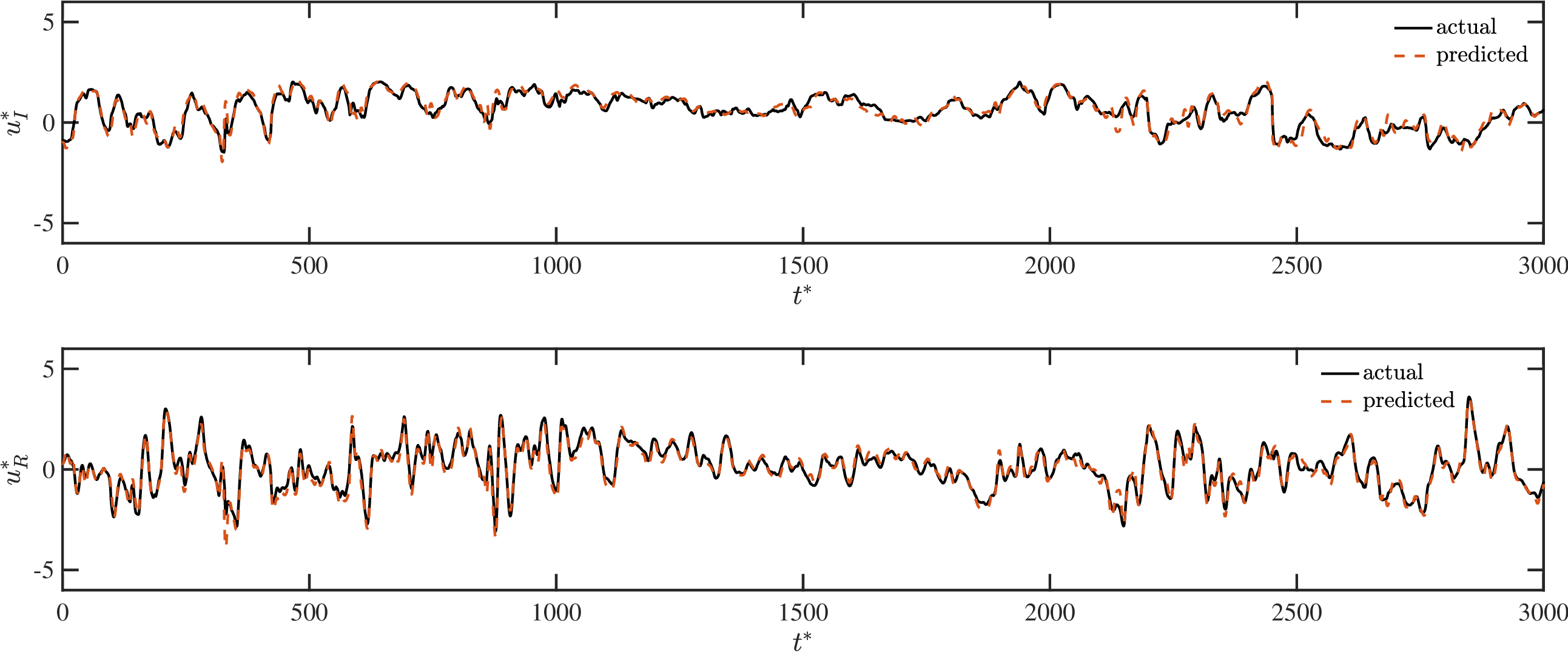}};
    \begin{footnotesize}
    \begin{scope}[x=(fig.south east),y=(fig.north west)]
      \node[fill=white] at (.51,.015) {$t^*$};
      \node[fill=white] at (.51,.53) {$t^*$};

      \node[fill=white,rotate=90] at (.01,.015+.28) {$u_R^*$};
      \node[fill=white,rotate=90] at (.01,.530+.28) {$u_I^*$};

      \fill[white] (.833,.37) rectangle ++(.09,.08);
      \fill[white] (.88,.880) rectangle ++(.09,.04);
      \fill[white] (.88,.920) rectangle ++(.07,.04);
    \end{scope}
    \end{footnotesize}
  \end{tikzpicture}

  \caption{Example of approximate decomposition of $u$ into informative
  $\tilde{u}_I$ (top) and residual $\tilde{u}_R$ (bottom) components
  using the ANN from \eq{eq:ANN_split} without the need of
  $\tauxf$.  Lines correspond to (solid black) the actual $u_I$ and $u_R$
  and (orange dashed) are the predicted value.\label{fig:ANN_split}}
\end{figure}

% ANN split
As a preliminary step to developing the forecasting models, we
use a feedforward artificial neural network (ANN) to separate $u$
into $u_I$ and $u_R$ without the need of $\tauxf$. This step is
required to make the models predictive, as in a practical case, the
future of $\taux$ is unknown and cannot be used to obtained the
informative and residual components. The model is given by
\begin{equation}\label{eq:ANN_split}
  [\tilde{u}_I(\boldsymbol{x}_0, t) , \tilde{u}_R(\boldsymbol{x}_0, t)] =
  \text{ANN}_{\text{I,R}}(u(\boldsymbol{x}_0, t),u(\boldsymbol{x}_0, t-\delta t),\dots, u(\boldsymbol{x}_0, t-p\delta t)),
\end{equation}
where the tilde in $\tilde{u}_I$ and $\tilde{u}_R$ denotes estimated
quantities, $\delta t^*=0.5$ and $p=1000$ is the number of time lags
considered.  Multiple time lags are required for predicting
$\tilde{u}_I$ and $\tilde{u}_R$, in the same manner as time series of
$u$ and $\tauxf$ were used to compute $u_I$. The
$\text{ANN}_{\text{I,R}}$ comprises 6 hidden layers with 50 neurons
per layer and ReLU activation functions. The roughly 700,000 samples
are divided into 80\% for training and 20\% for validation. The Adam
algorithm \citep{kingma2017adam} is used to find the optimum
solution. An example of the approximate decomposition from
\eq{eq:ANN_split} is shown in \fig{fig:ANN_split}.

% ANNs prediction
The three ANNs models trained to forecast $\tauxf$ are:
\begin{subequations}\label{eq:ANN_UIR}
  \begin{align}
  \tilde{\tau}_{x}^I(x_0,z_0,t+\Delta T)  &=
    \text{ANN}_I\left(\tilde{u}_I(\boldsymbol{x}_0, 
    t)\right),\label{eq:ANN_I}\\
    \tilde{\tau}_{x}^R(x_0,z_0,t+\Delta T) &=
    \text{ANN}_R\left(\tilde{u}_R(\boldsymbol{x}_0, t)\right),\label{eq:ANN_R}\\
    \tilde{\tau}_{x}^U(x_0,z_0,t+\Delta T) &=
    \text{ANN}_U\left(u(\boldsymbol{x}_0, t-\delta t),\dots, u(\boldsymbol{x}_0, t-p\delta t)\right).\label{eq:ANN_U}
\end{align}
\end{subequations}
Note that \eq{eq:ANN_I} and \eq{eq:ANN_R} use only one time step
of $\tilde{u}_I$ and $\tilde{u}_R$, respectively, while
\eq{eq:ANN_U} incorporates multiple time lags of $u$. This approach
is chosen because \eq{eq:ANN_split} (used to predict $\tilde{u}_I$
and $\tilde{u}_R$), also depends on multiple time lags of $u$. By
training \eq{eq:ANN_U} using the same time lags as
\eq{eq:ANN_split}, the predictions for $\tilde{\tau}_{x}^U$ rely on
a model that accesses an equivalent amount of information about past
states of the flow as do the models for predicting
$\tilde{\tau}_{x}^I$ and $\tilde{\tau}_{x}^R$. This ensures a fair
comparison among models.

\begin{figure}
  \centering
  \begin{tikzpicture}
    \node[anchor=south west] 
    (fig) at (0,0) {\ig[width=0.95\tw]{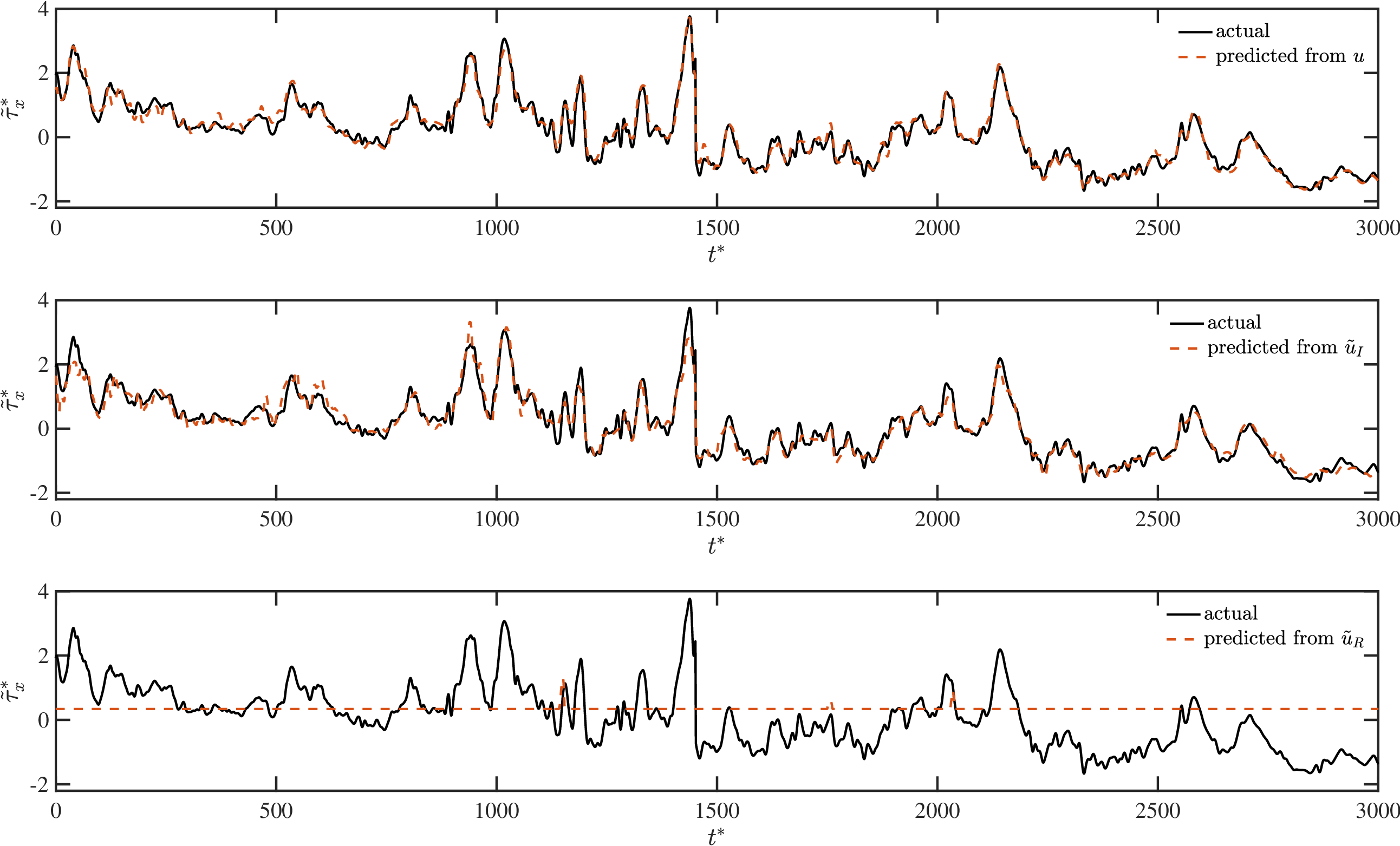}};
    \begin{footnotesize}
    \begin{scope}[x=(fig.south east),y=(fig.north west)]
      \node[fill=white] at (.51,.015) {$t^*$};
      \node[fill=white] at (.51,.35) {$t^*$};
      \node[fill=white] at (.51,.685) {$t^*$};

      \node[fill=white,rotate=90] at (.01,.015+.17) {$\tau_x^*$};
      \node[fill=white,rotate=90] at (.01,.350+.17) {$\tau_x^*$};
      \node[fill=white,rotate=90] at (.01,.685+.17) {$\tau_x^*$};
    \end{scope}
    \end{footnotesize}
  \end{tikzpicture}
\caption{Temporal reconstruction of the wall shear stress using
    ANNs in \eq{eq:ANN_UIR} trained with $u$ (top), $\tilde{u}_I$ (center), and
    $\tilde{u}_R$ (bottom) as inputs, respectively. (solid black)
    corresponds to the actual wall shear stress, (dashed orange) correspons to
    the ANN reconstruction.\label{fig:ANN}}
\end{figure}
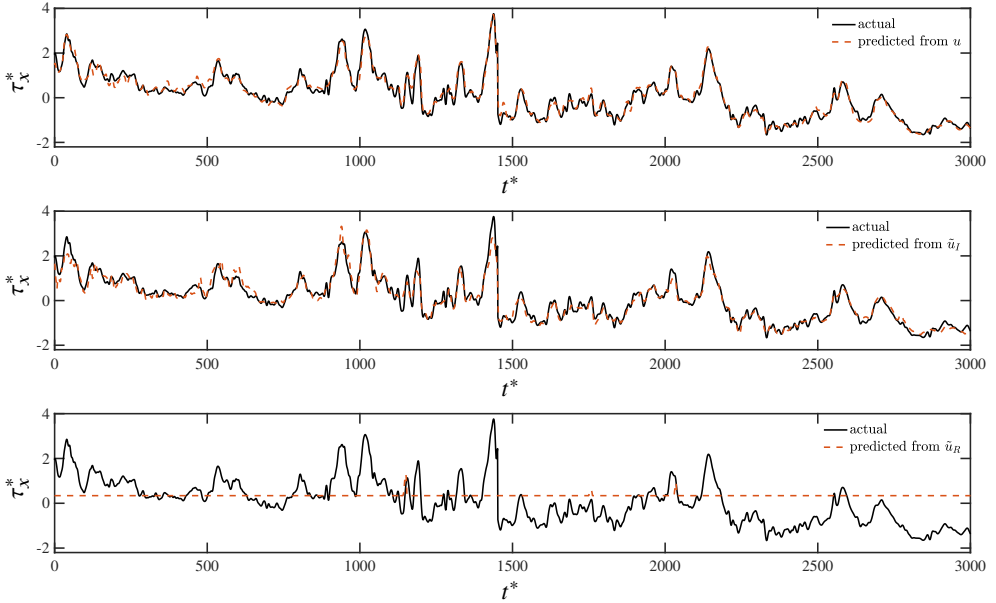

% results ANN prediction
The forecasting of the wall shear stress by the three models is
illustrated in \fig{fig:ANN}. The results indicate that the
predictions based on $u$ and $\tilde{u}_I$ are comparable, with
relative mean-squared errors of 18\% and 22\%, respectively. The
marginally larger error from the model using $\tilde{u}_I$ as input
arises from inaccuracies within the ANN responsible for decomposing
$u$ into $\tilde{u}_I$ and $\tilde{u}_R$. In a perfect scenario, the
forecasting error using either $u$ or $\tilde{u}_I$ as input would
be identical, implying that $\tilde{u}_I$ contains all the
information in $u$ to make predictions. In contrast, the model that
utilises the residual component $\tilde{u}_R$ fails to accurately
predict the wall shear stress (roughly by 100\% error), yielding
values that are nearly constant and close to the time-average of
$\tilde{\tau}_{x}$. These findings demonstrate that when $u$ is used
as input, the model extracts predictive information from
$\tilde{u}_I$, while $\tilde{u}_R$ provides no predictive value.

It is important to clarify that we are not advocating for the
separation of inputs into informative and residual components as a
standard practice for training models. Instead, our goal is to
illustrate that the training process of a model implicitly
discriminates between these components, supporting our claim that
all the necessary information for reduced-order modelling is
encapsulated in $u_I$.  An interesting consequence of this property
is that the characteristics and structure of $\tilde{u}_R$ are not
useful for understanding the predictive capabilities of the model;
instead, they help to discern which factors are irrelevant. For
further discussion on the role of information in predictive
modelling, the reader is referred to
\citet{lozanoduran2022,yuan2024}.

\begin{figure}
    \centering 
    \ig[width=.95\tw]{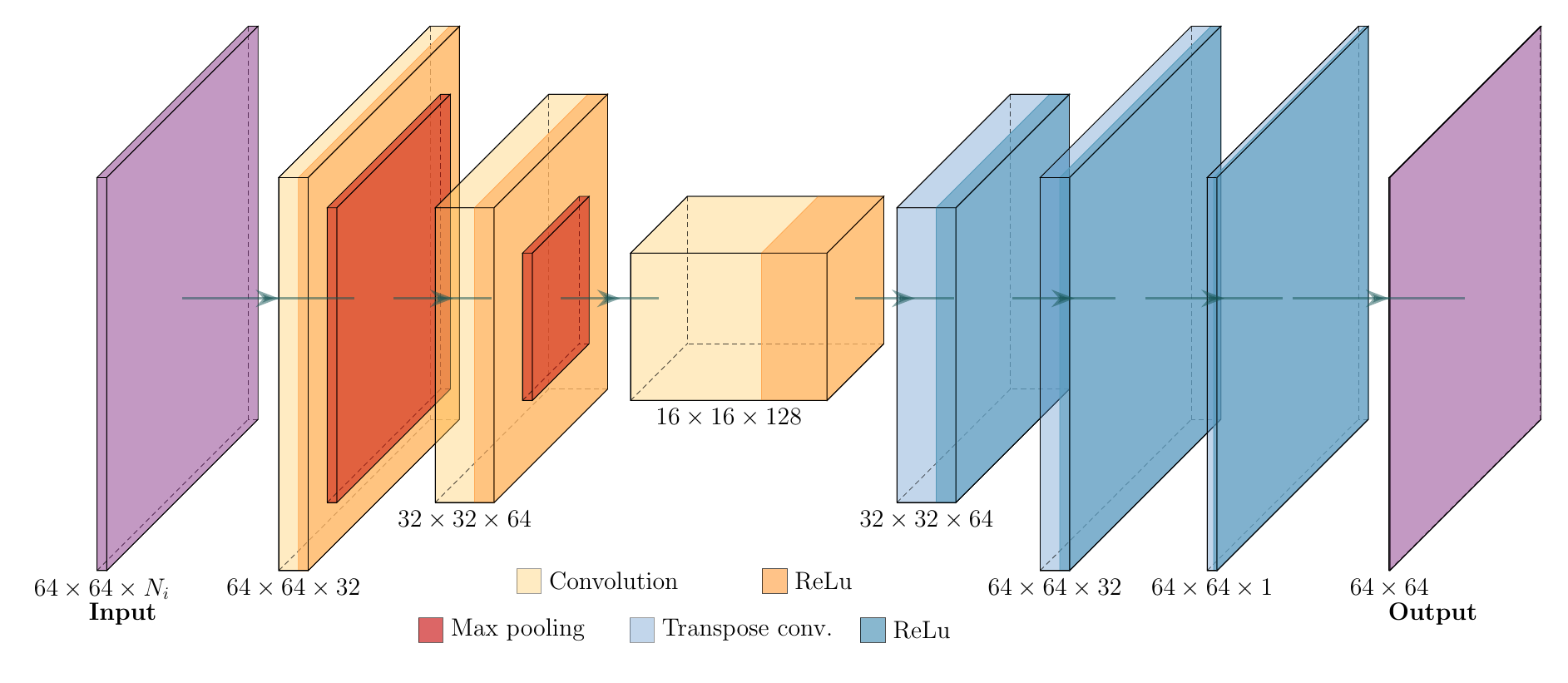}
    \caption{Schematic of the architecture for the (temporal) CNNs in \eq{eq:CNN_split} and
    \eq{eq:cnn}. 
    The numbers below the convolution and transpose convolution blocks
    correspond to the size of the filter and the number of channels
    applied.  For $\text{CNN}_{\text{I,R}}$ in \eq{eq:CNN_split} and 
    $\text{CNN}_U$ in \eq{eq:CNN_U}, $N_i
    = 500$; for $\text{CNN}_{I}$ and $\text{CNN}_{R}$
    from Eqs~\eqref{eq:CNN_I} and \eqref{eq:CNN_R}, $N_i = 1$.\label{fig:cnnarch}}
\end{figure}

\begin{figure}
  \centering 
  \begin{tikzpicture}
    \node[anchor=south west] (fig) at (0,0)
    {\ig[width=.9\tw]{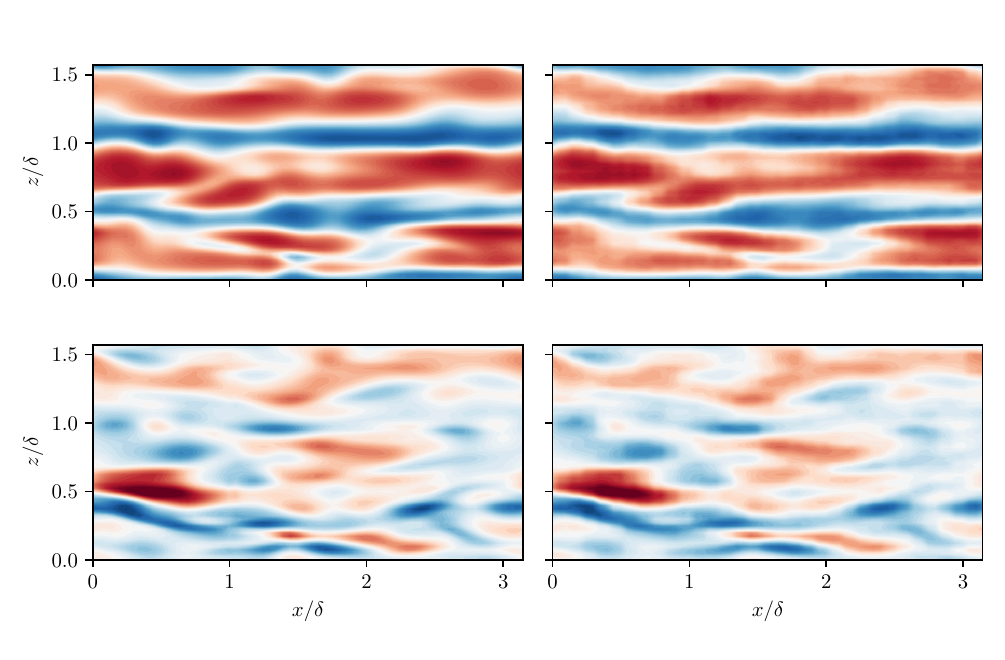}};
    \begin{scope}[x=(fig.south east), y=(fig.north west)]
      \node[anchor=north] at (1.02,.935) 
      {\ig[width=.088\tw]{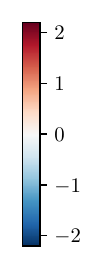}};
      \node[anchor=north] at (1.02,.523) 
      {\ig[width=.088\tw]{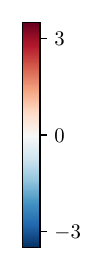}};
      \node[fill=white] at (.3 ,.08) {$x/\chh$};
      \node[fill=white] at (.75,.08) {$x/\chh$};
      \node[fill=white,rotate=90] at (.035,.32) {$z/\chh$};
      \node[fill=white,rotate=90] at (.035,.74) {$z/\chh$};
      \node at (.125,.85) {(a)};
      \node at (.575,.85) {(b)};
      \node at (.125,.44) {(c)};
      \node at (.575,.44) {(d)};
    \end{scope}
  \end{tikzpicture}
\caption{Example of approximate decomposition of $u$ into
    informative $\tilde{u}_I$ and residual $\tilde{u}_R$ components
    using the CNN from \eq{eq:CNN_split} without the need of
    $\tauxf$. (a) $u_I$, (b) $\tilde{u}_I$, (c) $u_R$, and (d) $\tilde{u}_R$.
    \label{fig:CNN_split}}
\end{figure}
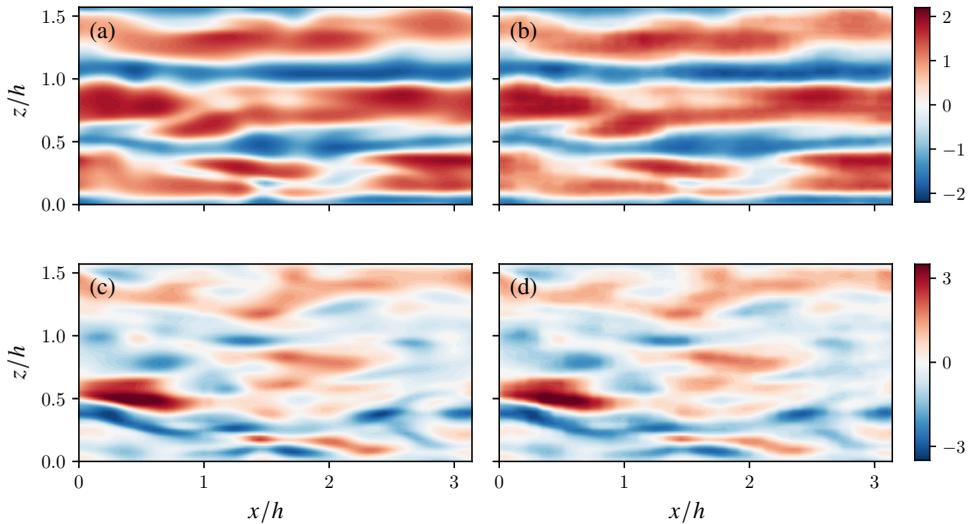

%-------------------------------------------------------------------------%
% CNN setup
Next, we reconstruct the spatially varying wall shear stress
$\tau_x(x,z,t+\Delta T)$ using $u(\boldsymbol{x}_{\text{ref}}, t)$,
where $\boldsymbol{x}_{\text{ref}} = [x, y_{\text{ref}}, z]$ and
$y_{\text{ref}}^*=10$. The steps followed are analogous to those
described above for the time signal prediction. First, we train a
model to approximately decompose $u(\boldsymbol{x}_{\text{ref}}, t)$
into its informative and residual parts without requiring
information about $\tau_x(x,z,t+\Delta T)$. To that end, we use a
temporal convolutional neural network (CNN)~\citep{long2015,guastoni2021} 
of the form:
\begin{equation}
[\tilde{u}_I(\boldsymbol{x}_\text{ref},t), \tilde{u}_R(\boldsymbol{x}_\text{ref},t)] =
\text{CNN}_\text{I,R}\left( u(\boldsymbol{x}_\text{ref},
t), u(\boldsymbol{x}_\text{ref},
t-\delta t),\dots,u(\boldsymbol{x}_\text{ref},
t-p\delta t) \right),\label{eq:CNN_split}
\end{equation}
where $p=500$ and $\delta t^*=0.5$. The CNN is designed to process
input data shaped as three-dimensional arrays, where dimensions
represent spatial coordinates and temporal slices. The CNN comprises
an image input layer, followed by three blocks consisting each of a
convolutional layer, batch normalization, and a ReLU activation
function. Spatial dimensions are reduced through successive max
pooling layers, while feature maps are subsequently upscaled back to
original dimensions via transposed convolutional layers with ReLU
activations.  Further details of the CNN are provided in
\fig{fig:cnnarch}. A total of 12000 snapshots are used, split into
training (80\%) and validation (20\%). An example of the approximate
decomposition from \eq{eq:CNN_split} is shown in
\fig{fig:CNN_split}. 
 
% reconstruction
The three models to predict the 2-D wall shear stress are
\begin{subequations}\label{eq:cnn}
\begin{align}
  \tilde{\tau}_{x}^I(x,z,t + \Delta T) &=
  \text{CNN}_I\left(\tilde{u}_I(\boldsymbol{x}_\text{ref},
  t)\right),\label{eq:CNN_I}\\
    \tilde{\tau}_{x}^R(x,z,t + \Delta T) &=
  \text{CNN}_R\left(\tilde{u}_R(\boldsymbol{x}_\text{ref},
  t)\right),\label{eq:CNN_R}\\
  \tilde{\tau}_{x}^U(x,z,t + \Delta T) &=
  \text{CNN}_U\left( u(\boldsymbol{x}_\text{ref},
  t), u(\boldsymbol{x}_\text{ref},
  t-\delta t),\dots,u(\boldsymbol{x}_\text{ref},
  t-p\delta t) \right),\label{eq:CNN_U}
\end{align}
\end{subequations}
Similarly to the previous case, the first two models only use one time
step for $\tilde{u}_I$ and $\tilde{u}_R$, respectively, whereas the
last model uses multiple times lags for $u$ (with $p=500$ and $\delta
t^*=0.5$).

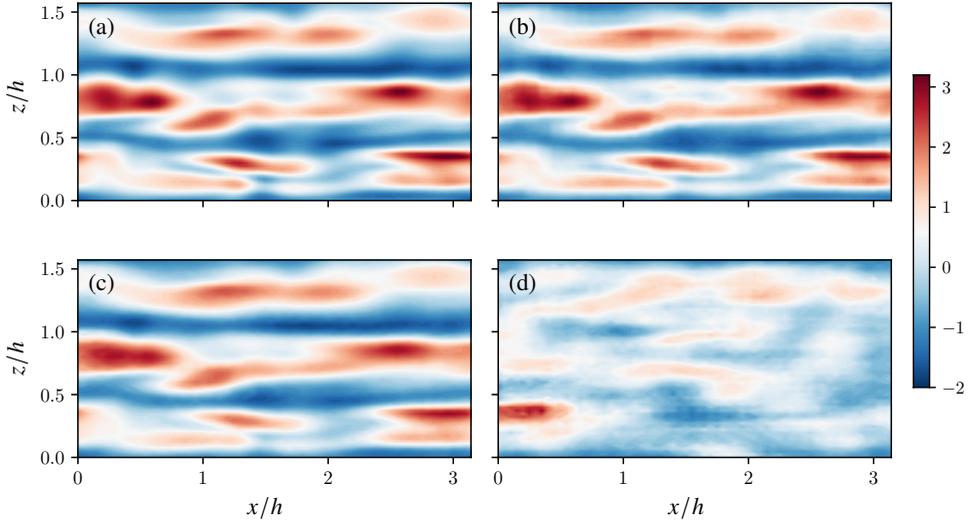
\begin{figure}
  \centering
  \begin{tikzpicture}
    \node[anchor=south west] (fig) at (0,0)
    {\ig[width=.9\tw]{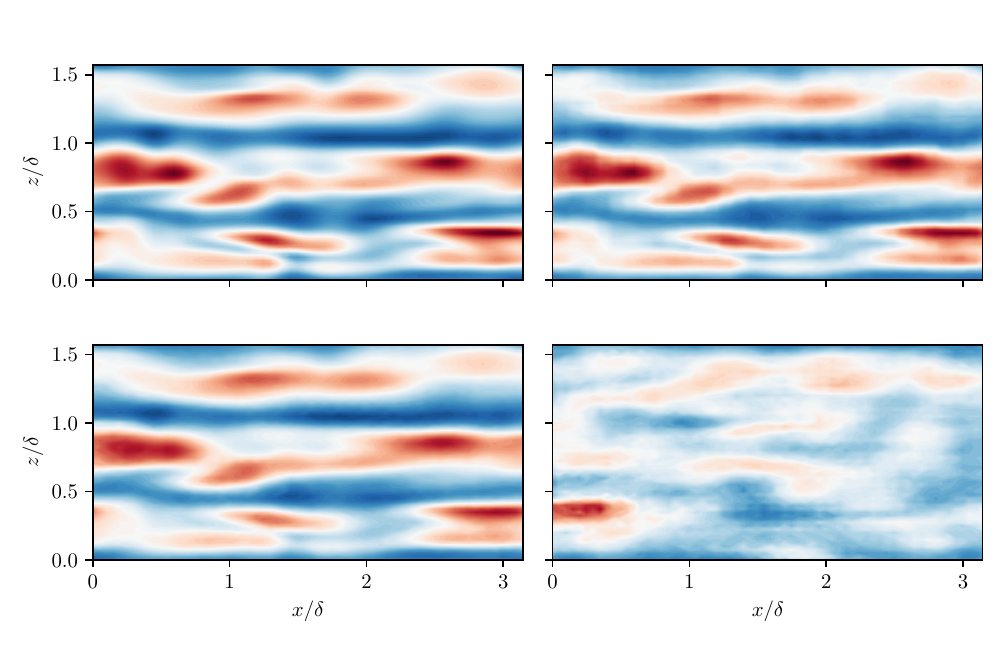}};
    \begin{scope}[x=(fig.south east), y=(fig.north west)]
      \node[anchor=north] at (1.02,.82) 
      {\ig[width=.088\tw]{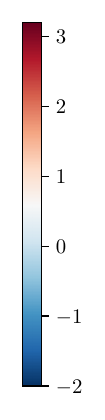}};
      \node[fill=white] at (.3 ,.08) {$x/\chh$};
      \node[fill=white] at (.75,.08) {$x/\chh$};
      \node[fill=white,rotate=90] at (.035,.32) {$z/\chh$};
      \node[fill=white,rotate=90] at (.035,.74) {$z/\chh$};
      \node at (.125,.85) {(a)};
      \node at (.575,.85) {(b)};
      \node at (.125,.44) {(c)};
      \node at (.575,.44) {(d)};
    \end{scope}
  \end{tikzpicture}
\caption{Example of spatial reconstruction of the instantaneous wall shear
    stress (a) using CNNs trained with $u$ (b), $\tilde{u}_I$
    (c), and $\tilde{u}_R$ (d) as inputs,
    respectively. \label{fig:CNN}}
\end{figure}

% results
The spatial reconstruction of the wall shear stress by the three
models is shown in \fig{fig:CNN} for one instant. Consistently with
our previous observations, the reconstructions using $u$ and
$\tilde{u}_I$ as inputs to the model are comparable in both
structure and magnitude, yielding relative mean-squared errors of
28\% and 30\%, respectively. Conversely, the CNN that utilises the
residual component $\tilde{u}_R$ is completely unable to predict the
2-D structure of the wall shear stress, yielding an average relative
error of 120\%. These results further reinforce the idea that models
rely on the informative component of the input to predict the output
variable, whereas the residual component is of no utility.  Finally,
it is worth noting that the CNNs used above have access to the 2-D
spatial structure of $u$ and $\taux$; however, the aIND method,
which was originally used to decompose the flow, only used pointwise
information. This along with the inability of $\tilde{u}_R$ to
predict the wall shear stress further confirms that the assumptions
of the aIND method hold reasonably well in this case.

%%%%%%%%%%%%%%%%%%%%%%%%%%%%%%%%%%%%%%%%%%%%%%%%%%%%%%%%%%%%%%%%%%%%%%%%%%%%%%%%
%% RESULTS - CONTROL
\subsection{Control: wall-shear stress reduction with opposition control} 
\label{sec:res:control}

% intro
We investigate the application of the {\method} to opposition control
in a turbulent channel flow~\citep{choi1994,hammond1998}. Opposition
control is a drag reduction technique based on blowing and sucking
fluid at the wall with a velocity opposed to the velocity measured at
some distance from the wall.  The hypothesis under consideration in
this section is that the informative component of the wall-normal
velocity is more impactful for controlling the flow compared to the
residual component.
The rationale behind this hypothesis is grounded in the
information-theoretic formulation of observability introduced by
\citet{lozanoduran2022}. This formulation defines the observability
of a variable ($\tauxf$) in terms of the knowledge gained from
another variable ($v$) as:
\begin{equation}\label{eq:obs}
    O_{v\to\tauxf} = \frac{I(\tauxf;v)}{H(\tauxf)}.
\end{equation}
The variable $\tauxf$ is said to be perfectly observable with respect
to $v$ when $O_{v\to\tauxf} = 1$, i.e. there is no uncertainty in the
state to be controlled conditioned to knowing the state of the
sensor. Conversely, $\tauxf$ completely unobservable when
$O_{u\to\tauxf} = 0$, i.e., the sensor does not have access to any
information about $\tauxf$. The greater the observability, the more
information is available for controlling the system. By substituting
Eqs.~\eqref{eq:cond2} and \eqref{eq:cond1} into \eq{eq:obs}, it
is easy to shows that $\tauxf$ is unobservable with respect to the
residual component ($O_{v_R\to\tauxf} = 0$), and perfectly observable
from the perspective of the informative component ($O_{v_I\to\tauxf} = 1$).

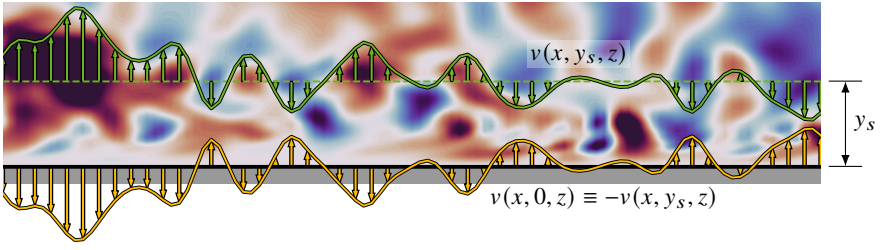
\begin{figure}
    \centering
    %\definecolor{vcon}{rgb}{0.8,0.7254901960784313,0.4549019607843137}
\definecolor{vcon}{HTML}{6EB621}
\begin{tikzpicture}[>={Latex[length=.15cm]}]
    \begin{footnotesize}

    \node[anchor=south west] (fig) at (0,0)
        {\ig[width=.8\tw,trim=0 .8cm 0 0,clip]{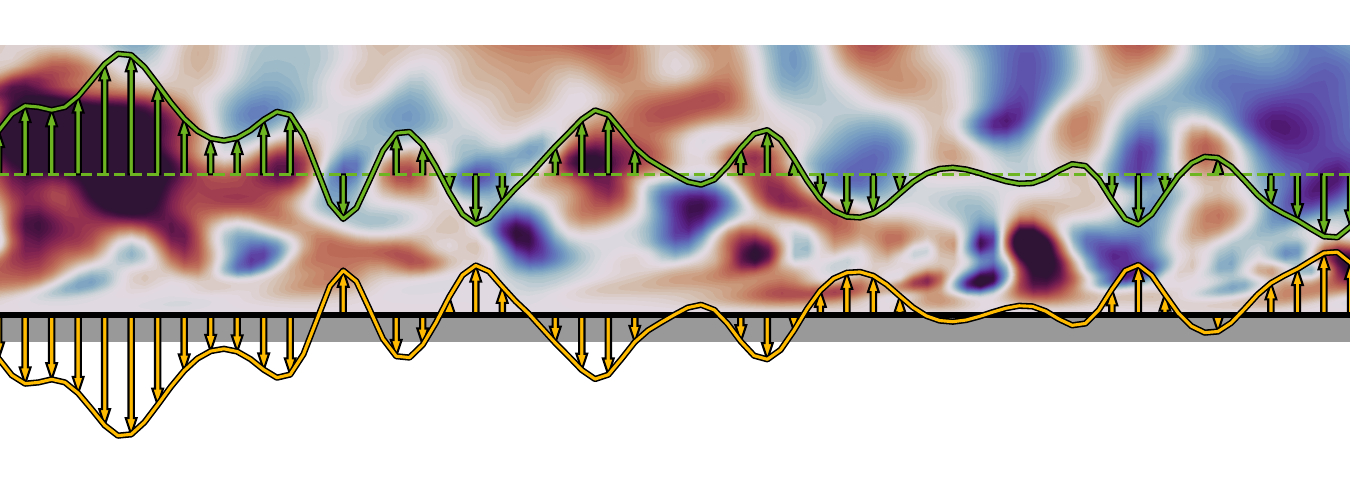}};

    \begin{scope}[x=(fig.south east),y=(fig.north west)]
        %\draw[green,ultra thin,step=.1] (0,0) grid (1,1); 
        \coordinate (c0) at (1.,.31);
        \coordinate (c1) at (1.,.605);

        \draw[thin] (c0)--+ (.04,0)
                    (c1)--+ (.04,0);

        \draw[<->] ($(c0)+(.02,0)$) -- ($(c1)+(.02,.0)$) 
        node [midway,anchor=west] {$y_s$};

        \node[black,fill=white,opacity=.3,text opacity=1,inner sep=1pt]
        at (.70,.7) {$v(x,y_s,z)$};

        \node[black,fill=white,opacity=.3,text opacity=1,inner sep=1pt]
        at (.73,.2) {$v(x,0,z) \equiv -v(x,y_s,z)$};
        
        %\fill[black!40!white] (.01,.05) rectangle (.99, 0);
        %\draw[black,thick] (.01,.05) -- (.99, .05);

    \end{scope}
    \end{footnotesize}
\end{tikzpicture}
    \caption{ Schematic of the opposition control.
    The contour corresponds to the instantaneous vertical velocity on
    a $(x,y)$-plane, the green line depicts the instantaneous velocity
    at the sensing plane, $y_s$, and the yellow line depicts the
    velocity at the wall imposed by the classical opposition control
    technique.
    Colormap ranges from (blue) $v^* = -1.8$ to (red) $1.8$.
    \label{fig:vcontrol}}
\end{figure}

% set up
\Fig{fig:vcontrol} shows a schematic of the problem setup for
opposition control in a turbulent channel flow.  The channel is as in
\S\ref{sec:res:tau:recons} but the wall-normal velocity at the wall is
replaced by $v(x,0,z,t) = f(v(x,y_s,z,t))$, where $y_s$ is the
distance to the \emph{sensing} plane, and $f$ is a user-defined
function.  In the original formulation by \citet{choi1994}, $f \equiv
-v(x,y_s,z,t)$, hence the name of \emph{opposition} control.  Here, we
set $y_s^* \approx 14$, which is the optimum wall distance reported in
previous works~\citep{chung2011,lozanoduran2022}.  Two Reynolds
numbers are considered, $\Rey_\tau= 180$ and $395$.

% problem statement with V_I and V_R
We split $v(x,y_s,z,t)$ into its informative ($v_I$) and residual
($v_R$) components to $\taux(x,z,t)$.  Three controllers are
investigated. In the first case, the function of the controller $f$ is
such that it only uses the informative component of $v(x,y_s,z, t)$,
namely $f(v(x,y_s,z,t)) \equiv -v_I(x,y_s,z,t)$. In the second case,
the controller uses the residual component $f(v(x,y_s,z,t)) \equiv
-v_R(x,y_s,z,t)$.  Finally, the third controller follows the original
formulation $f(v(x,y_s,z,t)) \equiv -v(x,y_s,z,t)$.

% approximation
This is a more challenging application of the {\method} due to the
dynamic nature of the control problem. When the flow is actuated, the
dynamics of the system change, and the controller should re-compute
$v_I$ (or $v_R$) for the newly actuated flow. This problem is
computationally expensive, and we resort to calculating an
approximation. The control strategy is implemented as follows:
\begin{enumerate}
    \item A simulation is performed with $f \equiv -v(x,y_s,z,t)$,
      corresponding to the original version of opposition control.
    \item The informative term ($v_I$) of $v(x,y_s,z,t)$ related to
      the wall shear stress $\taux(x,z,t)$ is extracted for $\Delta T
      = 0$.
    \item We find an approximation of the controller, such that
      $\tilde{v}_I = f(v) \approx -v_I$.  To obtain this
      approximation, we solve the minimisation problem
\begin{equation}
    \arg\min_{\tilde{v}_I} \quad \| v_I - \tilde{v}_I \|^2
    + \gamma \frac{I( \taux; \tilde{v}_R)}{H(\taux)}
\end{equation}
where $\gamma = 0.75$.  The approximated informative term is modelled
as a feed-forward artificial neural network with 3 layers and 8
neurons per layer.
      %A controller is developed to approximate $v_I$, denoted as
      %$\tilde{v}_I \equiv -f(v) \approx v_I$.
    \item Two new simulations are conducted using either $\tilde{v}_I$
      or $\tilde{v}_R = v - \tilde{v}_I$ for opposition control.
\end{enumerate}
Note that the devised controller can be applied in real time
(i.e., during simulation runtime), since the estimated information
component, $\tilde{v}_I(t)$, is computed using only information from
the present time instant, $v(t)$.

\begin{figure}
    \centering
    \begin{tikzpicture}
        \node[anchor=south west] (fig) at (0,0) 
        {\ig[width=.5\tw]{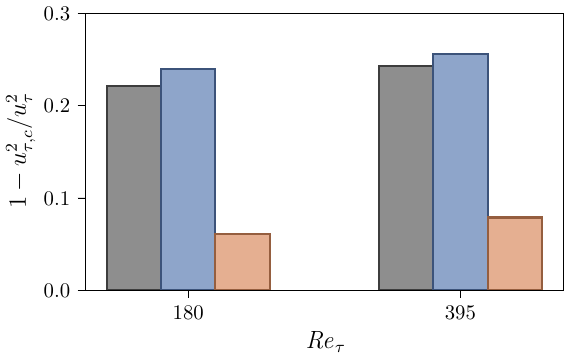}};
        \begin{scope}[x=(fig.south east), y=(fig.north west)]
            \node[fill=white] at (0.56,.06) {$\Rey_\tau$};
            \node[fill=white,rotate=90] at (0.04,.54) 
            {$1 - u_{\tau,c}^2 / u_{\tau}^2$};
        \end{scope}
    \end{tikzpicture}
    \caption{Drag reduction, computed as $1-u_{\tau,c}^2/u_{\tau}^2$,
      where $u_{\tau,c}$ is the friction velocity for the controlled
      case.
    Each colour corresponds to a different controller: (black) $f
    \equiv -v$; (blue) $f \equiv -\tilde{v}_I$; and (orange) $f \equiv
    -\tilde{v}_R$.
    \label{fig:opp_control}}
\end{figure}

% results
\Fig{fig:opp_control} summarises the drag reduction for the three
scenarios, namely: $f \equiv -v(x,y_s,z,t)$, $f \equiv
-\tilde{v}_I(x,y_s,z,t)$, and $f \equiv -\tilde{v}_R(x,y_s,z,t)$. The
original opposition control achieves a drag reduction of approximately
$22\%$ and $24\%$ for $\Rey_\tau=180$ and $\Rey_\tau=395$,
respectively. 
Similar reductions in drag using the same controller have been
documented in the literature~\citep{chung2011,luhar2014}.
The values show a marginal dependency on $\Rey_\tau$,
in agreement with previous studies~\citep{iwamoto2002}. Opposition
control based on $\tilde{v}_I$ yields a moderate increase in drag
reduction with a $24\%$ and $26\%$ drop for each $\Rey_\tau$,
respectively.  Conversely, the drag reduction is only up to $7\%$ for
the control based on the estimated residual velocity, $\tilde{v}_R$.
Note that $v_I$ is the component of $v$ with the highest potential to
modify the drag. Whether the drag increases or decreases depends on
the specifics of the controller. On the other hand, the residual
component $v_R$ is expected to have a minor impact on the drag. As
such, one might anticipate a 0\% drag reduction by using
$v_R$. However, the approximation $\tilde{v}_R$ retains some
information from the original velocity for intense values of the
latter, which seems to reduce the drag on some occasions.  Simulations
using $f \equiv - k \tilde{v}_R$ --with $k$ adjusted to $f \sim \|
v(x,y_s,z,t) \|^2$-- were also conducted, yielding no additional
improvements in the drag reduction beyond 8\%.
It is also interesting to note that, after performing steps
(i)-(iv) of the control strategy, the informative content in $v$
substantially increases (from $E_I^v \approx 0.1$ to $E_I^v \approx
0.8$).  This phenomenon exposes the dynamic nature of the control
problem highlighted above.

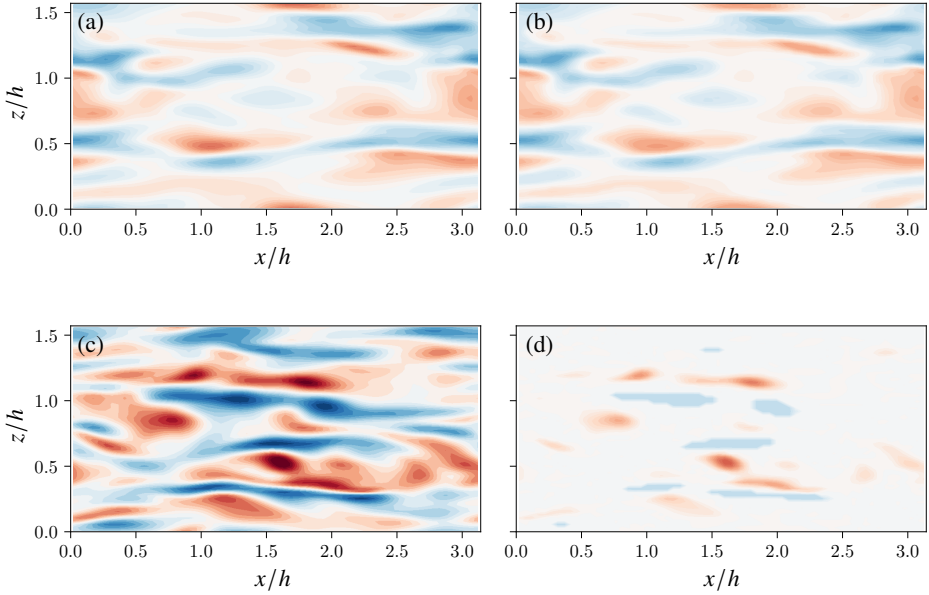
\begin{figure}
    \centering
    \begin{tikzpicture}
        \node[anchor=south west] (fig) at (0,0)
        {\ig[width=.95\tw]{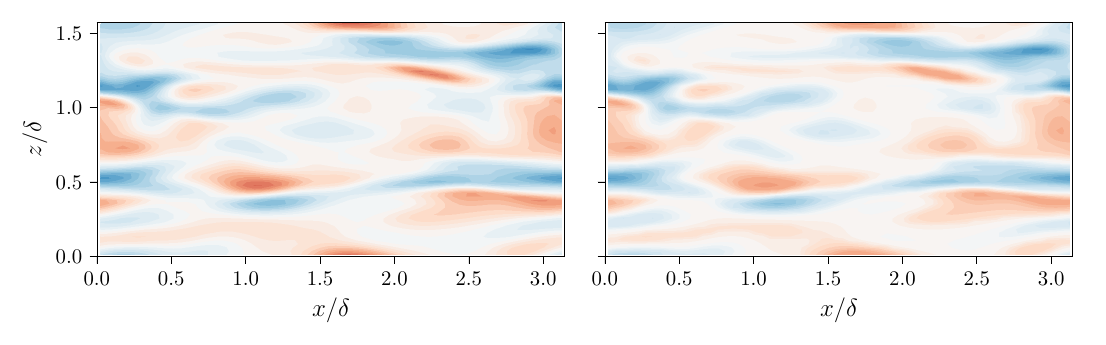}};
        \begin{scope}[x=(fig.south east), y=(fig.north west)]
            \node at (.115,.85) {(a)};
            \node at (.57,.85) {(b)};
            \node[fill=white] at (.30,.12) {$x/\chh$};
            \node[fill=white] at (.755,.12) {$x/\chh$};
            \node[rotate=90,fill=white] at (.04,.60) {$z/\chh$};
        \end{scope}
        \phantomsubcaption\label{fig:opp_control_vsI}
        \phantomsubcaption\label{fig:opp_control_vwI}
    \end{tikzpicture}
    \begin{tikzpicture}
        \node[anchor=south west] (fig) at (0,0)
        {\ig[width=.95\tw]{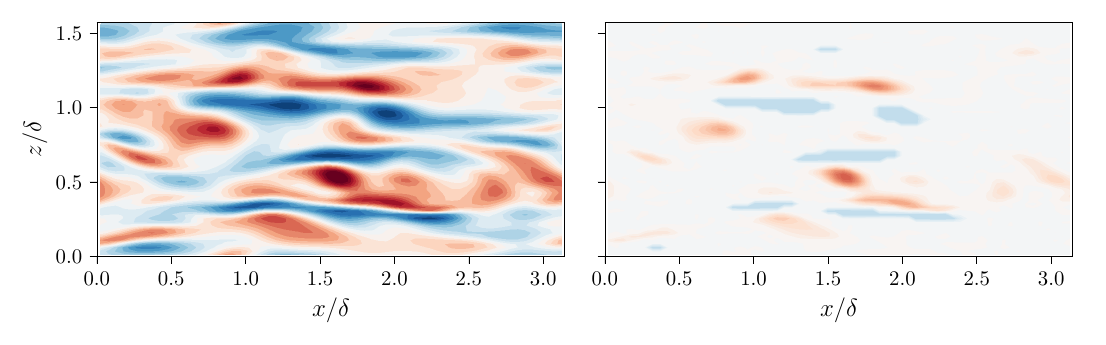}};
        \begin{scope}[x=(fig.south east), y=(fig.north west)]
            \node at (.115,.85) {(c)};
            \node at (.57,.85) {(d)};
            \node[fill=white] at (.30,.12) {$x/\chh$};
            \node[fill=white] at (.755,.12) {$x/\chh$};
            \node[rotate=90,fill=white] at (.04,.60) {$z/\chh$};
        \end{scope}
        \phantomsubcaption\label{fig:opp_control_vsR}
        \phantomsubcaption\label{fig:opp_control_vwR}
    \end{tikzpicture}
    \caption{(a) Wall normal velocity in the sensing plane, $v(x,y_s,z)$ and 
    (b) the minus velocity imposed at the wall for the case 
    $f\equiv -\tilde{v}_I$.
    (c,d) Same as (a,b) but for the case $f\equiv -\tilde{v}_R$.
    Contours range from (blue) $v^* = -0.9$ to (red) $v^* = 0.9$.
    \label{fig:opp_control_V}}
\end{figure}

% flow structure
\Fig{fig:opp_control_vsI,fig:opp_control_vsR} show the wall-normal
velocity in the sensing plane for the controlled cases at $\Rey_\tau =
180$ with $f \equiv -\tilde{v}_I$ and $f \equiv -\tilde{v}_R$,
respectively. Larger velocity amplitudes are observed in
\fig{fig:opp_control_vsR} compared to \fig{fig:opp_control_vsI},
indicating that higher Reynolds stresses are expected, which aligns
with a larger average wall shear stress. On the other hand,
\fig{fig:opp_control_vwI,fig:opp_control_vwR} display the negative
wall-normal velocity imposed at the boundary for the cases with $f
\equiv -\tilde{v}_I$ and $f \equiv -\tilde{v}_R$, respectively. The
informative component, $\tilde{v}_I$, closely resembles the original
velocity but with smaller amplitudes at extreme events of $v$.  This
appears to play a slightly beneficial role in drag
reduction. Conversely, \fig{fig:opp_control_vwR} shows that the
estimated residual component is negligible except for large values of
$v$. This is responsible for the smaller reduction in the mean
drag. Although not shown, similar flow structures are observed for
$\Rey_\tau = 395$, and the same discussion applies.  
In summary, we have utilised an example of opposition control in a turbulent 
channel to demonstrate the utility of IND.
However, it is important to emphasise that the primary focus of this section is 
not on the real-time applicability or the performance of the control in this 
specific case. 
Instead, the main message we aim to convey is more fundamental: the informative 
component of the variable measured by the sensor holds the essential information 
needed to develop successful control strategies, while the residual component is 
not useful in this regard.
%
% running the controller

%%%%%%%%%%%%%%%%%%%%%%%%%%%%%%%%%%%%%%%%%%%%%%%%%%%%%%%%%%%%%%%%%%%%%%%%%%%%%%%%
\section{Conclusions}\label{sec:conclusions}

% method summary
We have presented {\method}, a method for decomposing a flow field
into its informative and residual components relative to a target
field.  The informative field contains all the information necessary
to explain the target variable, contrasting with the residual
component, that holds no relevance to the target variable. The
decomposition of the source field is formulated as an optimisation
problem based on mutual information. To alleviate the computational
cost and data requirements of {\method}, we have introduced an
approximate solution, referred to as {\amethod}. This approach still
ensures that the informative component retains the information about
the target, by minimising the mutual information between the residual
and the target in a pointwise manner.

The {\method} is grounded in the fundamental principles of information
theory, offering key advantages over other methods. As such, it is
invariant under shifting, rescaling, and, in general, non-linear
$\mathcal{C}^1$-diffeomorphism transformations of the source and target
variables~\citep{kaiser2002}. The method is also fully non-linear
and does not rely on simplifications such as the Gaussianity of the
variables. This makes IND a suitable tool for studying turbulent
phenomena, which are intrinsically non-linear. In contrast, other
linear correlation-based methods, such as LSE and EPOD, are not well
equipped to capture non-linearities in the flows. Additionally, we
have shown that the pointwise formulation of method (aIND)
represents a cost-effective and memory-efficient implementation IND
without sacrificing performance compared to correlation-based
methods. This approach also allows for the
assimilation of experimental data.

% physics
The method has been applied to study the information content of the
velocity fluctuations in relation to the wall shear stress in a
turbulent channel flow at $\Rey_\tau = 180$. Our findings have
revealed that streamwise fluctuations contain more information about
the future wall shear stress than the cross-flow velocities. The
energy of the informative streamwise velocity peaks at $y^* \approx
10$, slightly below the well-known peak for total velocity, while the
residual component peaks at $y^* \approx 30$. This suggests that the
peak observed in the total velocity fluctuations results from both
active and inactive velocities, with `active' referring to motions
connected to changes in the wall shear stress. Further investigation
of the coherent structure of the flow showed that the informative
velocity consists of smaller near-wall high- and low-velocity streaks
collocated with vertical motions (i.e., sweeps and ejections). The
spanwise informative velocity is weak, except close to the wall within
the bottom root of the streamwise rolls. This informative streak-roll
structure is embedded within a larger-scale streak-roll structure from
the residual velocity, which bears no information about the wall-shear
stress for the considered time scale. We have also shown that
ejections propagate information about the wall stress further from the
wall than sweeps, while extreme values of the wall shear stress are
attributed to sweeps in close proximity to the wall.

% modeling
The utility of {\method} for reduced-order modelling was demonstrated
in the prediction of the wall shear stress in a turbulent channel
flow.  The objective was to estimate the 2-D wall shear stress
in the future, after $\Delta T^*=25$, by measuring the streamwise
velocity in a wall-parallel plane at $y^* \approx 10$ as input. The
approach was implemented using a fully convolutional neural network as
the predictor. Two cases were considered, using either the informative
or the residual velocity component as input, respectively. We have
shown that the model can make accurate predictions when the
observability between the input and the target is maximum as it is
the case when the input is the informative component of the
velocity. The main discrepancies were localised in regions with high
wall shear stress values. This outcome aligns with our prior analysis,
which indicated that extreme wall-shear stress events are produced by
short-time near-wall sweeps not captured in the input plane. In
contrast, the residual velocity component offers no predictive power
for wall shear stress, as it has no observability of the wall
shear stress, meaning that it lacks any information relevant to the
latter.  This example in reduced-order modelling reveals that models
achieving the highest performance are those that utilise input
variables with the maximum amount of information about the output.

% control
Finally, we have investigated the application of {\method} for drag
reduction in turbulent channel flows at $\Rey_\tau = 180$ and
$\Rey_\tau = 395$. The strategy implemented involved blowing/suction
via opposition control. To this end, the no-transpiration boundary
condition at the wall was replaced with the wall-normal velocity
measured in the wall-parallel plane at $y^*=14$. We explored the use
of three wall-normal velocities: the total velocity (i.e., as
originally formulated in opposition control), its informative
component, and its residual component. The largest reduction in drag
was achieved using the informative component of $v$, which performed
slightly better than the total velocity for both Reynolds numbers. The
residual component was shown to yield the poorest results. The
application to drag reduction demonstrated here illustrates that the
informative component of $v$ contains the essential information needed
for effective flow control. This paves the way for using {\method} to
devise enhanced control strategies by isolating the relevant
information from the input variables while disregarding the irrelevant
contributions.

% closing
We conclude this work by highlighting the potential of {\method} as a
post-processing tool for gaining physical insight into the
interactions among variables in turbulent flows. Nonetheless, it is
also worth noting that the approach relies on the mutual information
between variables, which requires estimating joint probability density
functions. This entails a data-intensive process that could become a
constraint in cases where the amount of numerical or experimental data
available is limited. Future efforts will be devoted to reducing the
data requirements of {\amethod} and extending its capabilities to
account for multi-variable and multiscale interactions among
variables.

%\backsection[Supplementary data]{\label{SupMat}Supplementary material and movies are available at \\https://doi.org/10.1017/jfm.2019...}

\backsection[Acknowledgements]{
The authors acknowledge the Massachusetts
Institute of Technology, SuperCloud, and Lincoln Laboratory
Supercomputing Center for providing HPC resources that have
contributed to the research results reported here.
The schematic of the CNN have been created using \href{https://doi.org/10.5281/zenodo.2526396}{PlotNeuralNet}.}

\backsection[Funding]{
 This work was supported by the National Science Foundation under Grant No. 
 2140775 and MISTI Global Seed Funds and UPM. 
  G.~A. was partially supported by the NNSA Predictive
  Science Academic Alliance Program (PSAAP; grant DE-NA0003993).}

\backsection[Declaration of interests]{The authors report no conflict of interest.}

\backsection[Data availability statement]{The code and examples 
of {\amethod} are openly available at
\href{https://github.mit.edu/Computational-Turbulence-Group/caudecIT}{INDtools}.}

\backsection[Author ORCIDs]{G. Arranz,
https://orcid.org/0000-0001-6579-3791; 
A. Lozano-Dur\'an, https://orcid.org/0000-0001-9306-0261}

%\newpage
\appendix

\section{Numerical implementation}\label{app:implementation}

\subsection{Solution for scalar variables using bijective functions}
\label{app:imp:bijective}

Here we provide the methodology to tackle the minimisation problem posed in
\eq{eq:problem_scalar}.
For convenience, we write \eq{eq:problem_scalar} again
\begin{equation}\label{eq:problem_app}
    \arg \min_{\cau,\mathcal{F}} 
    \quad 
	I(\var - \cau;\cau) + \gamma \| \var - \cau \|^2
	\qquad \mathrm{s.t.}\quad \q_{+} = \mathcal{F}(\cau).
\end{equation}
To solve \eq{eq:problem_app}, we note that there are 2 unknowns: $\cau$
and the function $\mathcal{F}$.
If we assume that $\mathcal{F}$ is invertible, namely
\begin{equation}
	\cau(t) = \mathcal{F}^{-1}(\q_+ (t)) \equiv \mathcal{B}(\q_+ (t)),
\end{equation}
then, \eq{eq:problem_app} can be recast as
\begin{equation}\label{eq:problem4}
    \arg \min_{\mathcal{B}} \quad 
	I( \var-\mathcal{B}(q_+);\mathcal{B}(q_+) )
	+ \gamma \| \var - \mathcal{B}(\q_+) \|^2,
\end{equation}
which can be solved by standard optimisation techniques upon the parametrisation
of the function $\mathcal{B}$.
 
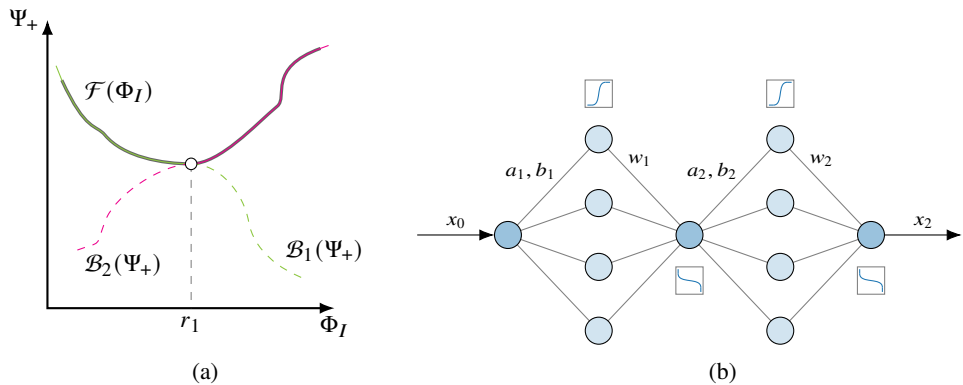
\begin{figure}
	\centering
	\begin{subfigure}{.4\tw}
		\begin{tikzpicture}[scale=3.8]

    \draw[<->, thick] (1,0) -- (0,0) -- (0,1);

    \draw[black!60!white, very thick] (.05,.79) .. controls ++(-65:.2) and ++(130:.05) .. (.2,.6) .. controls ++(130:-.1) and ++(0:-.1) .. (.5,.5)
                                                .. controls ++(0:.1) and ++(40:-.1) .. (.8,.7) .. controls ++(40:.05) and ++(20:-.2) .. (.95,.9);

    \draw[LimeGreen] (.03,.84) -- (.05,.79) .. controls ++(-65:.2) and ++(130:.05) .. (.2,.6) .. controls ++(130:-.1) and ++(0:-.1) .. (.5,.5);
    \draw[LimeGreen, dashed] (.5,.5) .. controls ++(0:.1) and ++(100:.1) .. (.7,.3) .. controls ++(100:-.1) and ++(160:.1) .. (.9,.1);
    \draw[magenta] (.5,.5) .. controls ++(0:.1) and ++(40:-.1) .. (.8,.7) .. controls ++(40:.05) and ++(20:-.2) .. (.95,.9) -- ++(20:.03);
    \draw[magenta, dashed] (.5,.5) .. controls ++(0:-.1) and ++(70:.1) .. (.2,.3) .. controls ++(70:-.1) and ++(20:.1) .. (.1,.2);

    \begin{footnotesize}
    \node[anchor=west] at (.1,.75) {$\mathcal{F}(\cau)$};
    \node[anchor=west] at (.8,.2)  {$\mathcal{B}_1(\q_+)$};
    \node[anchor=west] at (.1,.15) {$\mathcal{B}_2(\q_+)$};

    \node[anchor=north] at (1,0) {$\cau$};
    \node[anchor=east ] at (0,1) {$\q_+$};

    \draw [dashed, thin, black!50!white] (.5,.5) -- (.5,0) node [anchor=north, black] {$r_1$};

    \end{footnotesize}

    \draw [fill=white, thin] (.5,.5) circle (.02);

\end{tikzpicture}
		\caption{\label{fig:funcs}}
	\end{subfigure}~
	\begin{subfigure}{.6\tw}
		\tikzset{connections/.style={thin, black!50}}

\tikzset{neuron/.style = {circle, draw = black, fill, minimum size=#1,inner sep=0pt, outer sep=0pt},
	neuron/.default = 6pt  % size of the circle diameter 
}

\tikzset{pics/sigmoid/.style = { code = { % n args={5}{ code = {
        \draw (-.5,-.5) -- (-.3,-.5) edge[out=0,in=180] (.3,.5); 
        \draw (.3,.5) -- (.5,.5); 
        \draw[black!50] (-.6,-.6) rectangle (.6,.6);
    }
    }
}

\pgfmathsetmacro{\Rn}{2.5ex}      % radius of the neuron
\pgfmathsetmacro{\Nlayers}{2}   % number of hidden layers
\pgfmathsetmacro{\Nneurons}{4}  % number of neurons per layer

\pgfmathsetmacro{\hdis}{.7pt}   % vertical separation between neurons in the layer

\begin{tikzpicture}[scale=1.2]

    \begin{scriptsize}
    \foreach \i in {1,...,\Nlayers} {

        \coordinate (n\i) at ({2*(\i-1)},0); % coordinate of the logit neuron

    	\foreach \j in {1,...,\Nneurons} {
            
            % coordinate of each neuron inside the layer
            \coordinate (ln\i\j) at (2*\i-1,{(\j-1-(\Nneurons-1)/2)*\hdis});

            \draw[connections] (n\i) -- (ln\i\j); % from logit to layer
 
            \draw[connections] (ln\i\j) -- (2*\i,0); % from layer to next logit

            \node[neuron=\Rn,fill=C0!20] at (ln\i\j) {};

    	}
        \node[neuron=\Rn,fill=C0!45] (myn\i) at (n\i) {}; 

        \node[anchor=east] at ($(ln\i\Nneurons)+(-.4,-\hdis/2)$) {$a_\i,b_\i$}; 
        \node[anchor=west] at ($(ln\i\Nneurons)+(+.25,-\hdis/3)$) {$w_\i$}; 

        \pic[scale=.3,C0] at ($(ln\i\Nneurons)+(0,.5)$) {sigmoid};
        \pic[scale=.3,C0,rotate=90] at ($(n\i)+(2,-.5)$) {sigmoid};
    }
    \node[neuron=\Rn,fill=C0!45] (mynf) at (2*\Nlayers,0) {}; 
    
    \draw[->] (-1,0) -- (myn1) node [midway,above] {$x_0$};
    \draw[->] (mynf) --+(1,0) node [midway,above] {$x_\Nlayers$};

    \end{scriptsize}

\end{tikzpicture}
		\caption{\label{fig:dsf}}
	\end{subfigure}
	\caption{
	(a) Illustration of a surjective function, $\mathcal{F}(\cau)$, and its
	decomposition into
	two bijective functions: $\mathcal{B}_1(\q_+)$ and
	$\mathcal{B}_2(\q_+)$.
	(b) Example of a DSF architecture with 2 hidden layers and 4 neurons per
	hidden layer. The functions plotted within boxes are the activation 
    functions acting on the neurons.
    Adapted from \citet{huang2018}.}
\end{figure}

However, by imposing bijectivity we constrain the feasible
$\cau(t)$ solutions that satisfy $H(\q_+ | \cau) = 0$ and could lead to lower
%$\| \var - \cau\|^2$ 
values of the loss function than in the more lenient case, where $\mathcal{F}$
only needs to be surjective.
To circumvent this limitation, we recall that a surjective function with
$N-1$ local extrema points (points where the slope changes sign)
can be split into $N$ bijective functions (see \fig{fig:funcs}).
In particular, we define
\begin{equation}\label{eq:caut}
	\cau(t) = \mathcal{B}_i(\q_+(t)) \, | \, \var(t) \in [ r_{i-1}, r_i )\quad \forall i=1,\dots,N,
\end{equation}
where $r_i$ is the $i$th local extremum, such that $r_i > r_{i-1}$, $r_0 \rightarrow -\infty$,
and $r_N \rightarrow \infty$.

Therefore, the final form of the minimisation equation is
\begin{equation}\label{eq:sur_B}
    %\hat{\mathcal{B}}_i := \arg \min_{\mathcal{B}_{i}} \quad
    \arg \min_{\mathcal{B}_{i}} \quad
	I(\var-\mathcal{B}^{\cup}(\q_+);\mathcal{B}^{\cup}(\q_+))
	+ \gamma \sum_i^N \| \var - \mathcal{B}_i^0(\q_+;\var) \|^2,
\end{equation}
being
\begin{align*}
	\mathcal{B}_i^0(\q_+;\var) & = \begin{cases} \mathcal{B}_i, & \text{if}\quad\var(t) \in [ r_{i-1}, r_i ) \\ 0 & \text{otherwise} \end{cases} &
	\mathcal{B}^\cup(\q_+)     & = \sum_i^N \mathcal{B}_i^0(\q_+;\var),
\end{align*}
where the extrema ($r_i$) are unknowns to be determined in the minimisation 
problem, and $\gamma$ and $N$ are the only free parameters.
Once the functions $\mathcal{B}_i$ are computed, the informative component is obtained 
from
\begin{equation}
    {\var}_I(t) = {\mathcal{B}}_j(\q_+(t)) \quad | \quad
    j = \arg \min_{i} \quad \left( \var(t) - \mathcal{B}_i(\q_+(t)) \right)^2
\end{equation}
at every time step.

We use feed-forward networks to find $\mathcal{B}_i$,
as they are able to approximate any Borel-measurable function on a compact
domain \citep{hornik1989}.
In particular, we use the deep simgoidal flow (DSF) proposed by \citet{huang2018},
who proved that a feed-forward artificial neural network is a bijective
transformation if the activation functions are bijective and all the weights
are positive.
The details of the DSF architecture and the optimisation can be found
in Appendix~\ref{app:opt}.

One must emphasise that the current minimisation problem posed in 
\eq{eq:problem_scalar} differs from the classical flow reconstruction 
problem (e.g.: \citet{erichson2020}) where the maximum reconstruction of 
$\var$ is sought.
In those cases, we look for a function $\mathcal{G}(\q_+)$ that minimises 
$\| \var - \mathcal{G}(\q_+)\|^2$.
If the result is a non-bijective function, the constraint $H( \q_+ | \cau ) = 0$
will not be satisfied.

\subsection{Networks architecture and optimisation details}\label{app:opt}

The present algorithm uses DSF networks to approximate bijective functions.
This network architecture is depicted in \fig{fig:dsf}.
The DSF is composed of $L$ stacked sigmoidal transformations.
Each transformation produces the output,
\begin{equation}
	x_l = \sigma^{-1} \left( w_l^T \cdot \sigma ( a_l \cdot x_{l-1} + b_l) \right) \quad l = 1,\dots,L
\end{equation}
where $x_{l-1}$ is the input, $\sigma(y) = 1/(1 + e^{-y})$ is the logistic function,
$\sigma^{-1}$ is the inverse of $\sigma$, ${a}_l$ and ${b}_l$ are vectors with the
weights and biases of the decoder part of the $l$-layer, and $w_l$ is a vector
with the weights of the encoder part of the $l$-layer (see \fig{fig:dsf}).
In addition, the weights for each layer have to fulfil $0 < w_{l,i} < 1$, $\sum_i w_{l,i} = 1$, and $a_{l,i} > 0$, $i = 1,\dots,M$,
where $M$ is the number of neurons per layer.
These constraints are enforced via the softmax and the exponential activation functions for $w_l$ and $a_l$, respectively. Namely:
\begin{align*}
	w_{l,i} & = \frac{\exp({w^\prime_{l,i}})}{\sum_{i=1}^N \exp({w^\prime_{l,i}})} &
	a_{l,i} & = \exp({a^\prime_{l,i}}).
\end{align*}
More details on the DSF architecture can be found in \citet{huang2018}.

To compute the optimal weights and biases that yield the optimal
$\mathcal{B}_i$ that minimise \eq{eq:sur_B}, we use the Adam algorithm
\citep{kingma2017adam}.
This minimisation process requires all operations to be continuous and
differentiable.
To achieve that, we compute the mutual information using a kernel density estimator;
and the piecewise-defined functions $\mathcal{B}_i^0$ are made $\mathcal{C}^1$
continuous by applying the logistic function,
\begin{equation*}
	\mathcal{B}_i^0(\q_+;\var) = \mathcal{B}_i(\q_+) \sigma( k ( \var - \tilde{r}_{i-1} ) ) \sigma ( k (\tilde{r}_i - \var ) ),
\end{equation*}
where the parameter $k>0$ can be chosen to control the steepness of the function, and $\tilde{r}_{j} = r_j \pm \log ( p / (p-1) ) / k$, which
ensures $\mathcal{B}_i^0 = p \mathcal{B}_i$ at the boundaries.
 
In the present study, the first term in \eq{eq:problem_scalar} is
normalised with $\| \var \|^2$ and the second term is normalised with
$H(\var_I, \var_R)/2$.  Under this normalization, free
parameters $p=0.99$ and $k=500$ were determined to be adequate for
the optimization process. The number of bijective functions, $N$,
was selected to minimize \eq{eq:sur_B} while producing a
continuous mapping, as illustrated in Figs.~\ref{fig:val} and~\ref{fig:funcs}. 
In the study presented in Section~\ref{sec:results}, an $N$ value of 1 was found 
to be optimal. 
We also explored different values for the regularization
constant $\gamma$. For $N = 1$, similar mappings were achieved for
$0.5 \geq \gamma \geq 2$, and the results discussed in
Section~\ref{sec:results} were calculated with $\gamma = 1$. In
cases with $N \geq 1$, starting with a high $\gamma$ value,
approximately $10$, during initial iterations proved beneficial
for converging the solution. Subsequently, $\gamma$ was gradually
decreased to emphasise the minimisation of the first term in
\eq{eq:sur_B}. Currently, this adjustment is performed
manually, but future developments in aIND could automate this
process~\citep{groenendijk2021}. 
Finally, the DSF architecture was set to 3 layers with 12 neurons per layer.

%%%%%%%%%%%%%%%%%%%%%%%%%%%%%%%%%%%%%%%%%%%%%%%%%%%%%%%%%%%%%%%%%%%%%%%%%%%%%%%%
% VALIDATION AIND, LSE, EPOD
%%%%%%%%%%%%%%%%%%%%%%%%%%%%%%%%%%%%%%%%%%%%%%%%%%%%%%%%%%%%%%%%%%%%%%%%%%%%%%%%
\section{Validation of aIND and comparison with EPOD and
LSE}\label{app:validation}
\begin{figure}
  \centering
  \begin{tikzpicture}
    \node[anchor=north west] (f1) at (0,0) 
    {\ig[width=.8\tw,trim=0 .5cm 0 0,clip]{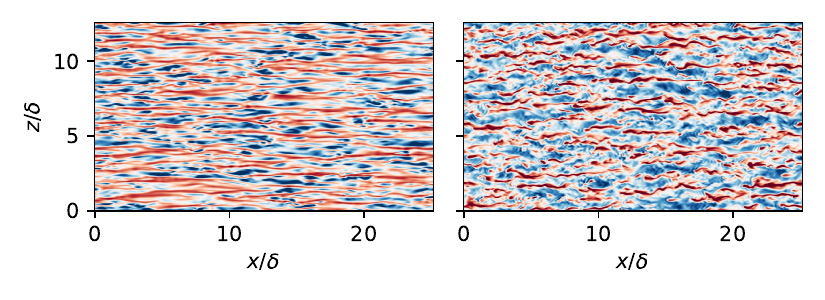}};
    \begin{scope}[x=(f1.north east),y=(f1.south west)]
      \node[below] at (.33,0) {$\Phi_I$};
      \node[below] at (.77,0) {$\Phi_R$};
      \node[above,fill=white] at (.33,1) {$x / \chh$};
      \node[above,fill=white] at (.77,1) {$x / \chh$};
      \node[rotate=90,fill=white] at (.042,.45) {$z / h$};
    \end{scope}
  \end{tikzpicture}
    \caption{Instantaneous velocities used to create synthetic
        examples: (left) informative field and (right) residual
        field. The velocities are extracted from a DNS of a turbulent
        channel flow at $\textit{Re}_\tau = 180$ at $y^+ \approx 5$
        and $40$, respectively. \label{fig:orig}}
\end{figure}
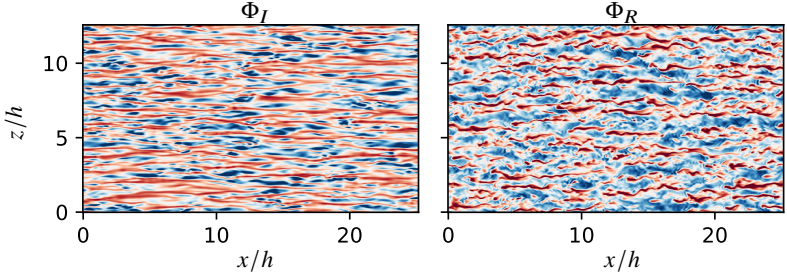

We include two additional validation cases of aIND applied to 2D
fields in a plane $\vec{x} = (x,z)$. 
These synthetic examples have an exact analytic solution
which enables to quantify the error produced by the different methods.
We consider the system:
\begin{align}
  &\text{source:}\quad \Phi(\vec{x},t) = \Phi_I(\vec{x},t) + \Phi_R(\vec{x},t),\\
  &\text{target:}\quad \Psi_+(\vec{x},t) = F(\Phi_I(\vec{x},t)),
\end{align}
where the fields ${\Phi}_I$ and ${\Phi}_R$ and the function $F$ are
given.  In particular, ${\Phi}_I$ and ${\Phi}_R$ are the velocity
fluctuations in the planes $y^* \approx 5$ and $40$, respectively, of
a turbulent channel flow with $\textit{Re}_\tau = 180$ in a domain
$8\pi\delta \times 2\delta \times 4\pi\delta$ in the streamwise,
wall-normal and spanwise directions, respectively.  Instantaneous
snapshots of the fields are shown in figure~\ref{fig:orig}.  To ensure
that the fields are independent (i.e.: $I({\Phi}_I, {\Phi}_R) = 0$),
the informative field is extracted at $y^* \approx 5$ from the bottom
wall, whereas the residual field is extracted at $y^* \approx 40$ from
the top wall at a shifted time step.

We compare aIND with the extended POD method (EPOD) proposed by
\citet{boree2003} and the spectral in space version of the LSE
presented by \citet{encinar2019}.  In the following sections we
provide a small overview of each method.

\subsection*{Extended POD}

The EPOD offers a linear decomposition of a source field,
$\Phi(\vec{x},t)$ into its correlated (C) and decorrelated (D)
contribution to a given target field such that
\begin{align}
  \Phi_C(\vec{x},t) &= \sum_n a^n_\Psi (t) U^n_\Phi(\vec{x}),\\
  \Phi_D(\vec{x},t) &= \Phi(\vec{x},t) - \Phi_C(\vec{x},t),
\end{align}
where $n$ is the number of modes, $a^n_\Psi$ is the temporal
coefficient of the $n$-th POD mode of the target field, $\Psi_+$, and
$U^n_\Phi$ in the $n$-th spatial mode.
The latter is computed as:
\begin{equation}
  U_\Phi^n(\vec{x}) = \frac{\langle a^n_\Psi
    \Phi(\vec{x},t)\rangle}{\langle a^n_\Psi a^n_\Psi \rangle},
\end{equation}
where the brackets denote temporal average. The EPOD decomposition has
the following properties~\citep{boree2003}:
\begin{itemize}
  \item the correlation between the original source field and the target field
    is the same as the correlation between the correlated field and the target
    field, namely
    \begin{equation*}
      \langle \Phi \Psi \rangle = \langle \Phi_C \Psi \rangle;
    \end{equation*}
  \item the decorrelated field is uncorrelated with the target field, i.e.,
    \begin{equation*}
      \langle \Phi_D \Psi \rangle = 0.
    \end{equation*}
\end{itemize}

Therefore, we define the EPOD informative component as the correlated
field ($\Phi_I^\text{EPOD} \equiv \Phi_C$) and EPOD residual component
as the decorrelated field ($\Phi_R^\text{EPOD} \equiv \Phi_D$).
In the following examples, the POD of the target field is obtained
using 300 snapshots and the informative field is reconstructed using
the 50 more energetic modes.

\subsection*{Spectral LSE}

The LSE, proposed by \citet{adrian1988}, provides the best mean square
linear estimate of the `response' field $\Phi(\vec{x},t)$ given the
`predictor' $\Psi_+(\vec{x},t)$~\citep{tinney2006}. 
Considering a collection of discrete spatial locations
$\vec{x}_i$,   
the  best linear estimate that minimises
\begin{equation}\label{eq:LSE}
  \arg \min_{\tilde{\Phi}}
  \langle (\Phi(\vec{x}_i,t) - \tilde{\Phi}(\vec{x}_i,t))^2 \rangle,
\end{equation}
is given by
\begin{equation}
  \tilde{\Phi}(\vec{x}_i, t) = L_{ij} \Psi_+(\vec{x}_j,t),
\end{equation}
where repeated indices implies summation.
%$L$ is a matrix whose elements are 
The entries of the matrix $L$ take the form~\citep{adrian1988}: 
\begin{equation}\label{eq:LSE_L}
  L_{ij} = \frac{ \langle \Phi(\vec{x}_i,t) \Psi(\vec{x}_m,t) \rangle_t }
  {\langle \Psi(\vec{x}_j,t) \Psi(\vec{x}_m,t) \rangle_t},
\end{equation}
where $\langle \cdot \rangle_t$ denotes temporal average.

From Eq.~\eqref{eq:LSE}, we define the LSE informative and residual
components as $\Phi_I^\text{LSE}(\vec{x},t) \equiv
\tilde{\Phi}(\vec{x},t)$ and $\Phi_R^\text{LSE}(\vec{x},t) \equiv
\Phi- \tilde{\Phi}(\vec{x},t)$, respectively.

In the following examples, we exploit the spatial periodicity of the
flow field. To that end, we adopt the approach by \citet{encinar2019}
and use a spatial Fourier basis to project the fields.  This procedure
is usually known as spectral linear stochastic estimation (SLSE).
Equation~\eqref{eq:LSE_L} becomes:
\begin{equation}\label{eq:SLSE}
  \widehat{L}(k_x, k_z) = \frac{\langle \widehat{\Psi}(k_x,
  k_z,t)\widehat{\Phi}^\dagger(k_x, k_z,t)
  \rangle_t}{\langle \widehat{\Phi}(k_x, k_z,t)\widehat{\Phi}^*(k_x,
  k_z,t)\rangle_t },
\end{equation}
where~$\widehat{(\cdot)}$~denotes the Fourier transform, $(\cdot)^\dagger$ is
the complex conjugate, and $k_x$, $k_z$ are the wave numbers in the $x$, $z$ 
directions, respectively.
It can be shown (see~\citet{tinney2006,encinar2019}) that optimal
estimator is
\begin{equation}
  {\Phi}_I(\vec{x}_i,t)  = \sum_{j} L(x_i -
  x_j, z_i - z_j)\Psi(x_j,z_j,t)
\end{equation}

%-------------------------------------%

\subsection*{Linear mapping}

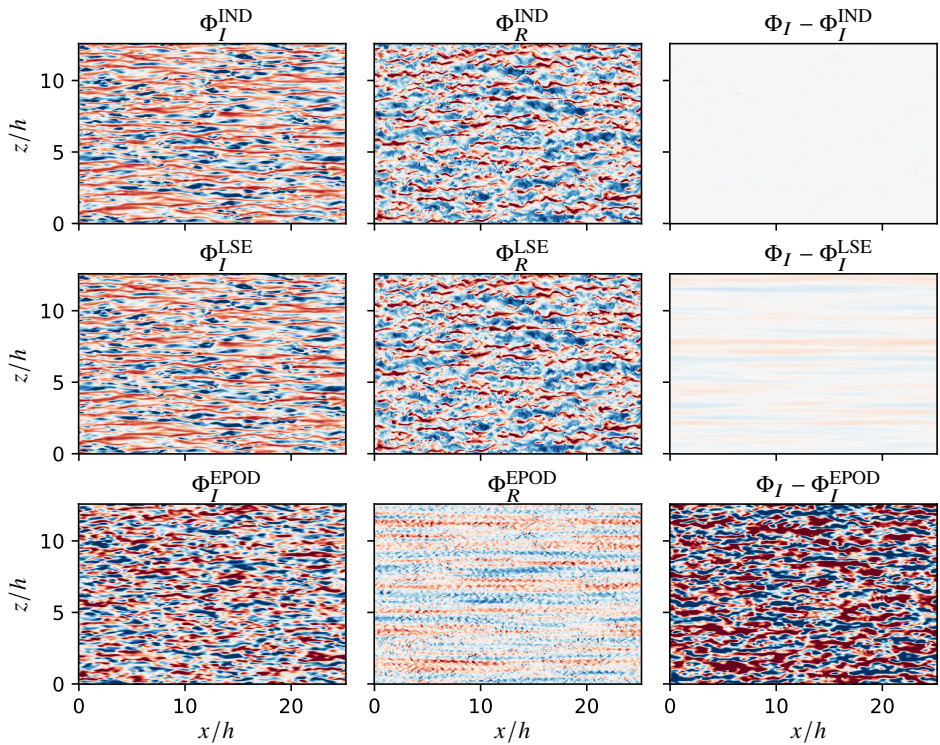
\begin{figure}
  \centering
  \begin{tikzpicture}
    \node[anchor=north west] (f1) at (0,0) 
    {\ig[width=.95\tw,trim=0 1.3cm 0 0,clip]{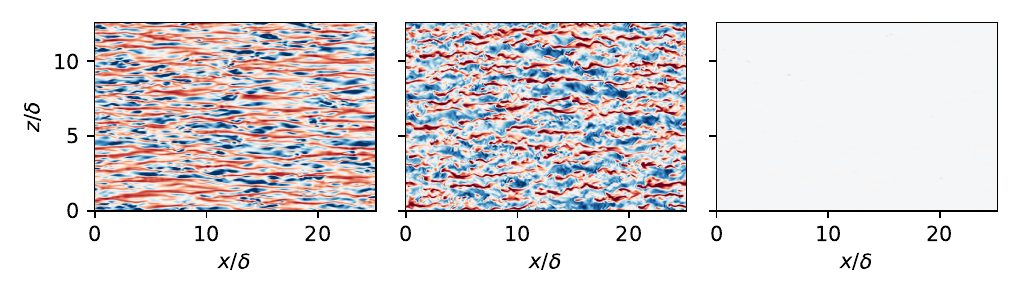}};
    \begin{scope}[x=(f1.north east),y=(f1.south west)]
      \node at (.25,.05) {$\Phi_I^{\text{IND}}$};
      \node at (.55,.05) {$\Phi_R^{\text{IND}}$};
      \node at (.85,.05) {$\Phi_I - \Phi_I^{\text{IND}}$};
      \node at (.04,.53) [rotate=90,fill=white] {$z/\chh$};
    \end{scope}
    \node[anchor=north west] (f2) at (f1.south west) 
    {\ig[width=.95\tw,trim=0 1.3cm 0 0,clip]{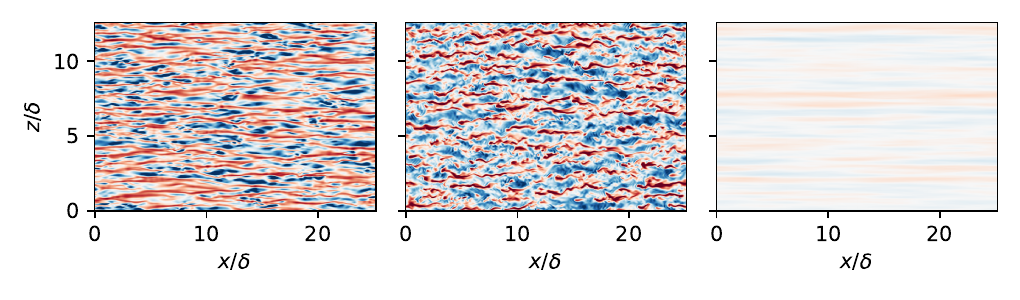}};
    \begin{scope}[shift=(f2.north west),x=(f2.north east),y=(f2.south west)]
      \node at (.25,.05) {$\Phi_I^{\text{LSE}}$};
      \node at (.55,.05) {$\Phi_R^{\text{LSE}}$};
      \node at (.85,.05) {$\Phi_I - \Phi_I^{\text{LSE}}$};
      \node at (.04,.53) [rotate=90,fill=white] {$z/\chh$};
    \end{scope}
    \node[anchor=north west] (f3) at (f2.south west) 
    {\ig[width=.95\tw]{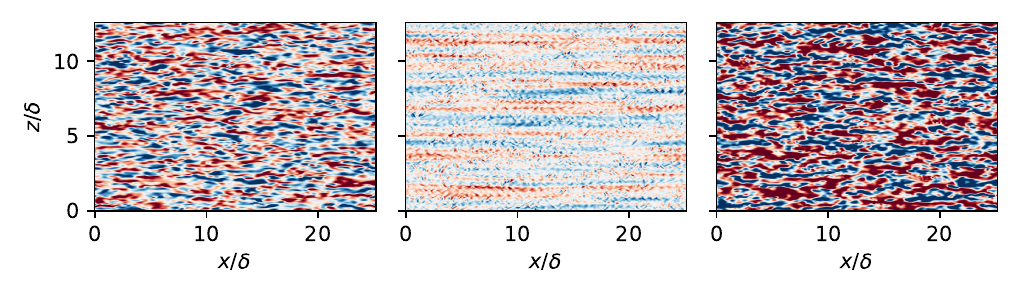}};
    \begin{scope}[shift=(f3.north west),x=(f3.north east),y=(f3.south west)]
      \node at (.25,.04) {$\Phi_I^{\text{EPOD}}$};
      \node at (.55,.04) {$\Phi_R^{\text{EPOD}}$};
      \node at (.85,.04) {$\Phi_I - \Phi_I^{\text{EPOD}}$};
      \node at (.04,.40) [rotate=90,fill=white] {$z/\chh$};
      \node at (.24,.85) [fill=white] {$x/\chh$};
      \node at (.54,.85) [fill=white] {$x/\chh$};
      \node at (.84,.85) [fill=white] {$x/\chh$};
    \end{scope}
  \end{tikzpicture}
    \caption{Instantaneous reconstructed fields of the
        informative and residual components for
        Eq~\eqref{eq:linear}. From top to bottom: aIND, SLSE and
        EPOD. For all rows, left column displays the reconstructed
        informative field, middle column displays the reconstructed
        residual field, and right column displays the error between
        the exact and the reconstructed informative
        field.\label{fig:linear}}
\end{figure}

As a first validation case, we consider a linear mapping function:
\begin{subequations}\label{eq:linear}
\begin{align}
  &\text{source:}\quad \Phi(\vec{x},t) = \Phi_I(\vec{x},t) + \Phi_R(\vec{x},t),\\
  &\text{target:}\quad \Psi_+(\vec{x},t) = \Phi_I(\vec{x},t)
\end{align}
\end{subequations}
The exact informative and residual fields are normalised such that
their standards deviations are $\langle \Phi_I \Phi_I \rangle = 1$,
and $\langle \Phi_R \Phi_R \rangle = 1$, respectively.  The
instantaneous reconstructed fields are displayed in
figure~\ref{fig:linear}.  To ease the comparison, the time instant is
the same as in figure~\ref{fig:orig}.

We can observe that aIND accurately reconstruct the informative and
residual fields.  SLSE is also able to reconstruct the mapping,
something expected since the mapping is linear.  On the contrary, EPOD
fails to obtain the correct informative/residual field despite the
linear character of the decomposition.  Instead, it tends to
reconstruct the original field, $\Phi$.

\subsection*{Non-linear mapping}

As a second validation case, we consider the non-linear mapping
function:
\begin{subequations}\label{eq:quat}
  \begin{align}
    &\text{source:}\quad \Phi(\vec{x},t) = \Phi_I(\vec{x},t) +
    \Phi_R(\vec{x},t),\label{eq:quatS} \\
    &\text{target:}\quad \Psi_+(\vec{x},t) = \Phi_I^2(\vec{x},t) - 0.2\Phi_I(\vec{x},t)\label{eq:quatT}
  \end{align}
\end{subequations}

\begin{figure}
  \centering
  \begin{tikzpicture}
    \node[anchor=north west] (f1) at (0,0) 
    {\ig[width=.95\tw,trim=0 1.3cm 0 0,clip]{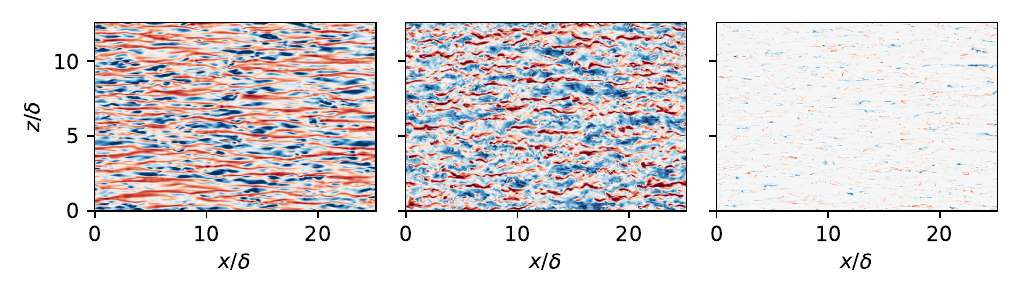}};
    \begin{scope}[x=(f1.north east),y=(f1.south west)]
      \node at (.25,.05) {$\Phi_I^{\text{IND}}$};
      \node at (.55,.05) {$\Phi_R^{\text{IND}}$};
      \node at (.85,.05) {$\Phi_I - \Phi_I^{\text{IND}}$};
      \node at (.04,.53) [rotate=90,fill=white] {$z/\chh$};
    \end{scope}
    \node[anchor=north west] (f2) at (f1.south west) 
    {\ig[width=.95\tw,trim=0 1.3cm 0 0,clip]{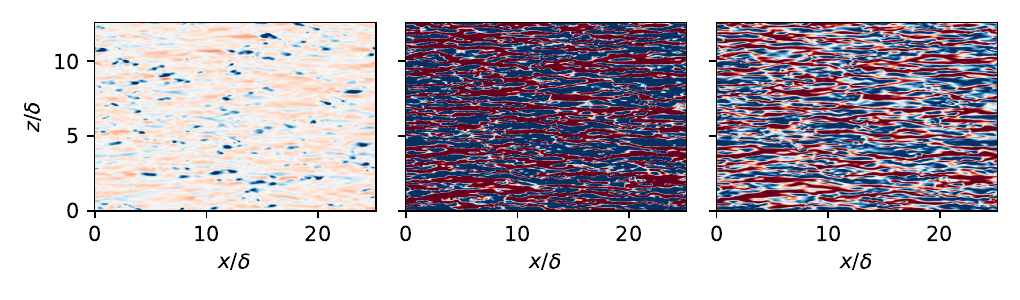}};
    \begin{scope}[shift=(f2.north west),x=(f2.north east),y=(f2.south west)]
      \node at (.25,.05) {$\Phi_I^{\text{LSE}}$};
      \node at (.55,.05) {$\Phi_R^{\text{LSE}}$};
      \node at (.85,.05) {$\Phi_I - \Phi_I^{\text{LSE}}$};
      \node at (.04,.53) [rotate=90,fill=white] {$z/\chh$};
    \end{scope}
    \node[anchor=north west] (f3) at (f2.south west) 
    {\ig[width=.95\tw]{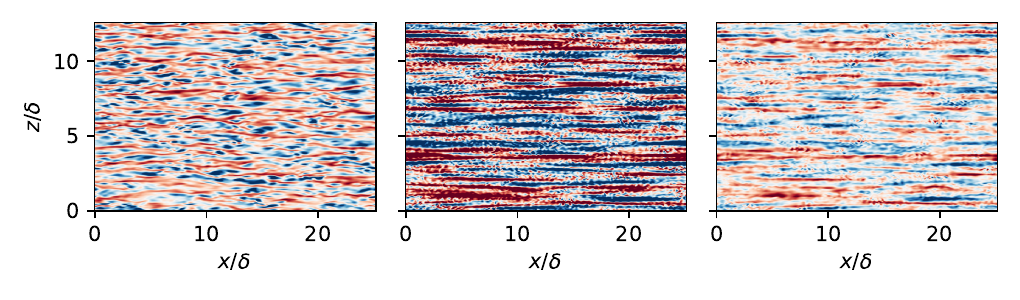}};
    \begin{scope}[shift=(f3.north west),x=(f3.north east),y=(f3.south west)]
      \node at (.25,.04) {$\Phi_I^{\text{EPOD}}$};
      \node at (.55,.04) {$\Phi_R^{\text{EPOD}}$};
      \node at (.85,.04) {$\Phi_I - \Phi_I^{\text{EPOD}}$};
      \node at (.04,.40) [rotate=90,fill=white] {$z/\chh$};
      \node at (.24,.85) [fill=white] {$x/\chh$};
      \node at (.54,.85) [fill=white] {$x/\chh$};
      \node at (.84,.85) [fill=white] {$x/\chh$};
    \end{scope}
  \end{tikzpicture}
    \caption{Instantaneous reconstructed fields of the
        informative and residual components for
        Eq.~\eqref{eq:quat}. From top to bottom: aIND, SLSE and
        EPOD. For all rows, left column displays the reconstructed
        informative field, middle column displays the reconstructed
        residual field, and right column displays the error between
        the exact and the reconstructed informative
        field.\label{fig:quat}}
\end{figure}

The exact informative and residual fields are normalised such that
their standards deviations are $\langle \Phi_I \Phi_I \rangle = 1$,
and $\langle \Phi_R \Phi_R \rangle = 0.2$, respectively.  The
instantaneous reconstructed fields are displayed in
figure~\ref{fig:quat} at the same time instant as in
figure~\ref{fig:orig}.

In this case, SLSE fails to correctly split the flow into the
informative and residual fields. The same applies to EPOD: although
the reconstruction of the informative resembles the original (due to
higher correlation between the original and the informative terms from
Eq.~\eqref{eq:quatS}), the error $\Phi_I - \Phi_I^{\text{EPOD}}$ is
significant everywhere. A similar error is observed for the residual
field, which is not correctly identified by EPOD. The aIND, similar to
the previous example, accurately reconstructs the informative and
residual fields. The small discrepancies in $\Phi_I -
\Phi_I^{\text{IND}}$ occur at the locations where $\Phi_I \approx 0$
and stem from the approach followed to compute $\Phi_I^{\text{IND}}$. Note that aIND accurately
reconstructs the analytical mapping.
\begin{figure}
  \centering
  \ig[width=.9\tw]{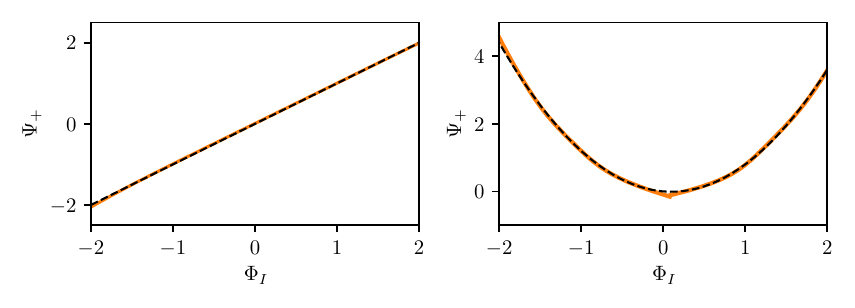}
    \caption{Mapping of (left) the linear example from
        Eq.~\eqref{eq:linear} and the (right) non-linear example from
        Eq.~\eqref{eq:quat}.  Dashed line corresponds to the analytic
        solution, solid line corresponds to the solution computed with
        aIND.\label{fig:mapping}}
\end{figure}

%%%%%%%%%%%%%%%%%%%%%%%%%%%%%%%%%%%%%%%%%%%%%%%%%%%%%%%%%%%%%%%%%%%%%%%%%%%%%%%%
%% METHODOLOGY - ADDITIONAL APPLICATIONS - GAUSSIAN
\section{Analytical solution for Gaussian distributions}\label{app:gaus}

\providecommand{\mat}[1]{\bm{\mathsf{#1}}}
\providecommand{\cM}[1]{\Sigma(#1)}
\providecommand{\ccM}[2]{\Sigma(#1 , #2)}
\providecommand{\csM}[2]{\Sigma(#1 \oplus #2)}
\subsection{Solution for Gaussian variables}

For the special case in which all the variables in $\Cau, \Q_+$ are jointly
normal distributed variables, we can write their mutual information
as~\citep{cover2006}:

\begin{equation}\label{eq:INDgaussI1}
    I(\Q_+; \Cau) = \frac{1}{2} \log \bk{ 
    \frac{|\cM{\Cau}| |\cM{\Q_+}|}{|\csM{\Q_+}{\Cau}|} }.
\end{equation}
In Eq.~\eqref{eq:INDgaussI1}, $\cM{\Cau}$ is the covariance matrix of $\Cau$ 
(and similarly for $\Q_+$), a square matrix whose $i,j$ entry is defined as:
\begin{equation*}
    \cM{\Cau}[i,j] = \langle \var_{I,i}(t) \var_{I,j}(t) \rangle_t
\end{equation*}
where $\var_{I,i}$ is the $i$-th element of $\Cau$.
The covariance matrix $\csM{\Q_+}{\Cau}$ can be written in block matrix form as: 
\begin{equation*}
    \csM{\Q_+}{\Cau} = \begin{bmatrix} \cM{\Q_+} & \ccM{ \Q_+}{\Cau} \\
    \ccM{\Q_+}{\Cau}^\top & \cM{\Cau} \end{bmatrix},
\end{equation*}
where $\ccM{\Q_+}{\Cau}$ is the cross-covariance matrix:
%$N_{\q} \times N_{\var}$ matrix 
%(where $N_{\q}$ is the number of points in $\Q_+$ and $N_{\var}$ is the number of
%points in $\Cau$):
%
\begin{equation*}
    \ccM{\Q_+}{\Cau}[i,j] = \langle \q_{+,i}(t) \var_{I,j}(t) \rangle_t.
\end{equation*}

The mutual information in Eq.~\eqref{eq:INDgaussI1} is maximized when
$|\csM{\Q_+}{\Cau}| = 0$, provided that $|\cM{\Cau}| \neq 0$.
Using the block determinant identity~\citep{johnson1985,barnett2009}:
\begin{subequations}
\begin{align}
    |\csM{\Q_+}{\Cau}| &= 
    |\cM{\Q_+}| | \cM{\Cau} - 
    \ccM{\Q_+}{\Cau}\cM{\Q_+}^{-1} \ccM{\Q_+}{\Cau}^\top | =\label{eq:covQ}\\
    &=
    |\cM{\Cau+}| | \cM{\Q_+} - 
    \ccM{\Q_+}{\Cau}^\top\cM{\Cau}^{-1} \ccM{\Q_+}{\Cau}|.\label{eq:covCau}
\end{align}
\end{subequations}
The second term in Eq.~\eqref{eq:covQ} [Eq.~\eqref{eq:covCau}] is the residual 
of a linear regression of $\Q_+$ on $\Cau$ [$\Cau$ on $\Q_+$]~\citep{barnett2009}.
Therefore, the mutual information in Eq.~\eqref{eq:INDgaussI1} is maximized when
$\Cau$ is a linear function of $\Q_+$ or vice-versa.
However, only when $\Q_+$ is a function of $\Cau$, $H( \Q_+ | \Cau ) = 0$, as
required by~Eq.~\eqref{eq:cond2}.

We assume that the number of elements in $\Var$, $N_\var$, is larger than that 
of $\Q_+$, $N_\q$,
so that, if we find % a $N_\var \times N_\q$ matrix:
\begin{equation}\label{eq:indgauss_map}
    \Cau = \mat{M} \Q_+, 
\end{equation}
We can find the inverse mapping
\begin{equation*}
    \Q_+ = \mat{M}^{-1} \Cau.
\end{equation*}

The mutual information in Eq.~\eqref{eq:cond1} can be expanded as 
\begin{equation}\label{eq:INDgaussR1}
    I(\Q_+; \Res) = \frac{1}{2} \log \bk{ 
    \frac{|\cM{\Res}| |\cM{\Q_+}|}{|\csM{\Q_+}{\Res}|} }
\end{equation}
which will be equal to zero for $|\csM{\Q_+}{\Res}| = 
|\cM{\Q_+}| |\cM{\Res}|$.
From the block determinant identity, this requires
\begin{equation}
    | \ccM{\Q_+}{\Res} \cM{\Q_+}^{-1} \ccM{\Q_+}{\Res}^\top | = 0.
\end{equation}
In a general scenario this requires:
\begin{equation*}
    \ccM{\Q_+}{\Res}[i,j] = 0,
\end{equation*}
namely:
\begin{equation}\label{eq:INDLSE}
    \langle \q_{+,i} (\var_j - \mat{M}[j,m]\q_{+,m}) \rangle_t \equiv 
    \langle \q_{+,i} \var_j \rangle_t 
    - \langle \q_{+,i} \mat{M}[j,m]\q_{+,m} \rangle_t
    = 0,
\end{equation}
where repeated indices imply summation. 
The solution to Eq.~\eqref{eq:INDLSE} is given by \cite{adrian1988} and it
correspond to the LSE:
\begin{equation*}
    \mat{M}[j,m] =  \frac{\langle \q_{+,i}\var_j \rangle_t} {\langle \q_{+,j}
    \q_{+,m} \rangle_t }.
\end{equation*}

Therefore, for the special case in which all variables involved are jointly
normal distributed variables, the solution to {\method} is LSE.
From the previous, it is straightforward to prove that the solution to 
{\amethod} when $\var, \q_+$ are jointly distributed is given by:
\begin{equation*}
    \cau(t) = \frac{\langle \q_+ \var \rangle_t}{\langle \q_+ \q_+ \rangle_t
    }\q_+(t).
\end{equation*}

We conclude by emphasizing that the similarity between {\method} and
higher-order versions of LSE does not extend to the most likely case where 
all the variables are not jointly normal distributed. In this scenario,
higher-order versions of LSE attempt to obtain a better reconstruction of $\Var$
using $\Q_+$, which will not fulfil the condition $H(\Q_+| \Cau) = 0$, as
discussed in the last paragraph of Appendix~\ref{app:imp:bijective}.

\section{Computation of $\Delta \vec{x}^{\max}$ for the turbulent channel flow}
\label{sec:optDx}

The {\amethod} requires the value of $\Delta \vec{x}_\square^{\max} =
(\Delta x_{\square}^{\max}, \Delta z_{\square}^{\max})$ for each
informative component $\square = u$, $v$ and $w$.
To that end, we calculate their relative energy as a function of
$\Delta x$, $\Delta z$ and the wall-normal distance:
\begin{align*}
  E_I^u(\Delta x,\Delta z, y) &= \frac{\|u_I^2\|}{\|u^2\|}, &
  E_I^v(\Delta x,\Delta z, y) &= \frac{\|v_I^2\|}{\|v^2\|}, &
  E_I^w(\Delta x,\Delta z, y) &= \frac{\|w_I^2\|}{\|w^2\|}.
\end{align*}
The parametric sweep is performed using data a channel flow at $\Rey =
180$ in a computational domain of size $\pi\chh \times 2\chh \times
\pi/2$ in the streamwise, wall-normal, and spanwise direction,
respectively.

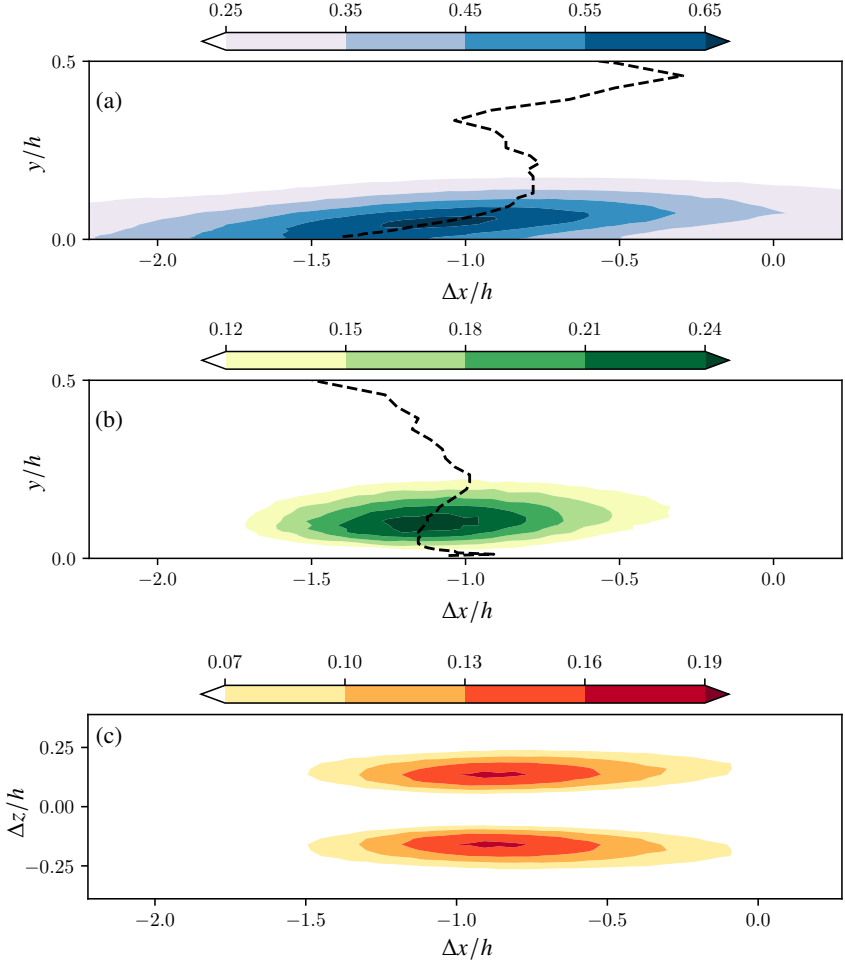
\begin{figure}
	\centering
        \begin{tikzpicture}
            \node[anchor=south west] (figu) at (0,0) 
            {\ig[width=.8\tw,trim=.39cm .30cm .37cm 0,clip]
            {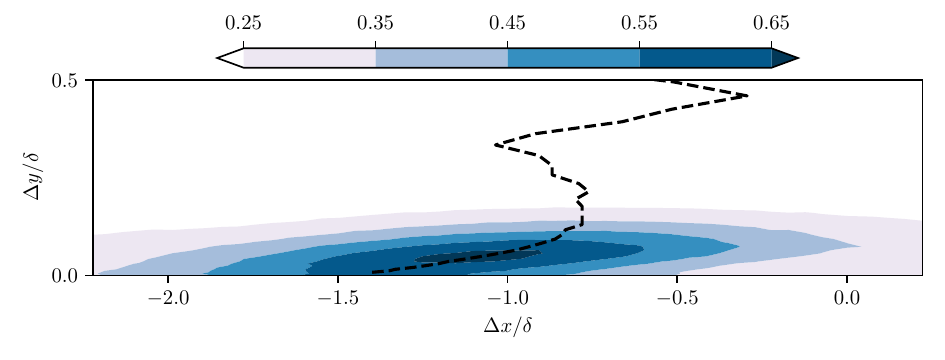}};
            \begin{scope}[shift={(figu.south west)},
                x=(figu.south east), y=(figu.north west)]
                \node at (.11,.63) {(a)};
                \node[fill=white] at (.54,.05) {$\Delta x / \chh $};
                \node[fill=white,rotate=90] at (.02,.47) {$y / \chh $};
            \end{scope}
        \phantomsubcaption
		\label{fig:ch_iso_u}
            \node[anchor=south west] (figv) at (0,-4.2cm)
            {\ig[width=.8\tw,trim=.39cm .30cm .37cm 0,clip]
            {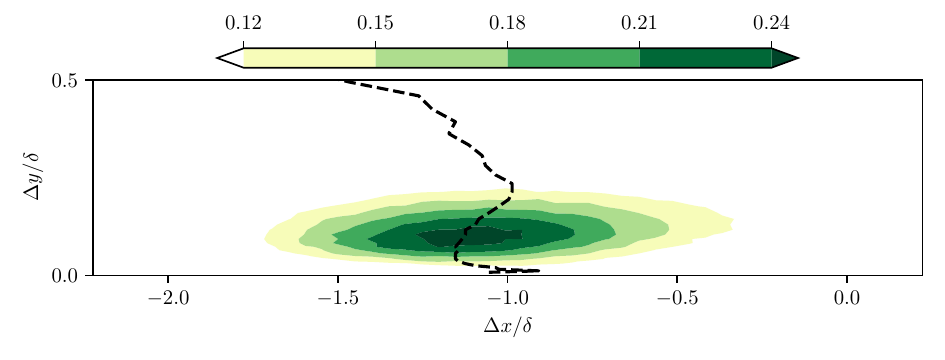}};
            \begin{scope}[shift={(figv.south west)},
                x=(figv.south east), y=(figv.north west)]
                \node at (.11,.63) {(b)};
                \node[fill=white] at (.54,.05) {$\Delta x / \chh $};
                \node[fill=white,rotate=90] at (.02,.47) {$y / \chh $};
            \end{scope}
        \phantomsubcaption
		\label{fig:ch_iso_v}
            \node[anchor=south west] (figw) at (0,-8.6cm) 
            {\ig[width=.8\tw,trim=.90cm .45cm .37cm 0,clip]
            {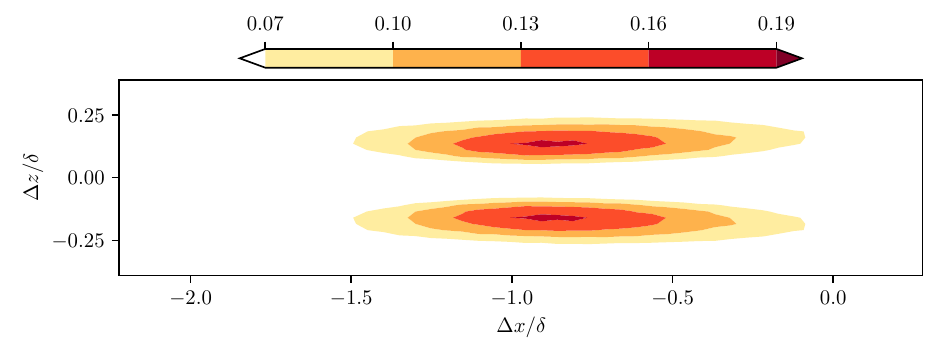}};
            \begin{scope}[shift={(figw.south west)},
                x=(figw.south east), y=(figw.north west)]
                \node at (.11,.68) {(c)};
                \node[fill=white] at (.54,+.04) {$\Delta x / \chh $};
                \node[fill=white,rotate=90] at (.0,.45) {$\Delta z / \chh $};
            \end{scope}
        \phantomsubcaption
		\label{fig:ch_iso_w}
        \end{tikzpicture}
	\caption{Informative regions quantifying the relative energy
          contained in the informative components $u_I$, $v_I$ and
          $w_I$ for $\Delta T^* = 25$.  (a) $E_I^u$ and (b) $E_I^v$ in
          the $\Delta z = 0$ plane; (c) $E_I^w$ in the plane
          $y^* \approx 6$.
          In (a,b) black dashed line corresponds to $\Delta
          x^{\max}(y)$.
          \label{fig:ch_iso}}
\end{figure}

\Fig{fig:ch_iso} displays $E_I^u$, $E_I^v$ and $E_I^w$ as functions of
$\Delta x$ and $\Delta z$.  Note that, due to the symmetry of the
flow, $E_I^{u}(\Delta x, \Delta z, y) = E_I^{u}(\Delta x, -\Delta z,
y)$ (similarly for $E_I^v$ and $E_I^w$).  For $E_I^u$ and $E_I^v$, the
maximum is always located at $\Delta z = 0$, which is the plane
displayed in \fig{fig:ch_iso_u,fig:ch_iso_v}.
For the spanwise component, the maximum value of $E_I^w$ is offset in
the spanwise direction and its location varies with $y$.
\Fig{fig:ch_iso_w} displays the horizontal section that contains its
global maximum, which is located at $y^* \approx 6$.
This offset is caused by the fact that $w$ motions travel in the
spanwise direction until they reach the wall and affect the wall shear
stress.

Close to the wall, we find high values of $E_I^u$, with a peak vale of
approximately $60\%$ at {$y^*\approx 8$}, and $\Delta x_u^{\max}(y)
\approx -\chh$, following an almost linear relationship with $y$.
Farther from the wall ($y > 0.2\chh$), $\Delta x_u^{\max}$ becomes
more or less constant, although it should be noted that, in this
region, the values of $E_I^u$ for a fixed $y$ are low and relatively
constant.
This may induce to some numerical uncertainty in the particular value
of $\Delta x_u^{\max}$, but the overall results are not affected.
In contrast, high values of $E_I^v$ are located in a compact region
further away from the wall ($y^*\approx 19$) and they tend to zero at the wall.
The values $\Delta x_v^{\max}(y)$ lies close to $-1.2\chh$ in this
region, following a negative linear relationship with $y$.
As before, $\Delta x_v^{\max}(y)$ remains relative constant in low
$E_I^v$ regions.
Finally, although not shown, $\Delta x_w^{\max}(y)$ and $\Delta
z_w^{\max}(y)$ lie in the interval $[-\chh, -0.7\chh]$ and
$\pm[0.1\chh, 0.2\chh]$, respectively, approaching zero at the wall.
Nevertheless, $E_w^I$ becomes negligible for $y > 0.2\chh$.

We close this section by noting that, although not explored in the present 
study, the $\Delta x^{\max}$ computed with aIND might correspond to potential 
locations for sensor placement, since it maximises the mutual information with 
the target variable \citep{lozanoduran2022}.

%%%%%%%%%%%%%%%%%%%%%%%%%%%%%%%%%%%%%%%%%%%%%%%%%%%%%%%%%%%%%%%%%%%%%%%%%%%%%%%%
%% APPENDIX - VALIDITY
\section{Validity of {\amethod} of $u$ with respect to $\taux$}
\label{app:validation_tauw}

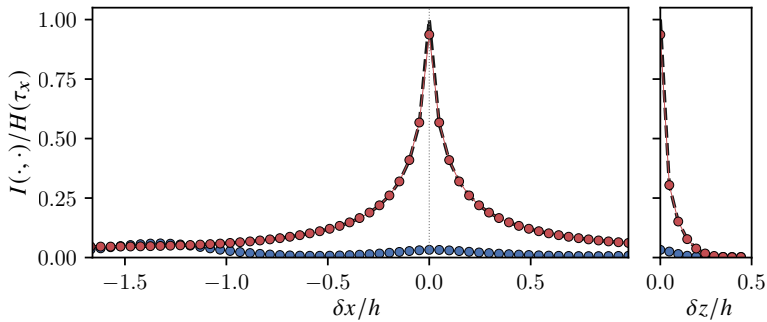
\begin{figure}
	\centering
    \begin{tikzpicture}
        \node[anchor=south west,inner sep=0pt]
        (fig) at (0,0) {\ig[width=.75\tw]{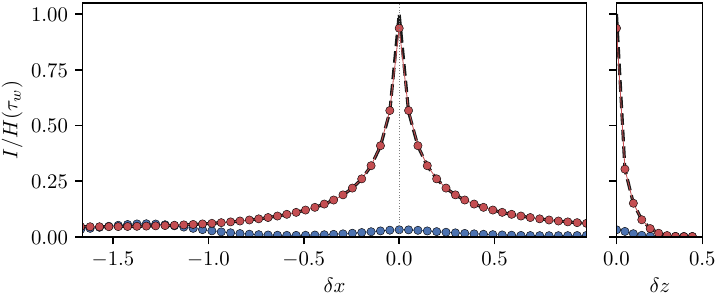}};
        \begin{scope}[x=(fig.south east), y=(fig.north west)]
            \node[fill=white] at (.46,.05) {$\delta x / \chh$};
            \node[fill=white] at (.92,.05) {$\delta z / \chh$};
            \node[fill=white,rotate=90] at (.02,.6) {$I(\cdot,\cdot)/H(\taux)$};
        \end{scope}
    \end{tikzpicture}
	\caption{Mutual information between the streamwise wall shear stress and
    (blue circles) the residual field, $I( u_R; \tauxf)(\delta \vec{x})$; 
    (red circles) the informative field$I( u_I; \tauxf)(\delta \vec{x})$.
    Dashed line correspond to $I(\tauxf;\tauxf)(\delta\vec{x})$;
	\label{fig:mutual_info}}
\end{figure}

% move to appendix
\Fig{fig:mutual_info} displays the mutual information between
$u_R(x_0, y_0, z_0)$ for $y_0^* \approx 10$, and $\tauxf(x_0-\Delta
x_u^{\max}-\delta x,z_0-\Delta z_u^{\max}-\delta z)$ as a function of
$\delta \vec{x} = [\delta x, \delta z]$, denoted as
$I(u_R;\tauxf)(\delta \vec{x})$.
The mutual information is normalised by the total Shannon
information of the wall shear stress, $H(\taux)$, such that $I( u_R;
\tauxf)(\delta \vec{x})/H(\taux) = 0$ means that $u_R$ contains
no information about the wall shear stress at $\delta \vec{x}$, and
$I( u_R; \tauxf)(\delta \vec{x})/H(\taux) = 1$ implies that $u_R$
contains all the information about $\tauxf(\delta \vec{x})$.  
Note that {\amethod} seeks to minimise $I(u_R;\tauxf)(\vec{0})$.
The results show that value of the $I( u_R; \tauxf)(\delta \vec{x})/H(\taux)$ 
remains always low, reaching a maximum of approximately 0.06 at 
$\delta x \approx -1.2\chh$ along the streamwise direction.  
Hence, we can conclude that the residual term contains a negligible amount 
of information about the wall shear stress at any point in the wall and 
{\amethod} is a valid approximation of {\method}.
For the sake of completeness, we also display in
\fig{fig:mutual_info} the mutual information between $u_I$ and
the wall shear stress.  Since $\tauxf = \mathcal{F}(u_I)$, the
mutual information $I( u_I; \tauxf)(\delta \vec{x})$ has to be
equal to $H(\taux)$ at $\delta \vec{x}=\boldsymbol{0}$, as
corroborated by the results. For larger distances, $I( u_I;
\tauxf)(\delta \vec{x})$ decays following the natural decay of 
$I( \tauxf; \tauxf)(\delta \vec{x})$, with values
below 0.1 after $|\delta \vec{x}| \approx \chh$.

\bibliographystyle{jfm}
\bibliography{bibliography}

\begin{thebibliography}{95}
\expandafter\ifx\csname natexlab\endcsname\relax\def\natexlab#1{#1}\fi
\def\au#1{#1} \def\ed#1{#1} \def\yr#1{#1}\def\at#1{#1}\def\jt#1{\textit{#1}}
  \def\bt#1{#1}\def\bvol#1{\textbf{#1}} \def\vol#1{#1} \def\pg#1{#1}
  \def\publ#1{#1}\def\arxiv#1{#1}\def\org#1{#1}\def\st#1{\textit{#1}}

\bibitem[Adrian(2007)]{adrian2007}
{\sc \au{Adrian, R.~J.}} \yr{2007}  \at{Hairpin vortex organization in wall
  turbulence}.  \jt{Phys. Fluids}  \bvol{19}~(4),  \pg{041301}.

\bibitem[Adrian {\em et~al.\/}(2000)Adrian, Meinhart \& Tomkins]{adrian2000}
{\sc \au{Adrian, R.~J.}, \au{Meinhart, C.~D.} \& \au{Tomkins, C.~D.}} \yr{2000}
   \at{Vortex organization in the outer region of the turbulent boundary
  layer}.  \jt{J. Fluid Mech.}  \bvol{422},  \pg{1--54}.

\bibitem[Adrian \& Moin(1988)]{adrian1988}
{\sc \au{Adrian, R.~J.} \& \au{Moin, P.}} \yr{1988}  \at{Stochastic estimation
  of organized turbulent structure: homogeneous shear flow}.  \jt{J. Fluid
  Mech.}  \bvol{190},  \pg{531--559}.

\bibitem[Baars \& Tinney(2014)]{baars2014}
{\sc \au{Baars, W.~J.} \& \au{Tinney, C.~E.}} \yr{2014}  \at{{Proper orthogonal
  decomposition-based spectral higher-order stochastic estimation}}.  \jt{Phys.
  Fluids}  \bvol{26}~(5),  \pg{055112}.

\bibitem[Bae \& Lee(2021)]{bae2021}
{\sc \au{Bae, H.~J.} \& \au{Lee, M.}} \yr{2021}  \at{Life cycle of streaks in
  the buffer layer of wall-bounded turbulence}.  \jt{Phys. Rev. Fluids}
  \bvol{6}~(6),  \pg{064603}.

\bibitem[Barnett {\em et~al.\/}(2009)Barnett, Barrett \& Seth]{barnett2009}
{\sc \au{Barnett, L.}, \au{Barrett, A.~B.} \& \au{Seth, A.~K.}} \yr{2009}
  \at{Granger causality and transfer entropy are equivalent for gaussian
  variables}.  \jt{Phys. Rev. Lett.}  \bvol{103},  \pg{238701}.

\bibitem[Betchov(1964)]{betchov1964}
{\sc \au{Betchov, R.}} \yr{1964}  \at{Measure of the intricacy of turbulence}.
  \jt{Phys. Fluids}  \bvol{7}~(8),  \pg{1160--1162}.

\bibitem[Blackwelder \& Kaplan(1976)]{blackwelder1976}
{\sc \au{Blackwelder, R.~F.} \& \au{Kaplan, R.~E.}} \yr{1976}  \at{On the wall
  structure of the turbulent boundary layer}.  \jt{J. Fluid Mech.}  \bvol{76},
  \pg{89--112}.

\bibitem[Bor\'ee(2003)]{boree2003}
{\sc \au{Bor\'ee, J.}} \yr{2003}  \at{Extended proper orthogonal decomposition:
  a tool to analyse correlated events in turbulent flows}.  \jt{Exp. Fluids}
  \bvol{35},  \pg{188--192}.

\bibitem[Brown \& Roshko(1974)]{brown1974}
{\sc \au{Brown, G.~L.} \& \au{Roshko, A.}} \yr{1974}  \at{On density effects
  and large structure in turbulent mixing layers}.  \jt{J. Fluid Mech.}
  \bvol{64}~(4),  \pg{775--816}.

\bibitem[Brunton {\em et~al.\/}(2020)Brunton, Noack \&
  Koumoutsakos]{brunton2019}
{\sc \au{Brunton, S.~L.}, \au{Noack, B.~R.} \& \au{Koumoutsakos, P.}} \yr{2020}
   \at{Machine learning for fluid mechanics}.  \jt{Annu. Rev. Fluid Mech.}
  \bvol{52}~(1),  \pg{477--508}.

\bibitem[Cerbus \& Goldburg(2013)]{cerbus2013}
{\sc \au{Cerbus, R.~T.} \& \au{Goldburg, W.~I.}} \yr{2013}  \at{Information
  content of turbulence}.  \jt{Phys. Rev. E}  \bvol{88},  \pg{053012}.

\bibitem[Cheng {\em et~al.\/}(2019)Cheng, Li, Lozano-Dur\'an \& Liu]{cheng2019}
{\sc \au{Cheng, C.}, \au{Li, W.}, \au{Lozano-Dur\'an, A.} \& \au{Liu, H.}}
  \yr{2019}  \at{Identity of attached eddies in turbulent channel flows with
  bidimensional empirical mode decomposition}.  \jt{J. Fluid Mech.}
  \bvol{870},  \pg{1037--1071}.

\bibitem[Choi {\em et~al.\/}(1994)Choi, Moin \& Kim]{choi1994}
{\sc \au{Choi, H.}, \au{Moin, P.} \& \au{Kim, J.}} \yr{1994}  \at{Active
  turbulence control for drag reduction in wall-bounded flows}.  \jt{J. Fluid
  Mech.}  \bvol{262},  \pg{75--110}.

\bibitem[Chung \& Talha(2011)]{chung2011}
{\sc \au{Chung, Y.~M.} \& \au{Talha, T.}} \yr{2011}  \at{Effectiveness of
  active flow control for turbulent skin friction drag reduction}.  \jt{Phys.
  Fluids}  \bvol{23}~(2),  \pg{025102}.

\bibitem[Corrsin \& Kistler(1954)]{corrsin1954}
{\sc \au{Corrsin, S.} \& \au{Kistler, A.~L.}} \yr{1954}  \bt{The free-stream
  boundaries of turbulent flows}. NACA Technical Note TN-3133.  \org{National
  Advisory Committee for Aeronautics}, adv. Conf. Rep. 3123.

\bibitem[Cover \& Thomas(2006)]{cover2006}
{\sc \au{Cover, T.~M.} \& \au{Thomas, J.~A.}} \yr{2006} {\em Elements of
  information theory\/}, 2nd edn.  \publ{Wiley}.

\bibitem[Del~{\'A}lamo {\em et~al.\/}(2006)Del~{\'A}lamo, Jim\'{e}nez,
  Zandonade \& Moser]{alamo2006}
{\sc \au{Del~{\'A}lamo, J.~C.}, \au{Jim\'{e}nez, J.}, \au{Zandonade, P.} \&
  \au{Moser, R.~D.}} \yr{2006}  \at{Self-similar vortex clusters in the
  turbulent logarithmic region}.  \jt{J. Fluid Mech.}  \bvol{561},
  \pg{329--358}.

\bibitem[Deshpande {\em et~al.\/}(2021)Deshpande, Monty \&
  Marusic]{deshpande2021}
{\sc \au{Deshpande, R.}, \au{Monty, J.~P.} \& \au{Marusic, I.}} \yr{2021}
  \at{Active and inactive components of the streamwise velocity in wall-bounded
  turbulence}.  \jt{J. Fluid Mech.}  \bvol{914},  \pg{A5}.

\bibitem[Encinar \& Jim\'enez(2019)]{encinar2019}
{\sc \au{Encinar, M.~P.} \& \au{Jim\'enez, J.}} \yr{2019}
  \at{Logarithmic-layer turbulence: A view from the wall}.  \jt{Phys. Rev.
  Fluids}  \bvol{4},  \pg{114603}.

\bibitem[Erichson {\em et~al.\/}(2020)Erichson, Mathelin, Yao, Brunton, Mahoney
  \& Kutz]{erichson2020}
{\sc \au{Erichson, N.~B.}, \au{Mathelin, L.}, \au{Yao, Z.}, \au{Brunton,
  S.~L.}, \au{Mahoney, M.~W.} \& \au{Kutz, J.~N.}} \yr{2020}  \at{Shallow
  neural networks for fluid flow reconstruction with limited sensors}.
  \jt{Proc. R. Soc. A}  \bvol{476}~(2238),  \pg{20200097}.

\bibitem[Farrell \& Ioannou(2012)]{farrell2012}
{\sc \au{Farrell, B.~F.} \& \au{Ioannou, P.~J.}} \yr{2012}  \at{Dynamics of
  streamwise rolls and streaks in turbulent wall-bounded shear flow}.  \jt{J.
  Fluid Mech.}  \bvol{708},  \pg{149--196}.

\bibitem[Ghaemi \& Scarano(2013)]{ghaemi2013}
{\sc \au{Ghaemi, S.} \& \au{Scarano, F.}} \yr{2013}  \at{Turbulent structure of
  high-amplitude pressure peaks within the turbulent boundary layer}.  \jt{J.
  Fluid Mech.}  \bvol{735},  \pg{381--426}.

\bibitem[Granero-Belinchon(2018)]{granero-belinchon_thesis}
{\sc \au{Granero-Belinchon, C.}} \yr{2018}  \at{{Multiscale Information
  Transfer in Turbulence}}. Theses, {Universit{\'e} de Lyon}.

\bibitem[Groenendijk {\em et~al.\/}(2021)Groenendijk, Karaoglu, Gevers \&
  Mensink]{groenendijk2021}
{\sc \au{Groenendijk, R.}, \au{Karaoglu, S.}, \au{Gevers, T.} \& \au{Mensink,
  T.}} \yr{2021} Multi-loss weighting with coefficient of variations.  \bt{In
  {\em Proceedings of the IEEE/CVF Winter Conference on Applications of
  Computer Vision (WACV)\/}},  \pg{pp. 1469--1478}.

\bibitem[Guastoni {\em et~al.\/}(2021)Guastoni, G\"{u}emes, Ianiro, Discetti,
  Schlatter, Azizpour \& Vinuesa]{guastoni2021}
{\sc \au{Guastoni, L.}, \au{G\"{u}emes, A.}, \au{Ianiro, A.}, \au{Discetti,
  S.}, \au{Schlatter, P.}, \au{Azizpour, H.} \& \au{Vinuesa, R.}} \yr{2021}
  \at{Convolutional-network models to predict wall-bounded turbulence from wall
  quantities}.  \jt{J. Fluid Mech.}  \bvol{928},  \pg{A27}.

\bibitem[Guerrero {\em et~al.\/}(2020)Guerrero, Lambert \& Chin]{guerrero2020}
{\sc \au{Guerrero, B.}, \au{Lambert, M.~F.} \& \au{Chin, R.~C.}} \yr{2020}
  \at{Extreme wall shear stress events in turbulent pipe flows: spatial
  characteristics of coherent motions}.  \jt{J. Fluid Mech.}  \bvol{904},
  \pg{A18}.

\bibitem[Hammond {\em et~al.\/}(1998)Hammond, Bewley \& Moin]{hammond1998}
{\sc \au{Hammond, E.~P.}, \au{Bewley, T.~R.} \& \au{Moin, P.}} \yr{1998}
  \at{Observed mechanisms for turbulence attenuation and enhancement in
  opposition-controlled wall-bounded flows}.  \jt{Phys. Fluids}  \bvol{10}~(9),
   \pg{2421--2423}.

\bibitem[Hornik {\em et~al.\/}(1989)Hornik, Stinchcombe \& White]{hornik1989}
{\sc \au{Hornik, K.}, \au{Stinchcombe, M.} \& \au{White, H.}} \yr{1989}
  \at{Multilayer feedforward networks are universal approximators}.  \jt{Neural
  Networks}  \bvol{2}~(5),  \pg{359--366}.

\bibitem[Huang {\em et~al.\/}(2018)Huang, Krueger, Lacoste \&
  Courville]{huang2018}
{\sc \au{Huang, C.-W.}, \au{Krueger, D.}, \au{Lacoste, A.} \& \au{Courville,
  A.}} \yr{2018} Neural autoregressive flows.  \bt{In {\em Proceedings of the
  35th International Conference on Machine Learning\/}},  \st{Proceedings of
  Machine Learning Research},  \vol{vol.~80},  \pg{pp. 2078--2087}.
  \publ{PMLR}.

\bibitem[Huang {\em et~al.\/}(1998)Huang, Shen, Long, Wu, Shih, Zheng, Yen,
  Tung \& Liu]{huang1998}
{\sc \au{Huang, N.~E.}, \au{Shen, Z.}, \au{Long, S.~R.}, \au{Wu, M.~C.},
  \au{Shih, H.~H.}, \au{Zheng, Q.}, \au{Yen, N.-C.}, \au{Tung, C.~C.} \&
  \au{Liu, H.~H.}} \yr{1998}  \at{The empirical mode decomposition and the
  hilbert spectrum for nonlinear and non-stationary time series analysis}.
  \jt{Proc. R. Soc. A}  \bvol{454}~(1971),  \pg{903--995}.

\bibitem[Hwang \& Sung(2018)]{hwang2018}
{\sc \au{Hwang, J.} \& \au{Sung, H.~J.}} \yr{2018}  \at{Wall-attached
  structures of velocity fluctuations in a turbulent boundary layer}.  \jt{J.
  Fluid Mech.}  \bvol{856},  \pg{958--983}.

\bibitem[Iwamoto {\em et~al.\/}(2002)Iwamoto, Suzuki \& Kasagi]{iwamoto2002}
{\sc \au{Iwamoto, K.}, \au{Suzuki, Y.} \& \au{Kasagi, N.}} \yr{2002}
  \at{Reynolds number effect on wall turbulence: toward effective feedback
  control}.  \jt{Int. J. Heat Fluid Flow}  \bvol{23}~(5),  \pg{678--689}.

\bibitem[Jim\'{e}nez(2018)]{jimenez2018}
{\sc \au{Jim\'{e}nez, J.}} \yr{2018}  \at{Coherent structures in wall-bounded
  turbulence}.  \jt{J. Fluid Mech.}  \bvol{842},  \pg{P1}.

\bibitem[Jim\'enez \& Hoyas(2008)]{jimenez2008}
{\sc \au{Jim\'enez, J.} \& \au{Hoyas, S.}} \yr{2008}  \at{Turbulent
  fluctuations above the buffer layer of wall-bounded flows}.  \jt{J. Fluid
  Mech.}  \bvol{611},  \pg{215--236}.

\bibitem[Johansson {\em et~al.\/}(1987)Johansson, Her \&
  Haritonidis]{johansson1987}
{\sc \au{Johansson, A.~V.}, \au{Her, J.-Y.} \& \au{Haritonidis, J.~H.}}
  \yr{1987}  \at{On the generation of high-amplitude wall-pressure peaks in
  turbulent boundary layers and spots}.  \jt{J. Fluid Mech.}  \bvol{175},
  \pg{119--142}.

\bibitem[Johnson \& Horn(1985)]{johnson1985}
{\sc \au{Johnson, C.~R.} \& \au{Horn, R.~A.}} \yr{1985} {\em Matrix
  analysis\/}.

\bibitem[Jovanovi{\'c} \& Bamieh(2005)]{jovanovic2005}
{\sc \au{Jovanovi{\'c}, M.~R.} \& \au{Bamieh, B.}} \yr{2005}  \at{Componentwise
  energy amplification in channel flows}.  \jt{J. Fluid Mech.}  \bvol{534},
  \pg{145--183}.

\bibitem[Kaiser \& Schreiber(2002)]{kaiser2002}
{\sc \au{Kaiser, A.} \& \au{Schreiber, T.}} \yr{2002}  \at{Information transfer
  in continuous processes}.  \jt{Physica D}  \bvol{166}~(1),  \pg{43--62}.

\bibitem[Kim(1985)]{kim1985}
{\sc \au{Kim, J.}} \yr{1985}  \at{Turbulence structures associated with the
  bursting event}.  \jt{Phys. Fluids}  \bvol{28},  \pg{52--58}.

\bibitem[Kim {\em et~al.\/}(1987)Kim, Moin \& Moser]{kim1987}
{\sc \au{Kim, J.}, \au{Moin, P.} \& \au{Moser, R.}} \yr{1987}  \at{Turbulence
  statistics in fully developed channel flow at low reynolds number}.  \jt{J.
  Fluid Mech.}  \bvol{177},  \pg{133--166}.

\bibitem[Kingma \& Ba(2017)]{kingma2017adam}
{\sc \au{Kingma, D.~P.} \& \au{Ba, J.}} \yr{2017} Adam: A method for stochastic
  optimization,  \arxiv{arXiv: 1412.6980}.

\bibitem[Kline {\em et~al.\/}(1967)Kline, Reynolds, Schraub \&
  Runstadler]{kline1967}
{\sc \au{Kline, S.~J.}, \au{Reynolds, W.~C.}, \au{Schraub, F.~A.} \&
  \au{Runstadler, P.~W.}} \yr{1967}  \at{The structure of turbulent boundary
  layers}.  \jt{J. Fluid Mech.}  \bvol{30}~(4),  \pg{741--773}.

\bibitem[Kutz {\em et~al.\/}(2016)Kutz, Fu \& Brunton]{kutz2016}
{\sc \au{Kutz, J.~N.}, \au{Fu, X.} \& \au{Brunton, S.~L.}} \yr{2016}
  \at{Multiresolution dynamic mode decomposition}.  \jt{SIAM J. App. Dyn. Sys.}
   \bvol{15}~(2),  \pg{713--735}.

\bibitem[Le~Clainche \& Vega(2017)]{leclainche2017}
{\sc \au{Le~Clainche, S.} \& \au{Vega, J.~M.}} \yr{2017}  \at{Higher order
  dynamic mode decomposition}.  \jt{SIAM J. App. Dyn. Sys.}  \bvol{16}~(2),
  \pg{882--925}.

\bibitem[Lee(2021)]{lee2021}
{\sc \au{Lee, T.-W.}} \yr{2021}  \at{Scaling of the maximum-entropy turbulence
  energy spectra}.  \jt{Eur. J. Mech. B Fluids}  \bvol{87},  \pg{128--134}.

\bibitem[Liang \& Lozano-Dur{\'a}n(2016)]{liang2016}
{\sc \au{Liang, X.~S.} \& \au{Lozano-Dur{\'a}n, A.}} \yr{2016}  \at{A
  preliminary study of the causal structure in fully developed near-wall
  turbulence}.  \jt{CTR - Proc. Summer Prog.}  \pg{pp. 233--242}.

\bibitem[Long {\em et~al.\/}(2015)Long, Shelhamer \& Darrell]{long2015}
{\sc \au{Long, J.}, \au{Shelhamer, E.} \& \au{Darrell, T.}} \yr{2015} Fully
  convolutional networks for semantic segmentation.  \bt{In {\em Proceedings of
  the IEEE Conference on Computer Vision and Pattern Recognition (CVPR)\/}}.

\bibitem[Lozano-Dur\'an \& Arranz(2022)]{lozanoduran2022}
{\sc \au{Lozano-Dur\'an, A.} \& \au{Arranz, G.}} \yr{2022}
  \at{Information-theoretic formulation of dynamical systems: Causality,
  modeling, and control}.  \jt{Phys. Rev. Res.}  \bvol{4},  \pg{023195}.

\bibitem[Lozano-Dur{\'a}n {\em et~al.\/}(2019)Lozano-Dur{\'a}n, Bae \&
  Encinar]{lozano2019b}
{\sc \au{Lozano-Dur{\'a}n, A.}, \au{Bae, H.~J.} \& \au{Encinar, M.~P.}}
  \yr{2019}  \at{Causality of energy-containing eddies in wall turbulence}.
  \jt{J. Fluid Mech.}  \bvol{882},  \pg{A2}.

\bibitem[Lozano-Dur{\'a}n {\em et~al.\/}(2012)Lozano-Dur{\'a}n, Flores \&
  Jim{\'e}nez]{lozano2012}
{\sc \au{Lozano-Dur{\'a}n, A.}, \au{Flores, O.} \& \au{Jim{\'e}nez, J.}}
  \yr{2012}  \at{The three-dimensional structure of momentum transfer in
  turbulent channels}.  \jt{J. Fluid Mech.}  \bvol{694},  \pg{100--130}.

\bibitem[Lozano-Dur\'an {\em et~al.\/}(2020)Lozano-Dur\'an, Giometto, Park \&
  Moin]{lozano2020}
{\sc \au{Lozano-Dur\'an, A.}, \au{Giometto, M.~G.}, \au{Park, G.~I.} \&
  \au{Moin, P.}} \yr{2020}  \at{Non-equilibrium three-dimensional boundary
  layers at moderate { Reynolds} numbers}.  \jt{J. Fluid Mech.}  \bvol{883},
  \pg{A20}.

\bibitem[Lozano-Dur\'an \& Jim\'enez(2014)]{lozano2014}
{\sc \au{Lozano-Dur\'an, A.} \& \au{Jim\'enez, J.}} \yr{2014}
  \at{Time-resolved evolution of coherent structures in turbulent channels:
  characterization of eddies and cascades}.  \jt{J. Fluid Mech.}  \bvol{759},
  \pg{432--471}.

\bibitem[Luhar {\em et~al.\/}(2014)Luhar, Sharma \& McKeon]{luhar2014}
{\sc \au{Luhar, M.}, \au{Sharma, A. S.} \& \au{McKeon, B. J.}} \yr{2014}
  \at{On the structure and origin of pressure fluctuations in wall turbulence:
  predictions based on the resolvent analysis}.  \jt{J. Fluid Mech.}
  \bvol{751},  \pg{38--70}.

\bibitem[Lumley(1967)]{lumley1967}
{\sc \au{Lumley, J.~L.}} \yr{1967}  \at{The structure of inhomogeneous
  turbulent flows}.  \jt{Atmospheric Turbulence and Radio Wave Propagation}
  \pg{pp. 166--178}.

\bibitem[Mart{\'\i}nez-S{\'a}nchez {\em
  et~al.\/}(2023)Mart{\'\i}nez-S{\'a}nchez, L{\'o}pez, Le~Clainche,
  Lozano-Dur{\'a}n, Srivastava \& Vinuesa]{martinez2023}
{\sc \au{Mart{\'\i}nez-S{\'a}nchez, A.}, \au{L{\'o}pez, E.}, \au{Le~Clainche,
  S.}, \au{Lozano-Dur{\'a}n, A.}, \au{Srivastava, A.} \& \au{Vinuesa, R.}}
  \yr{2023}  \at{Causality analysis of large-scale structures in the flow
  around a wall-mounted square cylinder}.  \jt{J. Fluid Mech.}  \bvol{967},
  \pg{A1}.

\bibitem[Materassi {\em et~al.\/}(2014)Materassi, Consolini, Smith \&
  De~Marco]{materassi2014}
{\sc \au{Materassi, M.}, \au{Consolini, G.}, \au{Smith, N.} \& \au{De~Marco,
  R.}} \yr{2014}  \at{Information theory analysis of cascading process in a
  synthetic model of fluid turbulence}.  \jt{Entropy}  \bvol{16}~(3),
  \pg{1272--1286}.

\bibitem[McKeon(2017)]{mckeon2017}
{\sc \au{McKeon, B.~J.}} \yr{2017}  \at{The engine behind (wall) turbulence:
  perspectives on scale interactions}.  \jt{J. Fluid Mech.}  \bvol{817},
  \pg{P1}.

\bibitem[McKeon \& Sharma(2010)]{mckeon2010}
{\sc \au{McKeon, B.~J.} \& \au{Sharma, A.~S.}} \yr{2010}  \at{A critical-layer
  framework for turbulent pipe flow}.  \jt{J. Fluid Mech.}  \bvol{658},
  \pg{336--382}.

\bibitem[Meinhart \& Adrian(1995)]{meinhart1995}
{\sc \au{Meinhart, C.~D.} \& \au{Adrian, R.~J.}} \yr{1995}  \at{On the
  existence of uniform momentum zones in a turbulent boundary layer}.
  \jt{Phys. Fluids}  \bvol{7}~(4),  \pg{694--696}.

\bibitem[Mezi\'{c}(2013)]{mezic2013}
{\sc \au{Mezi\'{c}, I.}} \yr{2013}  \at{Analysis of fluid flows via spectral
  properties of the koopman operator}.  \jt{Annu. Rev. Fluid Mech.}
  \bvol{45}~(1),  \pg{357--378}.

\bibitem[Moin \& Moser(1989)]{moin1989}
{\sc \au{Moin, P.} \& \au{Moser, R.~D.}} \yr{1989}  \at{Characteristic-eddy
  decomposition of turbulence in a channel}.  \jt{J. Fluid Mech.}  \bvol{200},
  \pg{471--509}.

\bibitem[Moisy \& Jim{\'e}nez(2004)]{moisy2004}
{\sc \au{Moisy, F.} \& \au{Jim{\'e}nez, J.}} \yr{2004}  \at{Geometry and
  clustering of intense structures in isotropic turbulence}.  \jt{J. Fluid
  Mech.}  \bvol{513},  \pg{111--133}.

\bibitem[Murray \& Ukeiley(2003)]{murray2003}
{\sc \au{Murray, N.} \& \au{Ukeiley, L.~S.}} \yr{2003}  \at{Estimation of the
  flow field from surface pressure measurements in an open cavity}.  \jt{{AIAA}
  J.}  \bvol{41}~(5),  \pg{969--972}.

\bibitem[Naguib {\em et~al.\/}(2001)Naguib, Wark \&
  Juckenh\"{o}fel]{naguib2001}
{\sc \au{Naguib, A.~M.}, \au{Wark, C.~E.} \& \au{Juckenh\"{o}fel, O.}}
  \yr{2001}  \at{{Stochastic estimation and flow sources associated with
  surface pressure events in a turbulent boundary layer}}.  \jt{Phys. Fluids}
  \bvol{13}~(9),  \pg{2611--2626}.

\bibitem[Panton(2001)]{panton2001}
{\sc \au{Panton, R.~L.}} \yr{2001}  \at{Overview of the self-sustaining
  mechanisms of wall turbulence}.  \jt{Prog. Aerosp. Sci.}  \bvol{37}~(4),
  \pg{341--383}.

\bibitem[Reynolds(1895)]{reynolds1895}
{\sc \au{Reynolds, O.}} \yr{1895}  \at{Iv. on the dynamical theory of
  incompressible viscous fluids and the determination of the criterion}.
  \jt{Philos. Trans. R. Soc. A}  \bvol{186},  \pg{123--164}.

\bibitem[Robinson(1991)]{robinson1991}
{\sc \au{Robinson, S.~K.}} \yr{1991}  \at{Coherent motions in the turbulent
  boundary layer}.  \jt{Annu. Rev. Fluid Mech.}  \bvol{23}~(1),  \pg{601--639}.

\bibitem[Rowley {\em et~al.\/}(2009)Rowley, Mezi\'{c}, Bagheri, Schlatter \&
  Henningson]{rowley2009}
{\sc \au{Rowley, C.~W.}, \au{Mezi\'{c}, I.}, \au{Bagheri, S.}, \au{Schlatter,
  P.} \& \au{Henningson, D.~S.}} \yr{2009}  \at{Spectral analysis of nonlinear
  flows}.  \jt{J. Fluid Mech.}  \bvol{641},  \pg{115--127}.

\bibitem[Schewe(1983)]{schewe1983}
{\sc \au{Schewe, G.}} \yr{1983}  \at{On the structure and resolution of
  wall-pressure fluctuations associated with turbulent boundary-layer flow}.
  \jt{J. Fluid Mech.}  \bvol{134},  \pg{311--328}.

\bibitem[Schmid(2010)]{schmid2010}
{\sc \au{Schmid, P.~J.}} \yr{2010}  \at{Dynamic mode decomposition of numerical
  and experimental data}.  \jt{J. Fluid Mech.}  \bvol{656},  \pg{5--28}.

\bibitem[Schmid(2022)]{schmid2022}
{\sc \au{Schmid, P.~J.}} \yr{2022}  \at{Dynamic mode decomposition and its
  variants}.  \jt{Annu. Rev. Fluid Mech.}  \bvol{54}~(1),  \pg{225--254}.

\bibitem[Schmid {\em et~al.\/}(2011)Schmid, Li, Juniper \& Pust]{schmid2011}
{\sc \au{Schmid, P.~J.}, \au{Li, L.}, \au{Juniper, M.~P.} \& \au{Pust, O.}}
  \yr{2011}  \at{Applications of the dynamic mode decomposition}.  \jt{Theor.
  Comput. Fluid Dyn.}  \bvol{25},  \pg{249--259}.

\bibitem[Schmidt(2020)]{schmidt2020bi}
{\sc \au{Schmidt, O.~T.}} \yr{2020}  \at{Bispectral mode decomposition of
  nonlinear flows}.  \jt{Nonlin. Dyn.}  \bvol{102},  \pg{2479--2501}.

\bibitem[Schmidt \& Schmid(2019)]{schmidt2019}
{\sc \au{Schmidt, O.~T.} \& \au{Schmid, P.~J.}} \yr{2019}  \at{A conditional
  space–time {POD} formalism for intermittent and rare events: example of
  acoustic bursts in turbulent jets}.  \jt{J. Fluid Mech.}  \bvol{867},
  \pg{R2}.

\bibitem[Shannon(1948)]{shannon1948}
{\sc \au{Shannon, C.~E.}} \yr{1948}  \at{A mathematical theory or
  communication}.  \jt{Bell Syst. Tech. J.}  \bvol{27}~(379--423),
  \pg{623--656}.

\bibitem[Shavit \& Falkovich(2020)]{shavit2020}
{\sc \au{Shavit, M.} \& \au{Falkovich, G.}} \yr{2020}  \at{Singular measures
  and information capacity of turbulent cascades}.  \jt{Phys. Rev. Lett.}
  \bvol{125},  \pg{104501}.

\bibitem[Sillero {\em et~al.\/}(2014)Sillero, Jim\'{e}nez \&
  Moser]{sillero2014}
{\sc \au{Sillero, J.~A.}, \au{Jim\'{e}nez, J.} \& \au{Moser, R.~D.}} \yr{2014}
  \at{Two-point statistics for turbulent boundary layers and channels at
  reynolds numbers up to $\delta^+ \approx 2000$}.  \jt{Phys. Fluids}
  \bvol{26}~(10),  \pg{105109}.

\bibitem[de~Silva {\em et~al.\/}(2016)de~Silva, Hutchins \& Marusic]{silva2016}
{\sc \au{de~Silva, C.~M.}, \au{Hutchins, N.} \& \au{Marusic, I.}} \yr{2016}
  \at{Uniform momentum zones in turbulent boundary layers}.  \jt{J. Fluid
  Mech.}  \bvol{786},  \pg{309--331}.

\bibitem[Sirovich(1987)]{sirovich1987}
{\sc \au{Sirovich, L.}} \yr{1987}  \at{Turbulence and the dynamics of coherent
  structures. {P}art {I}. {C} oherent structures}.  \jt{Quart. Appl. Math.}
  \bvol{45},  \pg{561--571}.

\bibitem[Smits {\em et~al.\/}(2011)Smits, McKeon \& Marusic]{smits2011}
{\sc \au{Smits, A.~J.}, \au{McKeon, B.~J.} \& \au{Marusic, I.}} \yr{2011}
  \at{High-{R}eynolds number wall turbulence}.  \jt{Annu. Rev. Fluid Mech.}
  \bvol{43}~(1),  \pg{353--375}.

\bibitem[Taira {\em et~al.\/}(2017)Taira, Brunton, Dawson, Rowley, Colonius,
  McKeon, Schmidt, Gordeyev, Theofilis \& Ukeiley]{taira2017}
{\sc \au{Taira, K.}, \au{Brunton, S.~L.}, \au{Dawson, S. T.~M.}, \au{Rowley,
  C.~W.}, \au{Colonius, T.}, \au{McKeon, B.~J.}, \au{Schmidt, O.~T.},
  \au{Gordeyev, S.}, \au{Theofilis, V.} \& \au{Ukeiley, L.~S.}} \yr{2017}
  \at{Modal analysis of fluid flows: An overview}.  \jt{AIAA J.}
  \bvol{55}~(12),  \pg{4013--4041}.

\bibitem[Taira {\em et~al.\/}(2020)Taira, Hemati, Brunton, Sun, Duraisamy,
  Bagheri, Dawson \& Yeh]{taira2020}
{\sc \au{Taira, K.}, \au{Hemati, M.~S.}, \au{Brunton, S.~L.}, \au{Sun, Y.},
  \au{Duraisamy, K.}, \au{Bagheri, S.}, \au{Dawson, S. T.~M.} \& \au{Yeh,
  C.-A.}} \yr{2020}  \at{Modal analysis of fluid flows: {A}pplications and
  outlook}.  \jt{AIAA J.}  \bvol{58}~(3),  \pg{998--1022}.

\bibitem[Tanogami \& Araki(2024)]{tanogami2024}
{\sc \au{Tanogami, T.} \& \au{Araki, R.}} \yr{2024}
  \at{Information-thermodynamic bound on information flow in turbulent
  cascade}.  \jt{Phys. Rev. Res.}  \bvol{6},  \pg{013090}.

\bibitem[Theodorsen(1952)]{theodorsen1952}
{\sc \au{Theodorsen, T.}} \yr{1952} Mechanisms of turbulence.  \bt{In {\em
  Proceedings of the $2^{nd}$ Midwestern Conference on Fluid Mechanics,
  1952\/}}.

\bibitem[Tinney {\em et~al.\/}(2006)Tinney, Coiffet, Delville, Hall, Jordan \&
  Glauser]{tinney2006}
{\sc \au{Tinney, C.~E.}, \au{Coiffet, F.}, \au{Delville, J.}, \au{Hall, A.~M.},
  \au{Jordan, P.} \& \au{Glauser, M.~N.}} \yr{2006}  \at{On spectral linear
  stochastic estimation}.  \jt{Exp. Fluids}  \bvol{41}~(5),  \pg{763--775}.

\bibitem[Towne {\em et~al.\/}(2018)Towne, Schmidt \& Colonius]{towne2018}
{\sc \au{Towne, A.}, \au{Schmidt, O.~T.} \& \au{Colonius, T.}} \yr{2018}
  \at{Spectral proper orthogonal decomposition and its relationship to dynamic
  mode decomposition and resolvent analysis}.  \jt{J. Fluid Mech.}  \bvol{847},
   \pg{821--867}.

\bibitem[Townsend(1961)]{townsend1961}
{\sc \au{Townsend, A.~A.}} \yr{1961}  \at{Equilibrium layers and wall
  turbulence}.  \jt{J. Fluid Mech.}  \bvol{11}~(1),  \pg{97--120}.

\bibitem[Trefethen {\em et~al.\/}(1993)Trefethen, Trefethen, Reddy \&
  Driscoll]{trefethen1993}
{\sc \au{Trefethen, L.~N.}, \au{Trefethen, A.~E.}, \au{Reddy, S.~C.} \&
  \au{Driscoll, T.~A.}} \yr{1993}  \at{Hydrodynamic stability without
  eigenvalues}.  \jt{Science}  \bvol{261}~(5121),  \pg{578--584}.

\bibitem[Wallace(2016)]{wallace2016}
{\sc \au{Wallace, J.~M.}} \yr{2016}  \at{Quadrant analysis in turbulence
  research: history and evolution}.  \jt{Annu. Rev. Fluid Mech.}  \bvol{48},
  \pg{131--158}.

\bibitem[Wallace {\em et~al.\/}(1972)Wallace, Eckelman \& Brodkey]{wallace1972}
{\sc \au{Wallace, J.~M.}, \au{Eckelman, H.} \& \au{Brodkey, R.~S.}} \yr{1972}
  \at{The wall region in turbulent shear flow}.  \jt{J. Fluid Mech.}
  \bvol{54},  \pg{39--48}.

\bibitem[Wang {\em et~al.\/}(2021)Wang, Chu, Lozano-Dur\'an, Helmig \&
  Weigand]{wang2021}
{\sc \au{Wang, W.}, \au{Chu, X.}, \au{Lozano-Dur\'an, A.}, \au{Helmig, R.} \&
  \au{Weigand, B.}} \yr{2021}  \at{Information transfer between turbulent
  boundary layers and porous media}.  \jt{J. Fluid Mech.}  \bvol{920},
  \pg{A21}.

\bibitem[Williams {\em et~al.\/}(2015)Williams, Kevrekidis \&
  Rowley]{williams2015}
{\sc \au{Williams, M.~O.}, \au{Kevrekidis, I.~G.} \& \au{Rowley, C.~W.}}
  \yr{2015}  \at{A {D}ata-{D}riven approximation of the {K}oopman operator: {E}
  xtending {D}ynamic {M}ode {D}ecomposition}.  \jt{J. Nonlinear Sci.}
  \bvol{25},  \pg{1307--1346}.

\bibitem[Yuan \& Lozano-Dur\'an(2024)]{yuan2024}
{\sc \au{Yuan, Y.} \& \au{Lozano-Dur\'an, A.}} \yr{2024}  \at{Limits to extreme
  event forecasting in chaotic systems}.  \jt{Physica D}  \bvol{467},
  \pg{134246}.

\bibitem[Zaki \& Wang(2021)]{zaki2021}
{\sc \au{Zaki, T.~A.} \& \au{Wang, M.}} \yr{2021}  \at{From limited
  observations to the state of turbulence: Fundamental difficulties of flow
  reconstruction}.  \jt{Phys. Rev. Fluids}  \bvol{6},  \pg{100501}.

\end{thebibliography}

\end{document}